\documentclass{aa}  
\usepackage{txfonts, rotate,psfig,subfigure,supertabular,lscape,times, graphicx}
\usepackage{txfonts}
\begin{document}
   \title{Non-radial motion in the TeV blazar S5 0716+714}

   \subtitle{The pc-scale kinematics of a BL Lac Object}

   \author{S. Britzen\inst{1}, V.A. Kam\inst{1}\fnmsep\thanks{Change of name: Veronika Meyer changed her name to Veronika Kam by marriage}, A. Witzel\inst{1}, I. Agudo\inst{2,1}, M.F. Aller\inst{3}, H.D. Aller\inst{3}, M. Karouzos\inst{1}\fnmsep\thanks{Member of the International Max Planck Research School (IMPRS) for Radio and Infrared Astronomy at the Universities of Bonn and Cologne}, A. Eckart\inst{4}, and J.A. Zensus\inst{1}}
   \institute{Max-Planck-Institut f\"ur Radioastronomie, Auf dem H\"ugel 69, 53121 Bonn, Germany\\ \email{sbritzen@mpifr-bonn.mpg.de} \and Instituto de Astrof\'{\i}sica de Andaluc\'{\i}a (CSIC), Apartado 3004, E-18080 Granada, Spain \and Astronomy Department, University of Michigan, Ann Arbor, MI 48109-1042, USA \and I. Physikalisches Institut K\"oln, Universit\"at zu K\"oln, Z\"ulpicher Str. 77, 50937 K\"oln, Germany}
   \date{Received \today; accepted \today}
\authorrunning{S. Britzen}
  \abstract
   {Flat-spectrum radio sources often show a core-jet structure on pc-scales. Individual jet components reveal predominantly outward directed motion. For the BL Lac object S5 0716+714 conflicting apparent velocities have been reported in the literature. This object is an intra-day variable source and suited to investigate a possible correlation between kinematic properties and flux-density variability on different timescales.}
   {We study the kinematics in the pc-scale jet of S5 0716+714 to determine the apparent speeds of the jet components based on a much improved data set. In addition, we search for correlations between the radio flux-density light curves and the morphological changes detected along the VLBI jet.}
   {We (re-)analyze 50 VLBI observations obtained with the VLBA at 5 different frequencies (5 -- 43 GHz) between 1992.73 and 2006.32. The data have been parameterized using circular Gaussian components. We analyze the jet component motion in detail taking care not only to account for motion in the radial but also in the orthogonal direction. We study the evolution of the jet ridge line and investigate the spectral properties of the individual components. We search for correlations between radio band light curves and the kinematic properties of the jet components.}
   {We present an alternative kinematic scenario for jet component motion in S5 0716+714. We present evidence for the apparent stationarity of jet components (with regard to their core separation) with time. Jet components, however,  do seem to move significantly non-radially with regard to their position angle and in
   a direction perpendicular to the major axis of the jet. We discuss a possible correlation between the long-term radio flux-density variability and apparent jet component motions.}
   {In S5 0716+714 an alternative motion scenario is proposed. With regard to the core separation, rather stationary components can fit the VLBI observations well. A new model to explain the observed motion with regard to the position angle is required. Based on the correlation between the longterm radio flux-density variability and the position angle evolution of a jet component, we conclude that a geometric contribution to the origin of the long-term variability might not be negligible. Subluminal motion has been reported for most of the TeV blazars. Our analysis also confirms this finding for the case of S5 0716+714. This result increases the number of TeV blazars showing apparent subluminal motion to 7.}
   \maketitle

\begin{table*}[htb] 
\begin{center}
\caption[]{Details of the observations presented in this paper (in epochs marked by a star, S5 0716+714 served as calibrator for observations targeted on NRAO 150).}
\label{observations}
\begin{tabular}{|c|c|c|r|}
\hline
Epoch & $\nu$ [GHz] & Array/Survey & Reference \\
\hline
1992.73  &  5.0  & VLBA CJF-Survey  & Britzen et al. (2007)  \\
1992.85  & 22.2  & Global array & Bach et al. (2005) \\
1993.71  & 22.2  & Global array & Bach et al. (2005)  \\
1994.21  &  8.4  & VLBA     & Bach et al. (2005)\\
1994.67  & 15.3  & VLBA MOJAVE/2cm Survey & Kellermann et al. (1998); Zensus et al. (2002)\\
1994.70  &  5.0  & VLBA CJF-Survey & Britzen et al. (2007)  \\
1995.15  &  22.2 & VLBA & Jorstad et al. (2001)   \\
1995.47  &  22.2 & VLBA & Jorstad et al. (2001)    \\
1995.65  &  8.4, 22.2  & VLBA    & Bach et al. (2005)   \\
1996.34  &  22.2  & VLBA & Jorstad et al. (2001)  \\
1996.53  & 15.3  & VLBA MOJAVE/2cm Survey & Kellermann et al. (1998); Zensus et al. (2002)\\
1996.63  &  5.0  & VLBA/CJF-Survey & Britzen et al. (2007)  \\
1996.82  & 15.3  & VLBA MOJAVE/2cm Survey & Kellermann et al. (1998); Zensus et al. (2002)\\
1996.90  &  22.2  & VLBA & Jorstad et al. (2001)  \\
1997.03  &  8.4  & VLBA & Fey et al. (2000) \\
1997.58  &  22.2  & VLBA & Jorstad et al. (2001)  \\
1997.93  &  8.4  &  VLBA & Ros et al. (2001) \\
1999.41  &  8.4  &  VLBA & Ros et al. (2001) \\
1999.55  & 15.3  & VLBA MOJAVE/2cm Survey & Kellermann et al. (1998); Zensus et al. (2002)\\
1999.89  &  5.0  &  VLBA/CJF-Survey  &  Britzen et al. (2007) \\
2000.82  &  5.0  &  VLBA+Effelsberg  & Bach et al. (2005)\\
2001.17  & 15.3  & VLBA MOJAVE/2cm Survey & Kellermann  et al. (1998); Zensus et al. (2002)\\
2002.48  &  8.4, 22.2, 43.2 & VLBA$^{*}$  &  Agudo et al. (2007) \\
2003.20  &  8.4, 22.2, 43.2 & VLBA$^{*}$  &  Agudo et al. (2007) \\
2003.49  &  8.4, 22.2, 43.2 & VLBA$^{*}$  &  Agudo et al. (2007) \\
2003.65  & 15.3  & VLBA MOJAVE/2cm Survey  & Kellermann et al. (2004)\\ 
2003.88  & 22.2, 43.2  &      VLBA  &  Agudo et al. (A\&A, 2008 in prep.) \\
2004.44  & 15.3  & VLBA MOJAVE/2cm Survey & Kellermann et al. (2004)\\
2004.60  & 15.3, 22.2, 43.2&  VLBA$^{*}$  & Agudo et al. (2007)  \\
2004.79  & 15.3  & VLBA MOJAVE/2cm Survey  & Kellermann et al. (2004)\\
2004.97  & 15.3, 22.2, 43.2&  VLBA$^{*}$  & Agudo et al. (2007)  \\
2005.01  & 15.3  & VLBA MOJAVE/2cm Survey  & Kellermann et al. (2004) \\
2005.09  & 15.3, 22.2, 43.2 & VLBA$^{*}$  & Agudo et al. (2007)  \\
2005.70  & 15.3  & VLBA MOJAVE/2cm Survey  & Kellermann et al. (2004)\\
2005.72  & 15.3  & VLBA MOJAVE/2cm Survey  & Kellermann et al. (2004) \\
2006.32  & 15.3  & VLBA MOJAVE/2cm Survey  & Kellermann et al. (2004)\\
\hline
\end{tabular}    
\end{center}
\end{table*}

\section{Introduction}
The BL Lac Object S5 0716+714 (redshift 0.31$\pm$0.08, Nilsson et al. 2008) shows flux-density variability at all wavelength ranges that have been investigated so far (e.g., Wagner \& Witzel 1995). It is still the only Active Galactic Nuclei (AGN) for which a simultaneous change in the mode of the radio/optical variability has been observed on intra-day time scales (Quirrenbach et al. 1991) and is thus so far the best candidate for an intrinsic origin of intra-day variability (IDV, Wagner \& Witzel 1995, and references therein).\\  
Several authors claim the detection of periodicity or nearly periodic variations in radio and optical (Quirrenbach et al. 1991), optical (Gupta et al., 2008) and X-ray light-curves (Gupta et al., 2009). Apparently correlated optical/radio quasi-periods of $\sim$1 day lasting over several days have been observed by Quirrenbach et al. (1991). Gupta et al. (2008) report a rough period of $\sim 3.0\pm0.3$ years and Gupta et al. (2009) find quasi-periodic oscillations (QPOs) of $\sim$30 minutes. As a possible origin of the X-ray QPOs, shocks propagating along the jet have been discussed.
Most recently, S5 0716+714 has been detected in the TeV-regime by the MAGIC collaboration (Teshima et al. 2008, Anderhub et al. 2009).\\
S5 0716+714 has been investigated with Very Long Baseline Interferometry (VLBI) at various radio frequencies with different arrays. On parsec-scales it reveals a North-South directed jet with several embedded components (e.g., Eckart et al. 1987). On kpc-scales, the jet is more East-West oriented with some evidence for halo-emission (e.g., Antonucci et al. 1986). A rather large range in proper motions (0.05 mas/yr -- 1.11 mas/yr) has been reported by different authors based on the investigation of an increasing core separation with time modeled for jet components in this source (e.g., Eckart et al. 1987, Witzel et al. 1988, Schalinski et al. 1992, Gabuzda et al. 1998, Jorstad et al. 2001, P\'erez-Torres et al. 2004, Kellermann et al. 2004, Bach et al. 2005, Rastorgueva et al. 2008). Based on these results it is not clear, whether S5 0716+714 reveals slow apparent velocities typical for BL Lac objects (e.g., Britzen et al. 2007) or faster apparent velocities more typical for quasars (e.g., Kellermann et al. 2004).
\\ Based on the now available improved database with multi-frequency observations at 5 different frequencies, covering a time span of $\sim$14 years, we can address the question concerning the kinematics in the pc-scale jet of S5 0716+714 in more detail. This source belongs to the most rapidly variable radio sources (e.g., Wagner \& Witzel 1995) and we can - in addition to the kinematics - investigate whether these sources show evidence for a possible correlation between morphological changes and flux-density variability.\\
In an earlier publication of the results relying on the 15 GHz data only, we presented a description of preliminary results (Britzen et al. 2006). A detailed description of the data reduction, analysis, and investigation is presented in Meyer (2007).\\
The current paper is organized as follows: we first give an overview of the data that have been (re-)investigated; we then describe the model-fitting results and discuss the model-component kinematics; we present and discuss a new motion scenario; and we investigate a possible correlation between the total flux-density data and the VLBI kinematics, and concentrate in the discussion on the implications and resulting consequences.\\
Throughout this paper we use a Hubble constant of  H$_{0} = 71$km s$^{-1}$Mpc$^{-1}$. 

\section{The observations, data reduction and analysis}
\subsection{The observations}
S5 0716+714 has been observed in 34 VLBA observations and in 2 Global array VLBI observations between 1992.73 and 2006.32. These observations have been performed at 5.0, 8.4, 15.3, 22.2, and 43.2 GHz. 12 of the VLBA epochs have been obtained by the MOJAVE / 2 cm-VLBA survey group (MOJAVE, e.g., Lister \& Homan 2005, 2cm Survey, e.g., Kellermann et al. 2004). 4 observations were taken from the CJF-survey (Britzen et al. 2007). S5 0716+714 was used as calibrator source for observations targeted at NRAO 150 (Agudo et al. 2007) in 6 epochs. Part of the data presented here have already been analyzed  and discussed by Bach et al. (2005). All of the above mentioned data-sets have been re-analyzed by us. \\ 
In Table ~\ref{observations} we summarize the observations presented in this paper. In column (1) we list the epoch, in column (2) the frequency of the observation, column (3) lists the array and survey-type, and in column (4) we note references to the original publication of the individual data sets. We refer to the references in column (4) for more details concerning the observations and data reduction.

\begin{figure*}[htb]
\begin{center}
\subfigure[]{\includegraphics[clip,width=4.8cm,angle=-90]{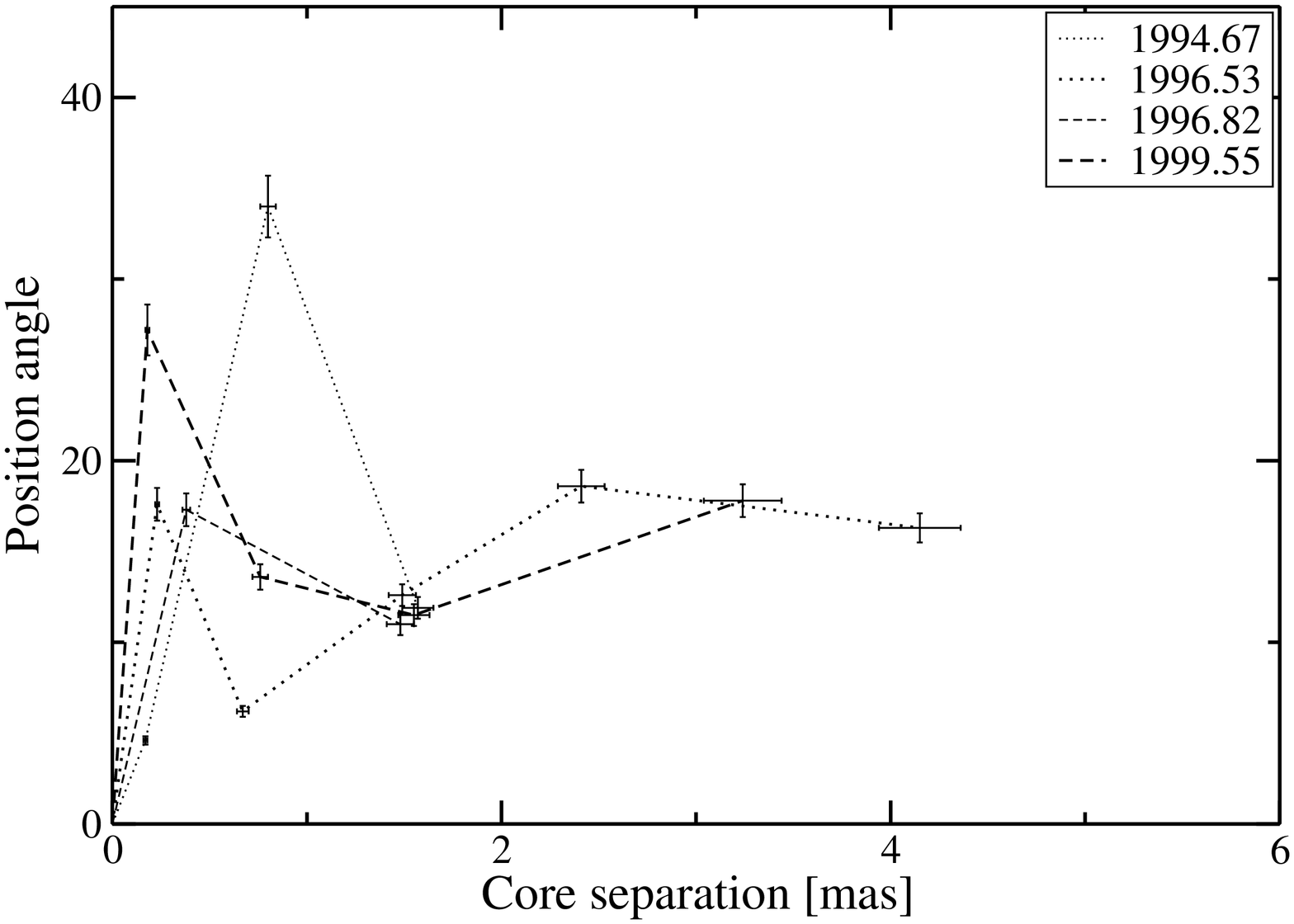}}
\subfigure[]{\includegraphics[clip,width=4.8cm,angle=-90]{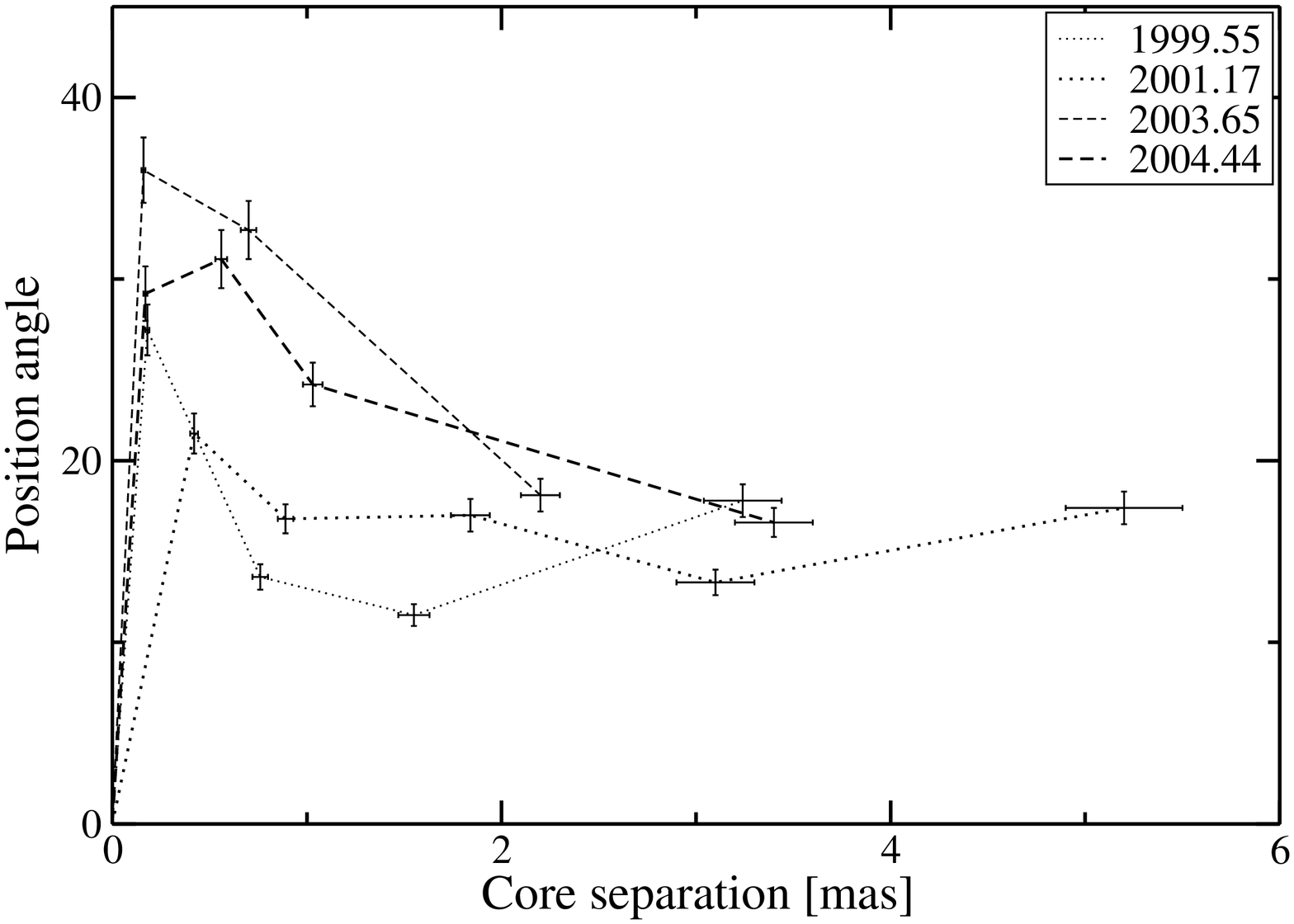}}\\
\subfigure[]{\includegraphics[clip,width=4.8cm,angle=-90]{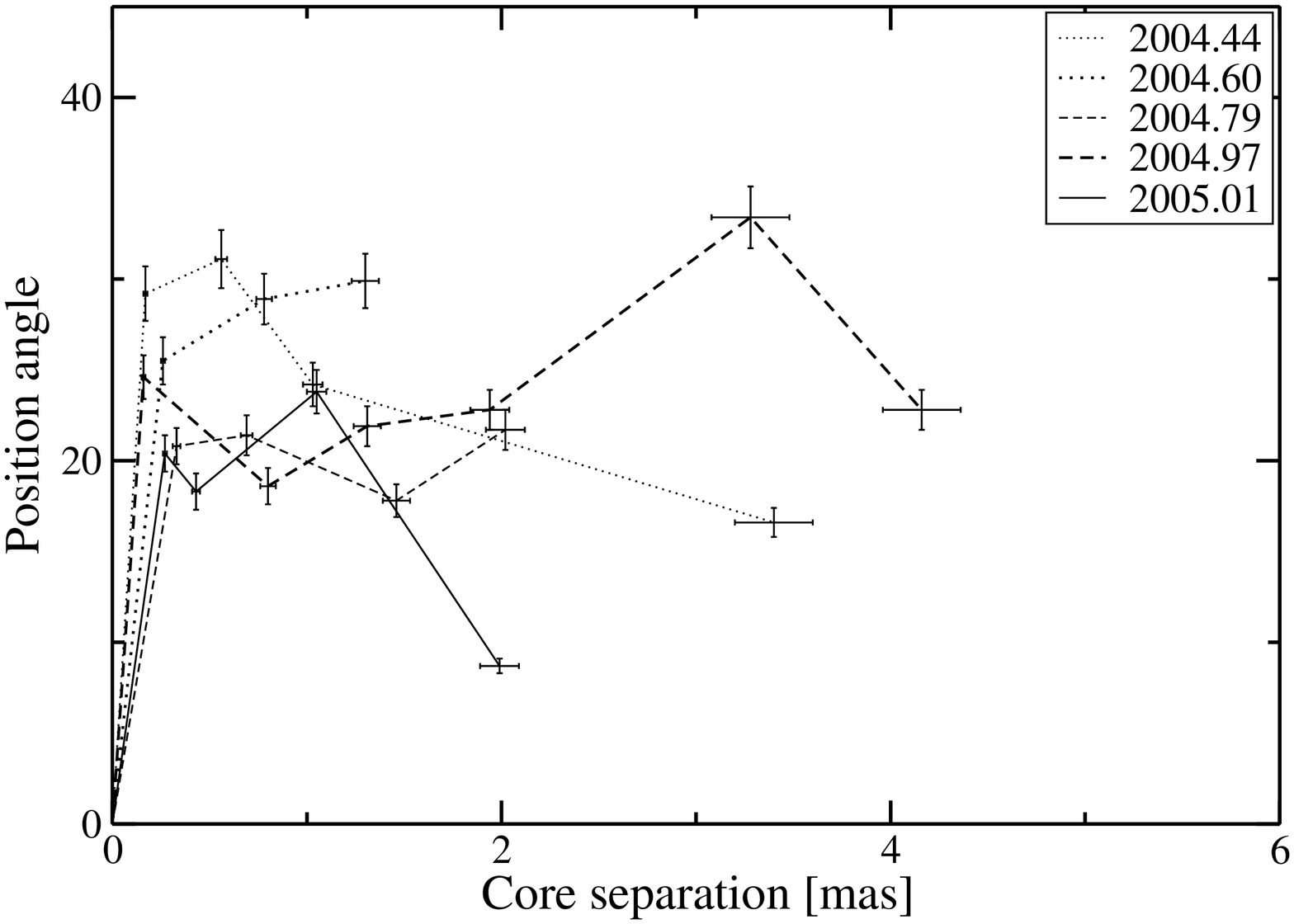}}
\subfigure[]{\includegraphics[clip,width=4.8cm,angle=-90]{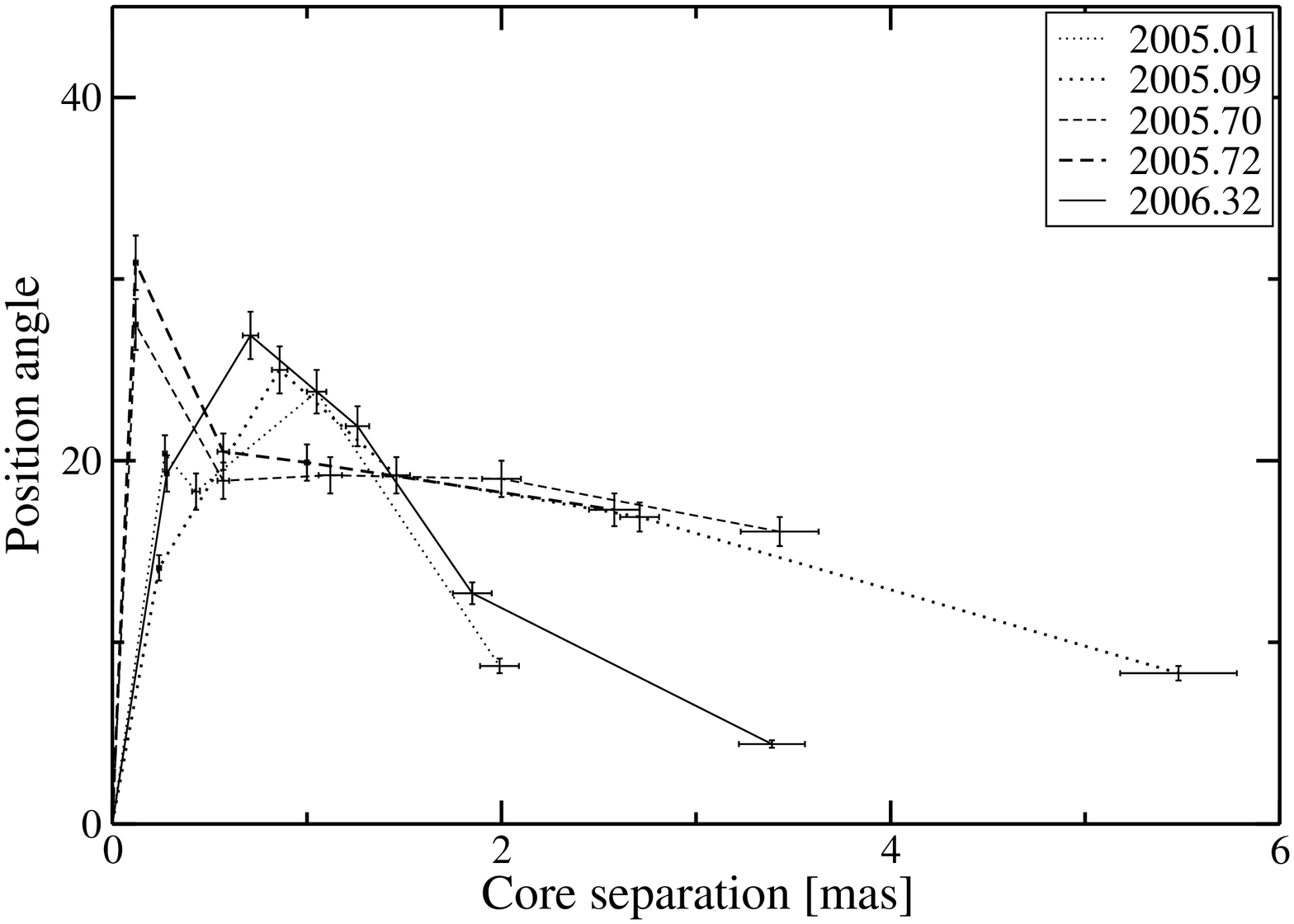}}\\
\subfigure[]{\includegraphics[clip,width=4.8cm,angle=-90]{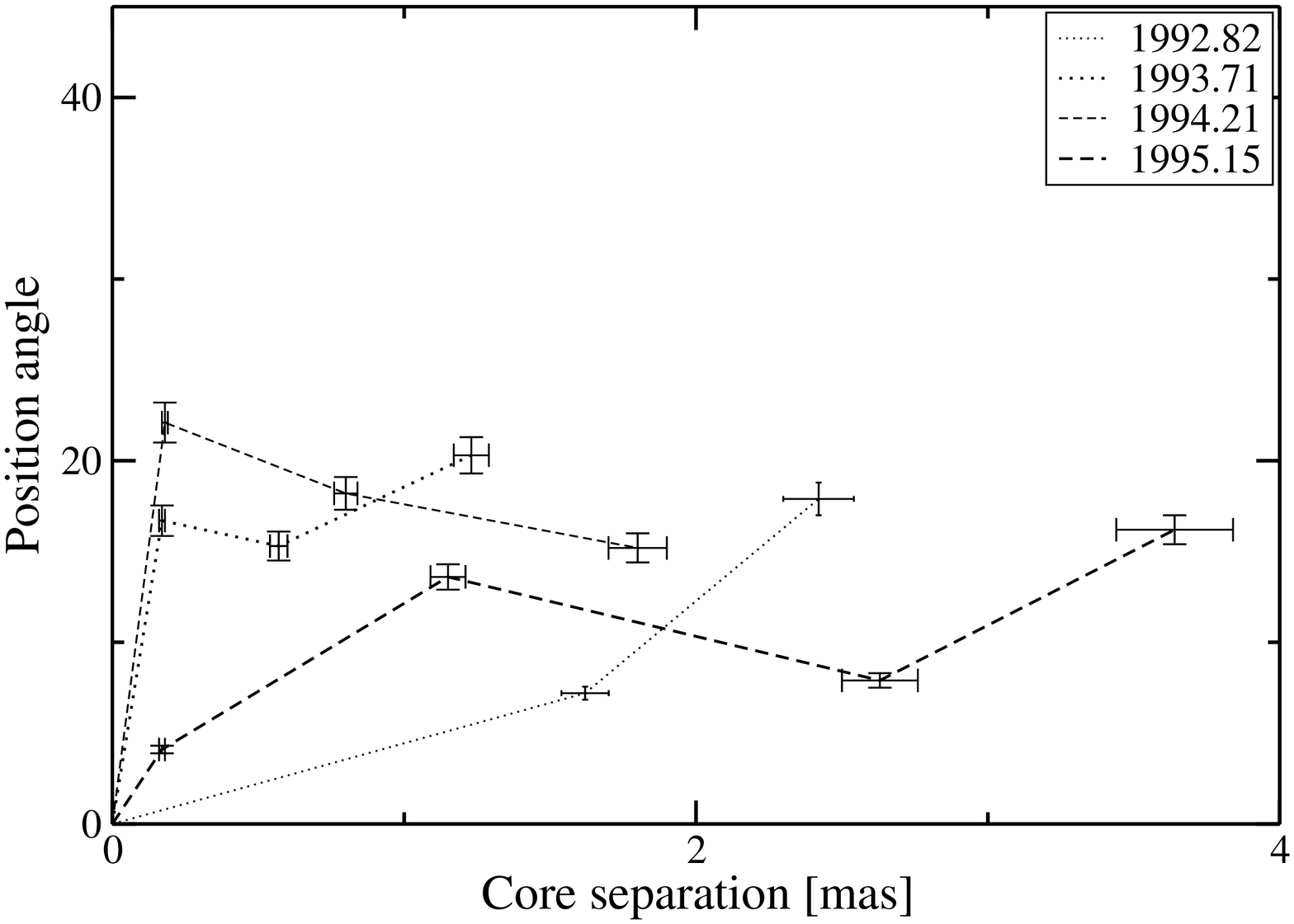}}
\subfigure[]{\includegraphics[clip,width=4.8cm,angle=-90]{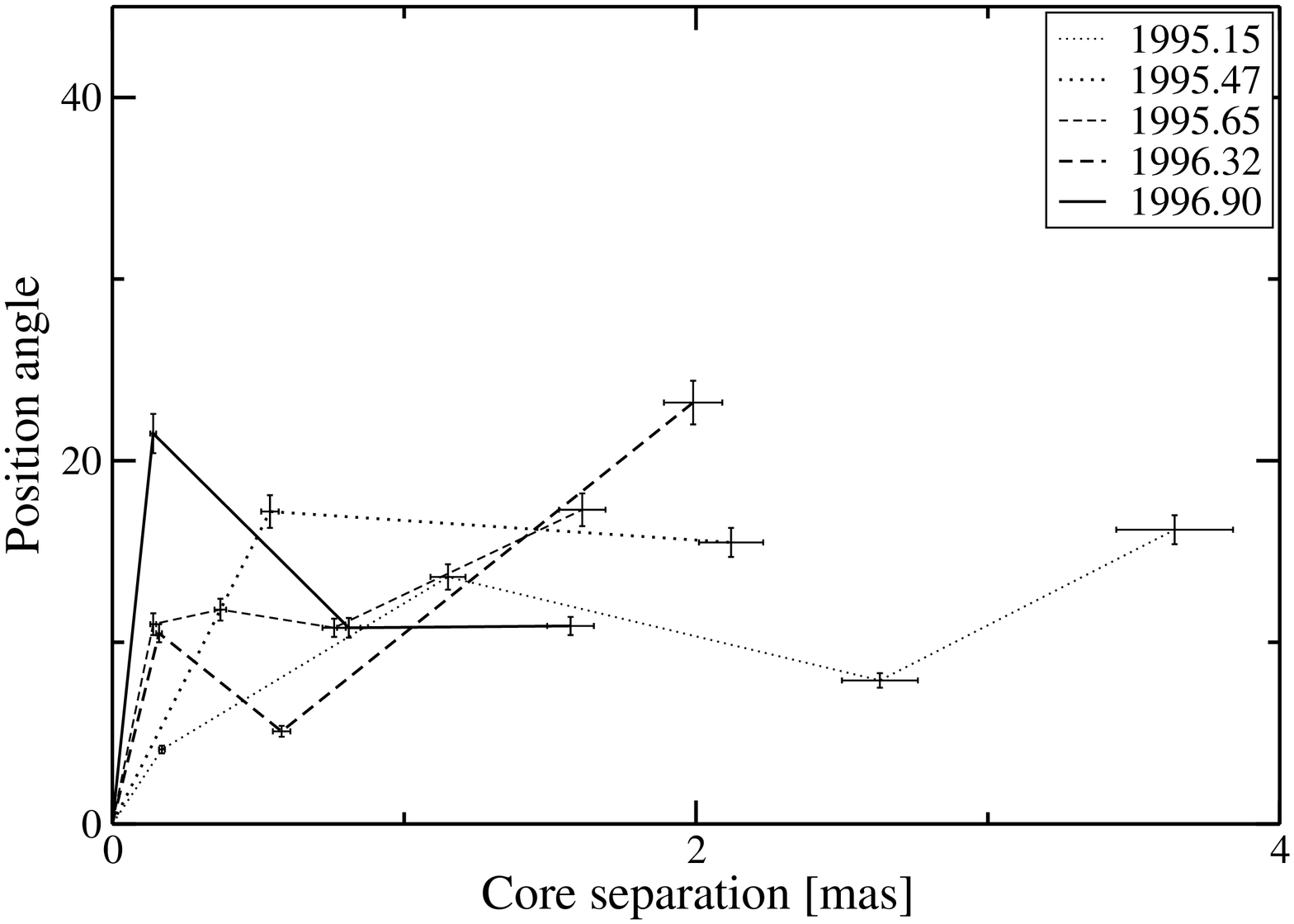}}\\
\subfigure[]{\includegraphics[clip,width=4.8cm,angle=-90]{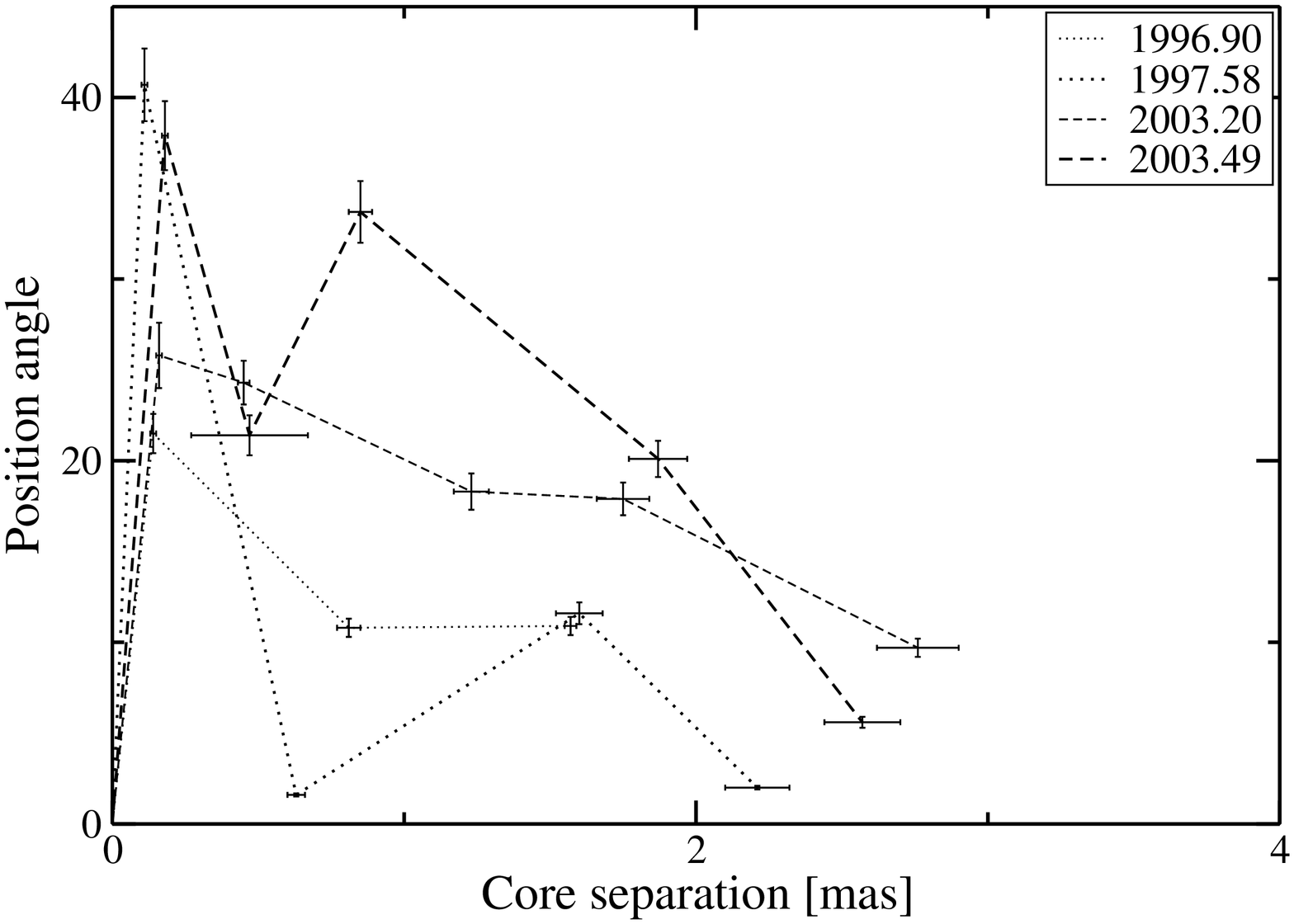}}
\subfigure[]{\includegraphics[clip,width=4.8cm,angle=-90]{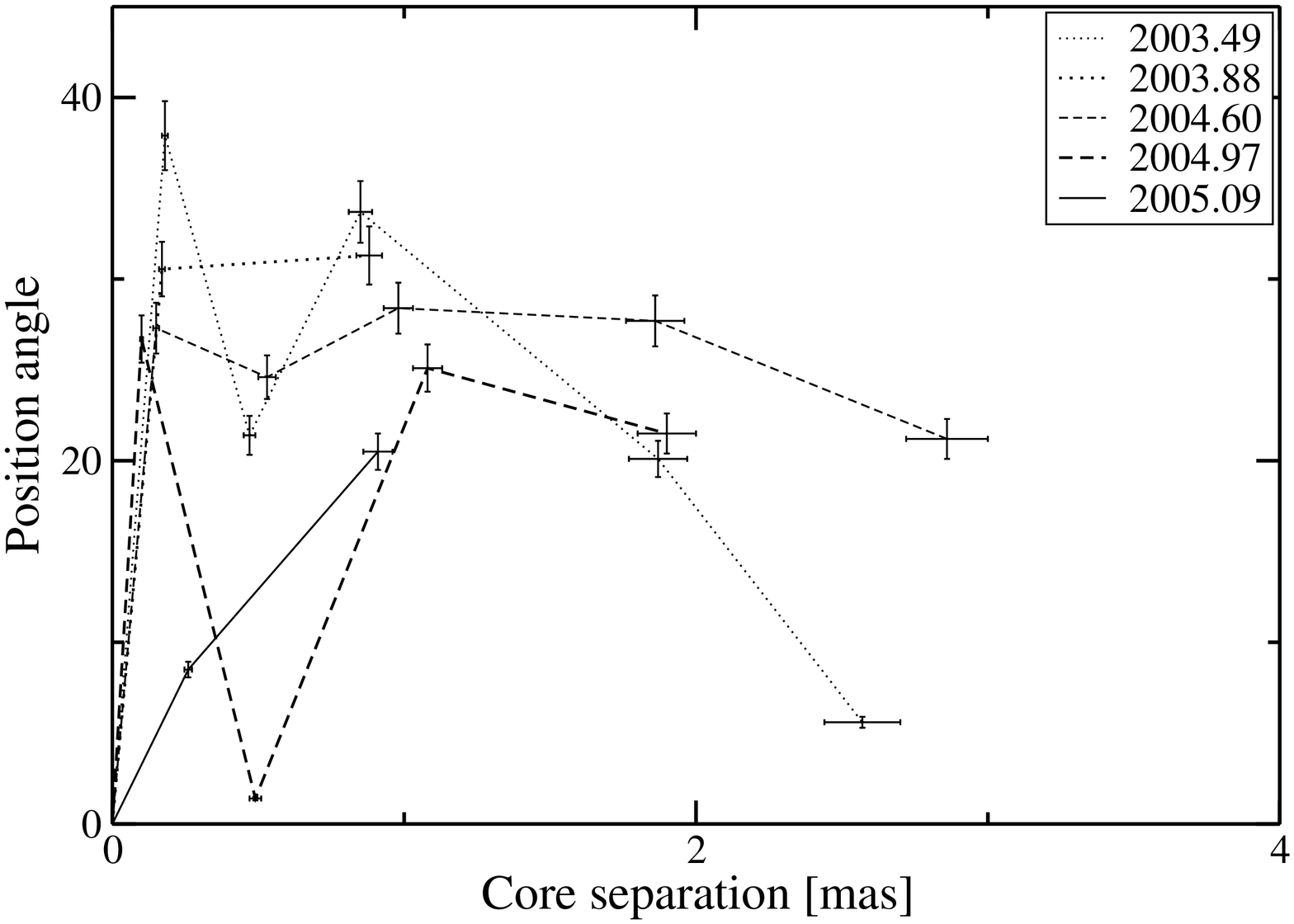}}
\end{center}
\caption{The jet ridge line (as defined by all component positions per epoch) is shown at 15 GHz in (a)--(d) and at 22 GHz in (e)--(h).}
\label{jetridgeline}
\end{figure*}
\clearpage

\subsection{Model-fitting}
\begin{figure*}[htb]
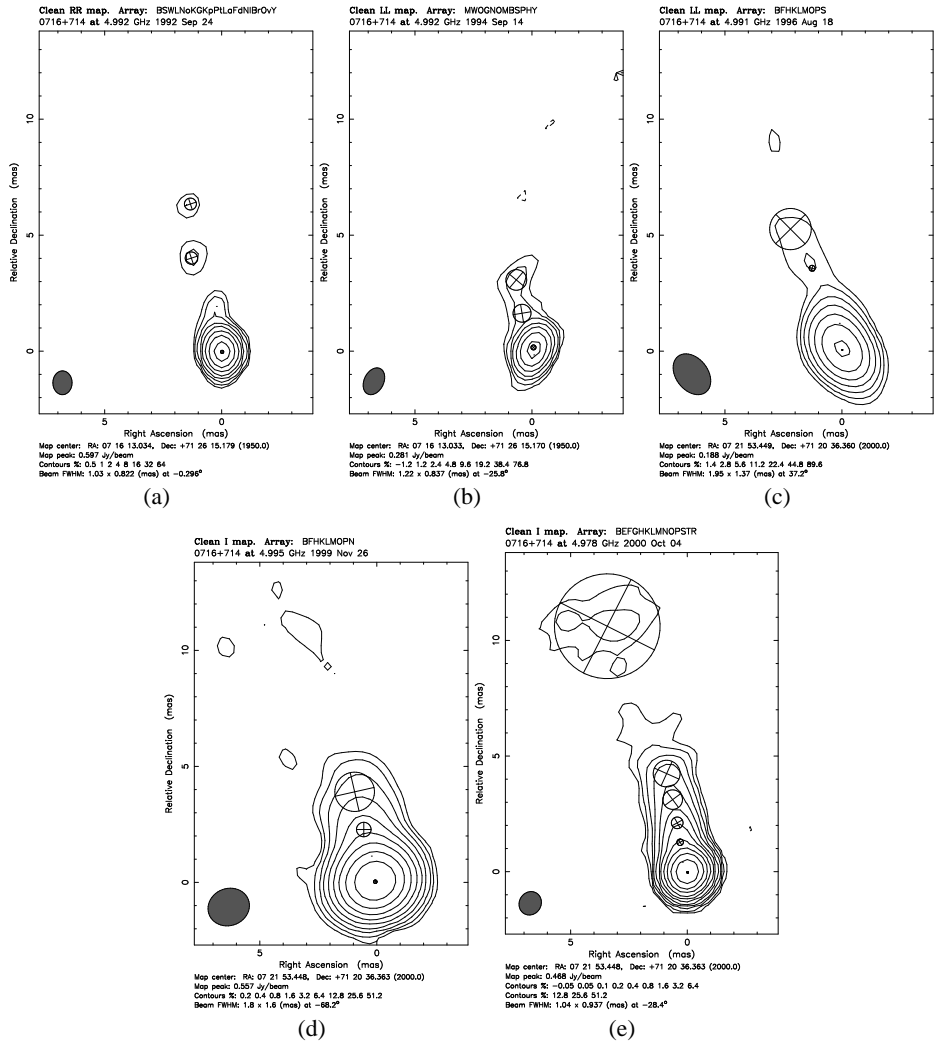

\begin{center}
\subfigure[]{\includegraphics[clip,width=4cm]{10875f7a.ps}}
\subfigure[]{\includegraphics[clip,width=4cm]{10875f7b.ps}}
\subfigure[]{\includegraphics[clip,width=4cm]{10875f7c.ps}}\\
\subfigure[]{\includegraphics[clip,width=4cm]{10875f7d.ps}}
\subfigure[]{\includegraphics[clip,width=4cm]{10875f7e.ps}}
\end{center}
\caption{5 GHz VLBI images of S5 0716+714. Model-fits convolved with the interferometric beam and with the residual map added to them are shown with the Gaussian model components superimposed. The following figures show similar images obtained at different frequencies.}
\label{maps5}
\end{figure*}

\begin{figure*}[htb]
\begin{center}
\subfigure[]{\includegraphics[clip,width=4cm]{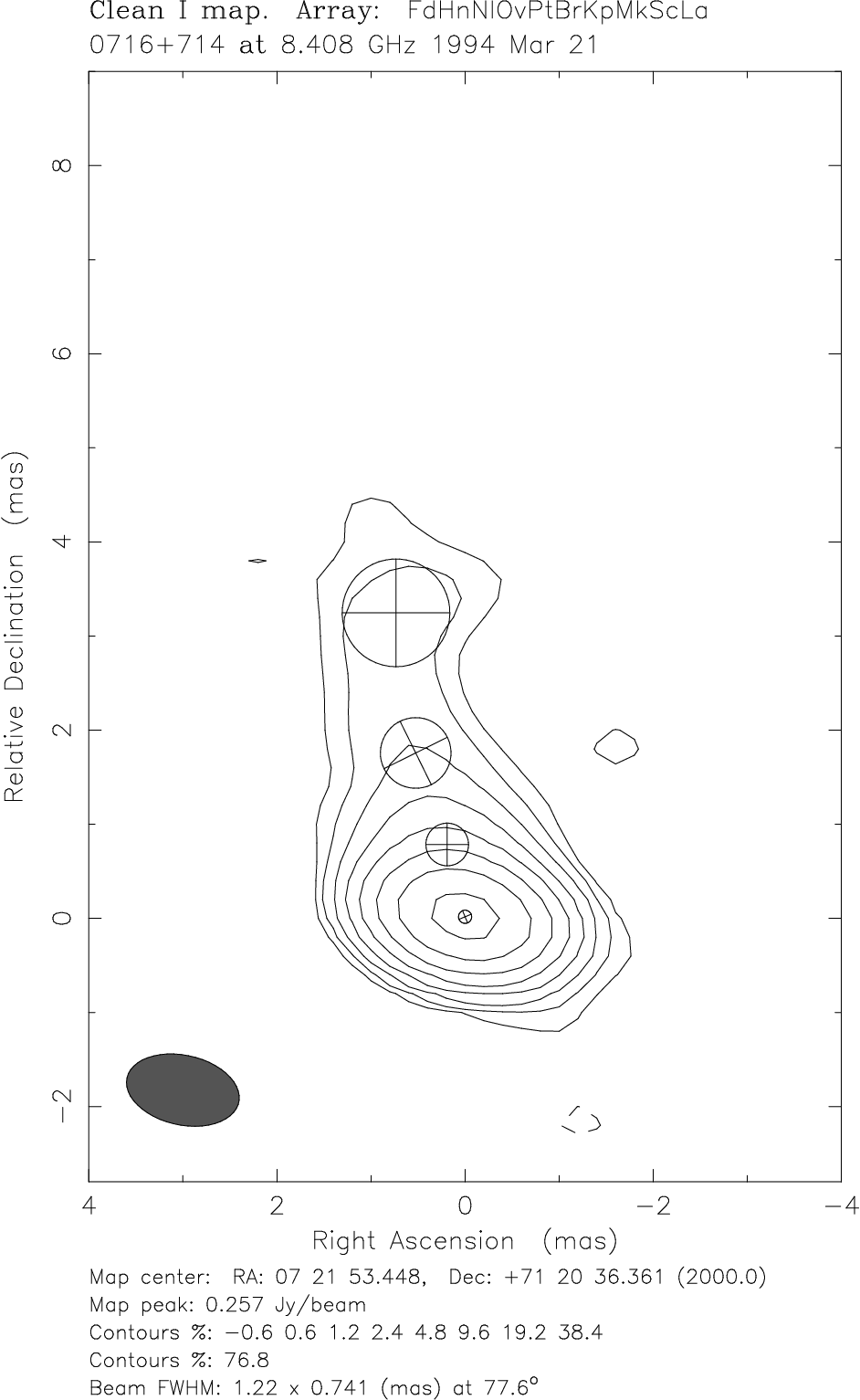}}
\subfigure[]{\includegraphics[clip,width=4cm]{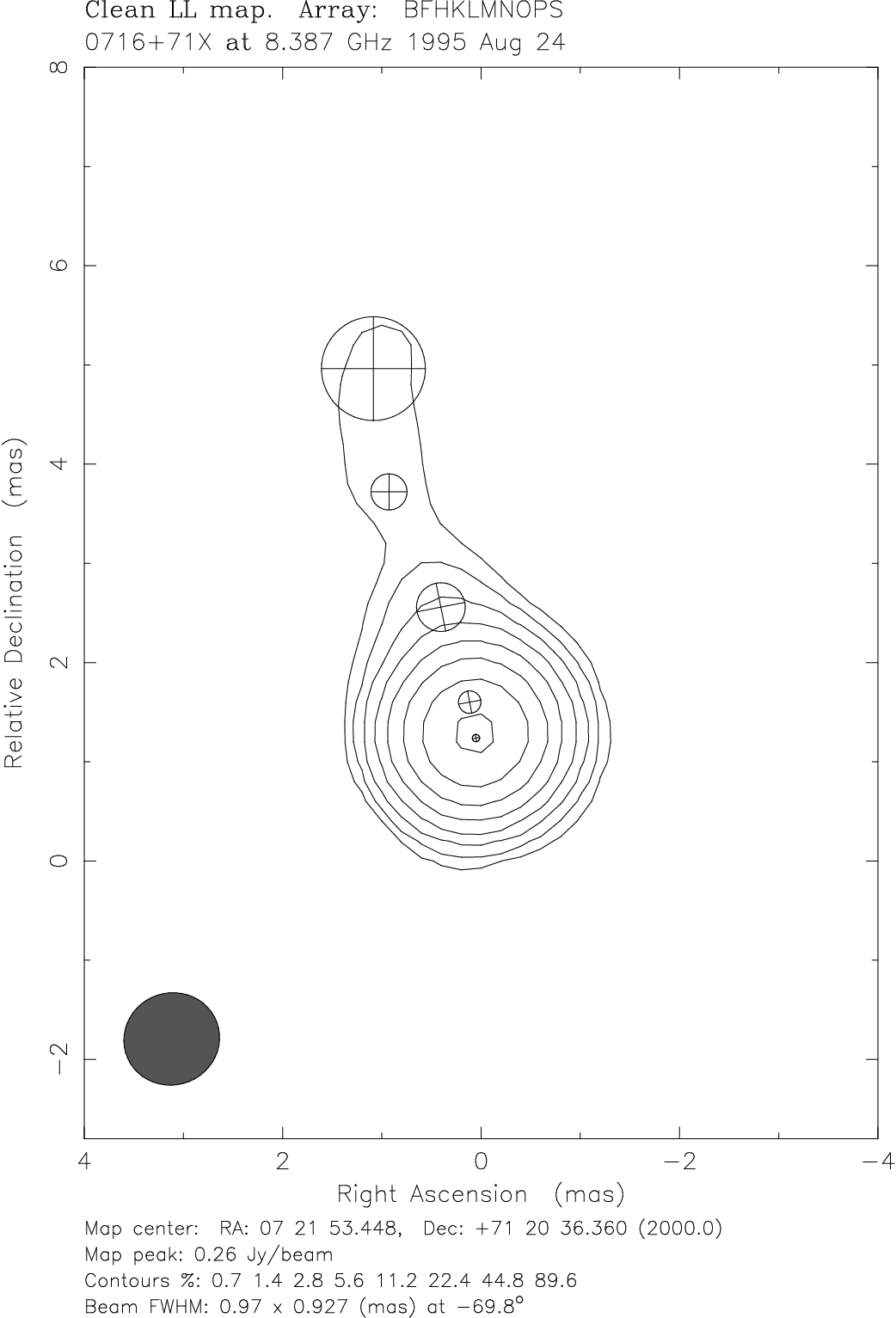}}
\subfigure[]{\includegraphics[clip,width=4cm]{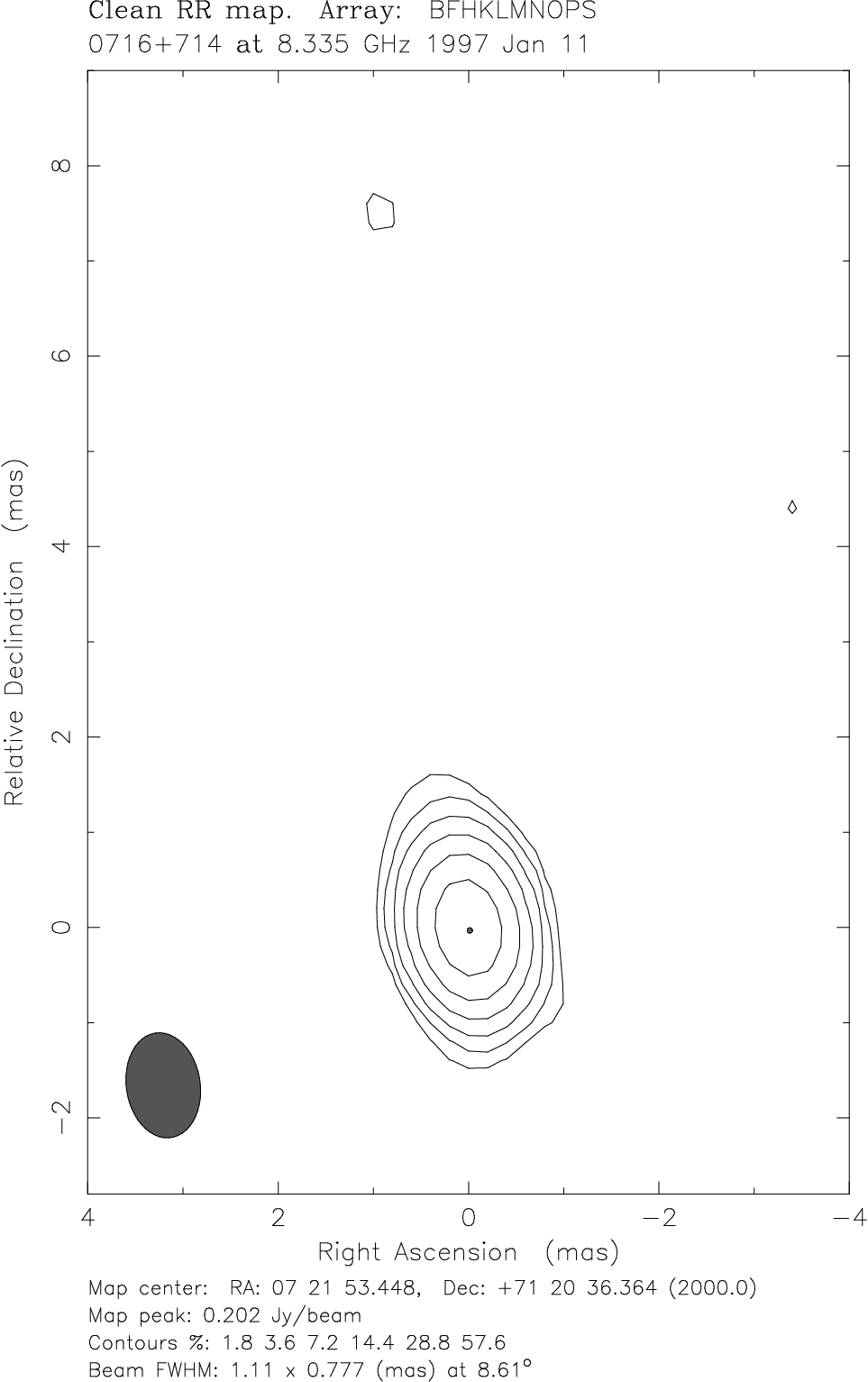}}\\
\subfigure[]{\includegraphics[clip,width=4cm]{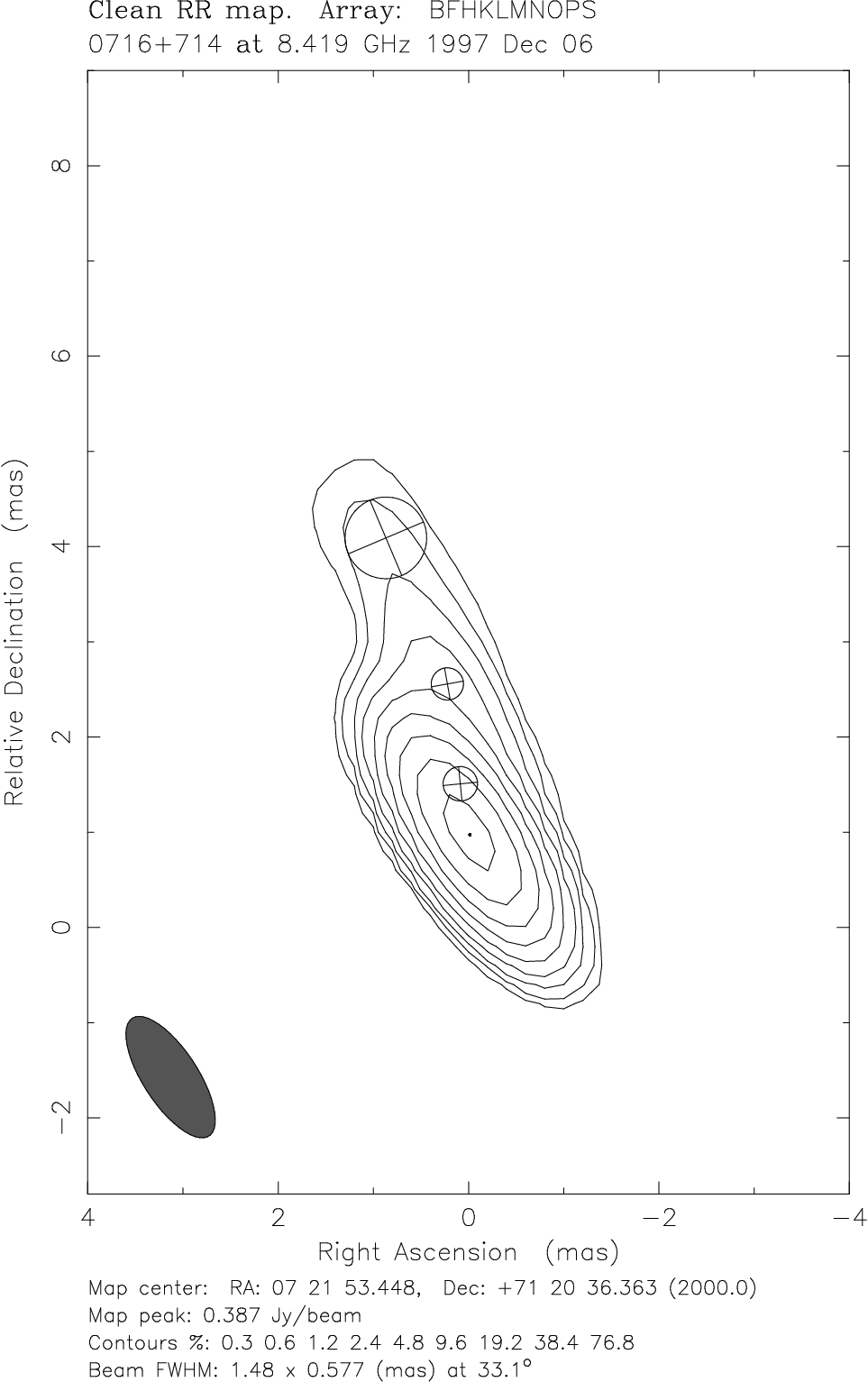}}
\subfigure[]{\includegraphics[clip,width=4cm]{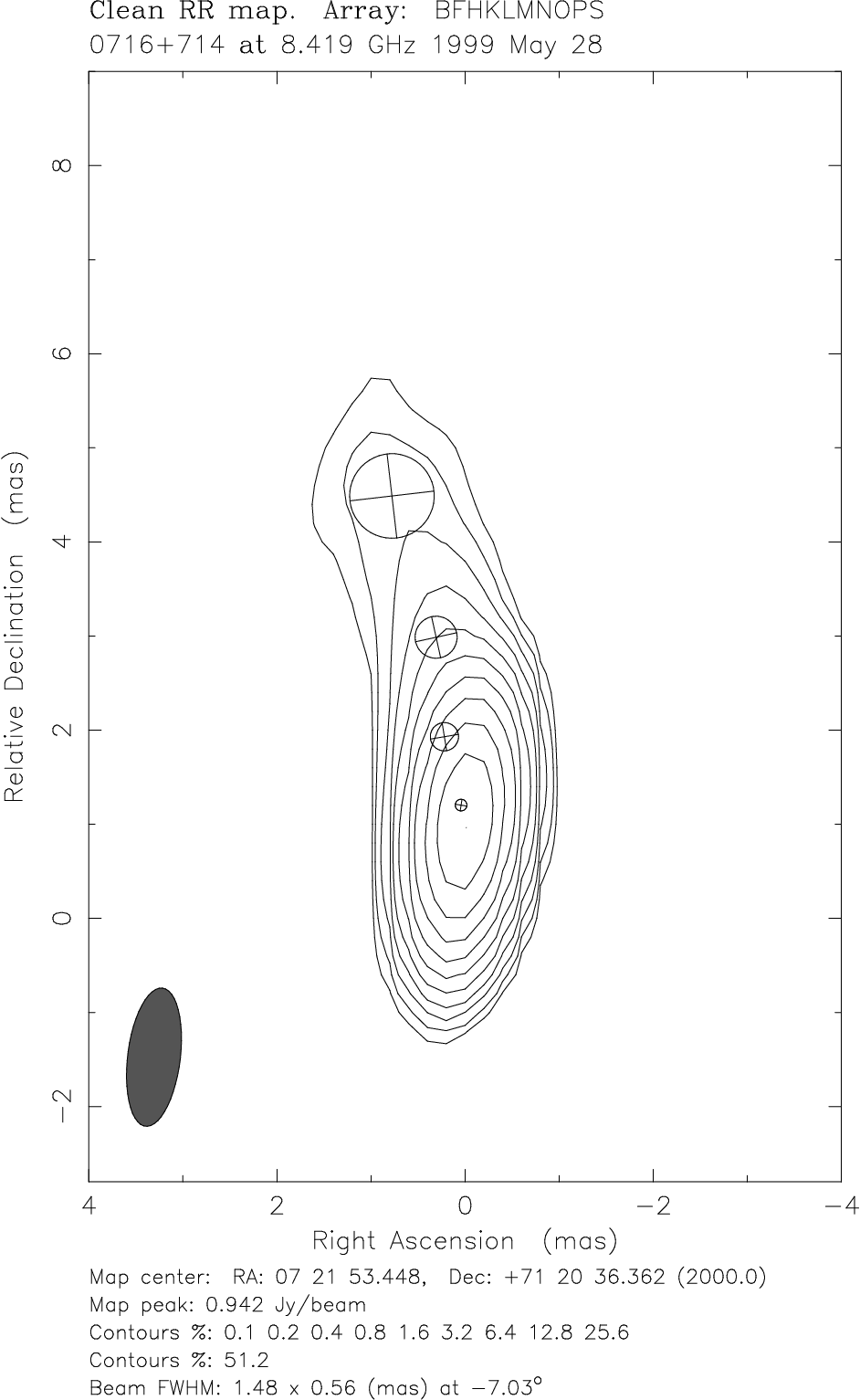}}
\subfigure[]{\includegraphics[clip,width=4cm]{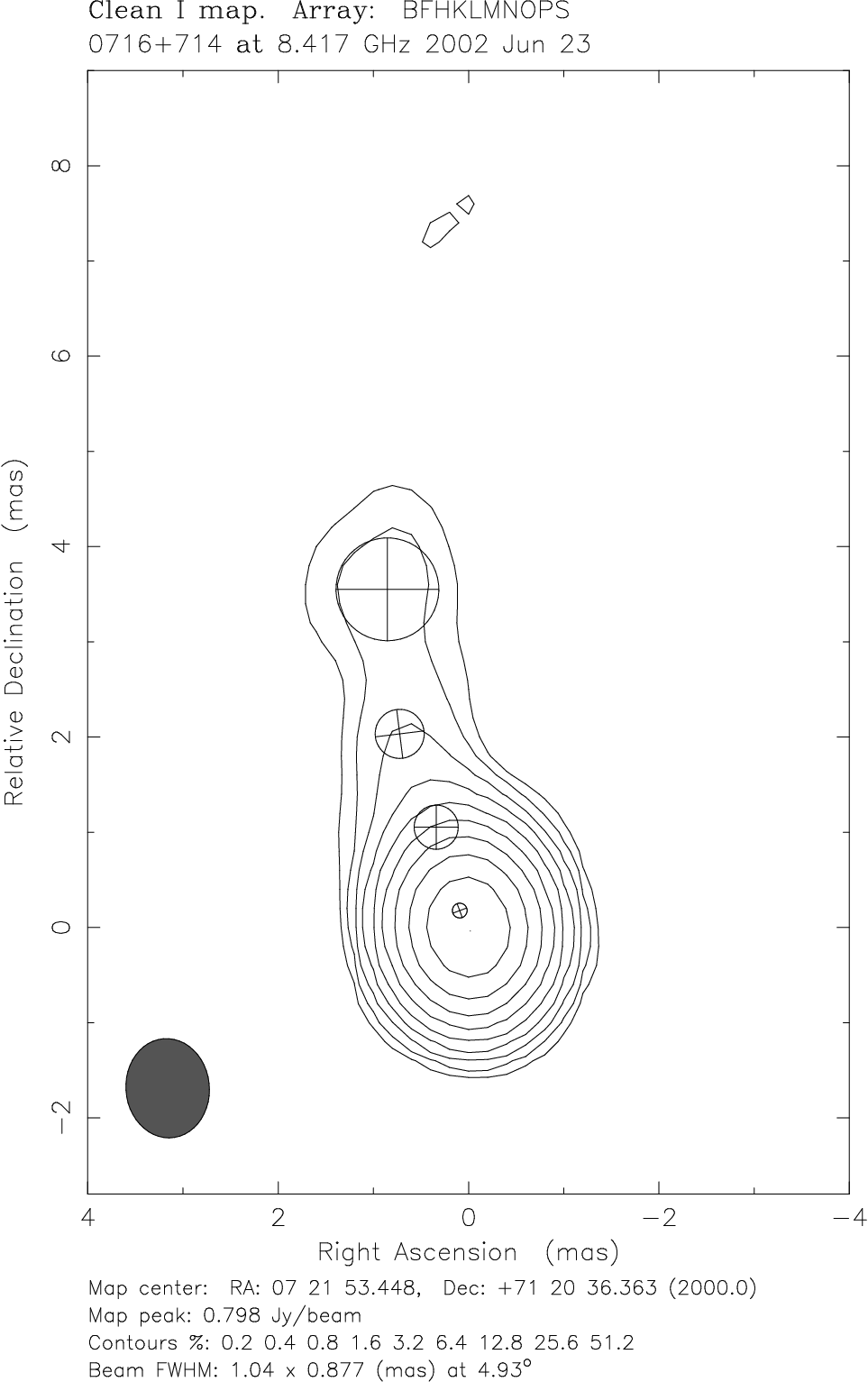}}\\
\subfigure[]{\includegraphics[clip,width=4cm]{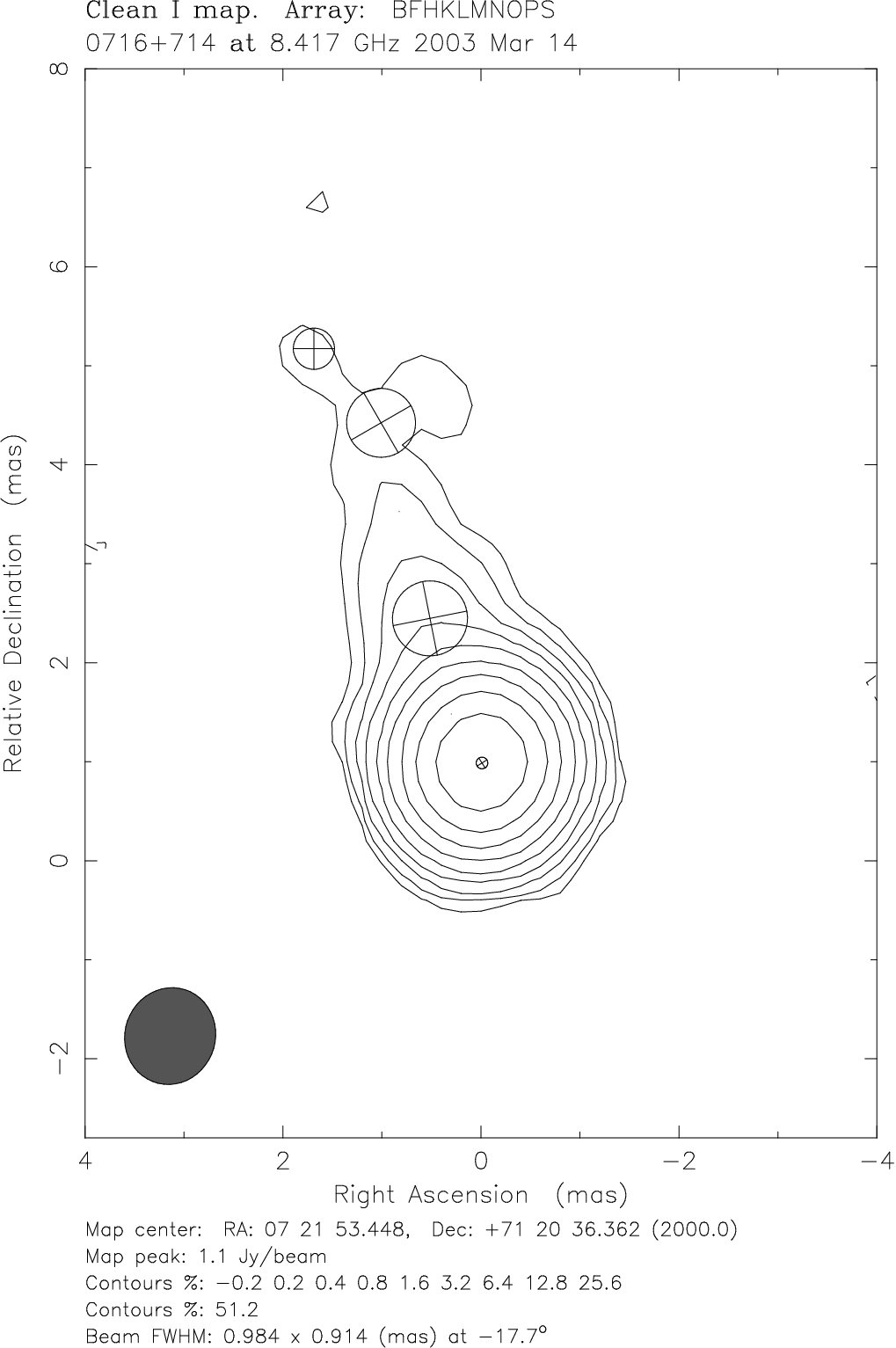}}
\subfigure[]{\includegraphics[clip,width=4cm]{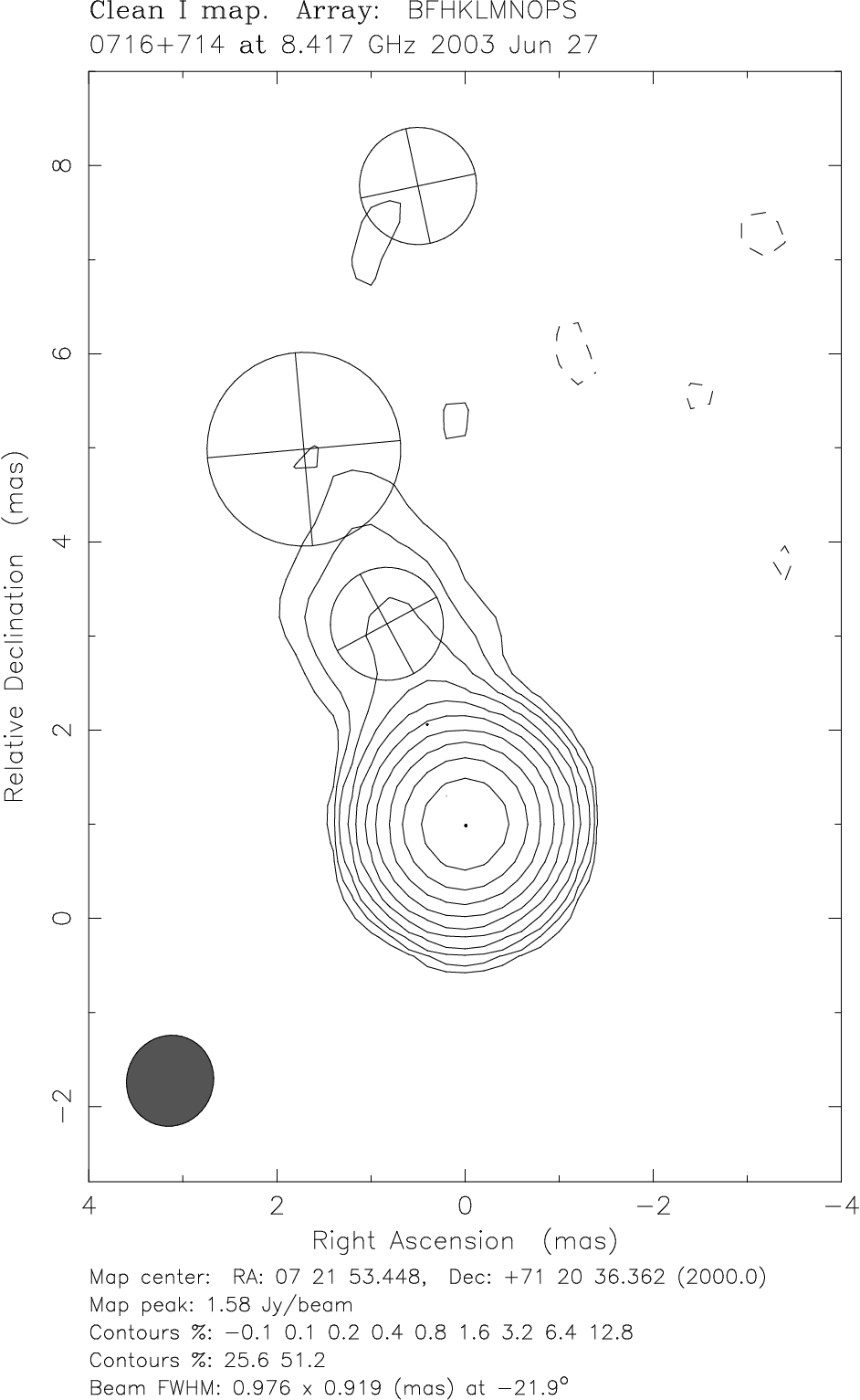}}
\end{center}
\caption{Images of S5 0716+714 at 8 GHz.}
\label{maps8}
\end{figure*}

\begin{figure*}[htb]
\begin{center}
\subfigure[]{\includegraphics[clip,width=4cm]{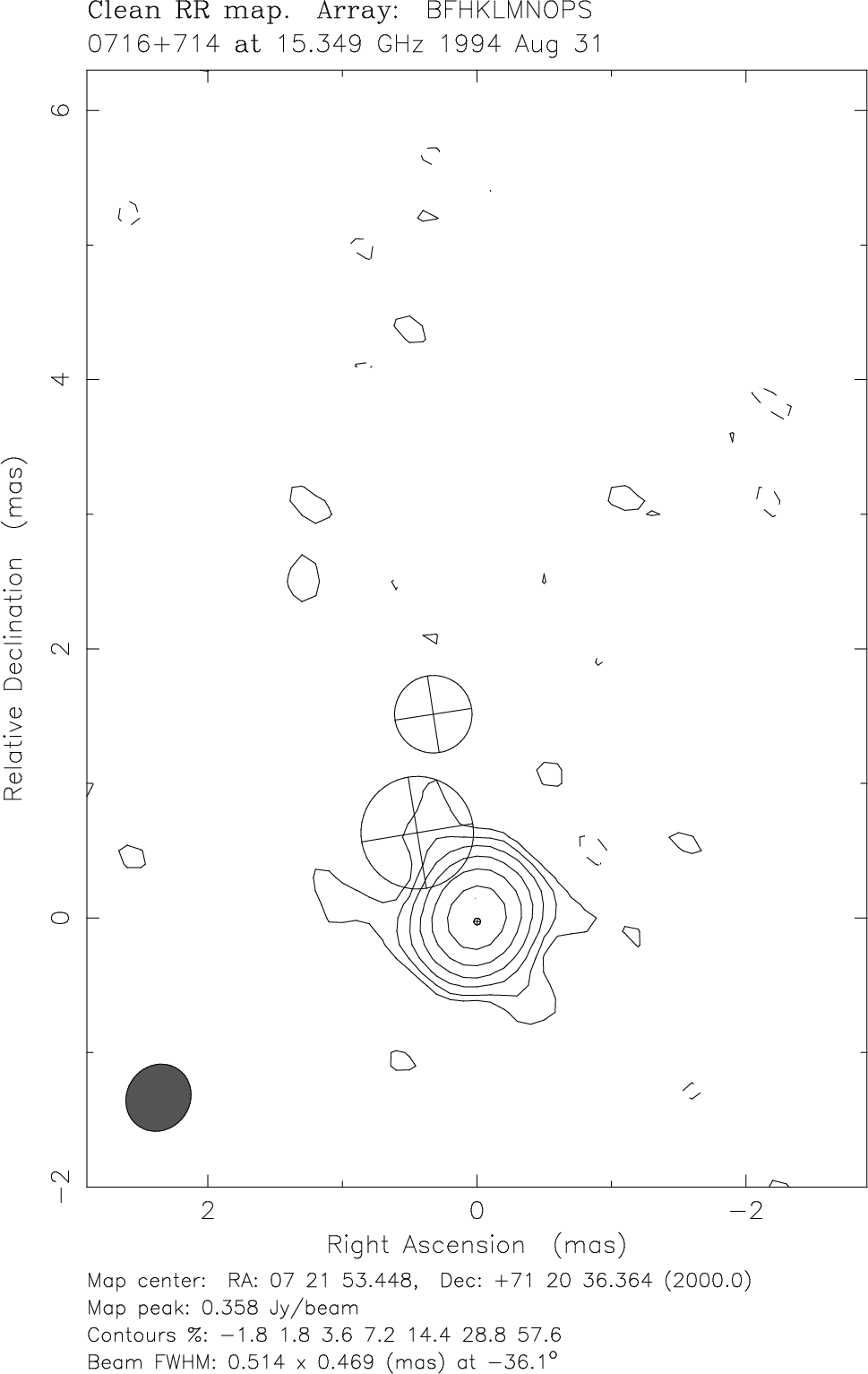}}
\subfigure[]{\includegraphics[clip,width=4cm]{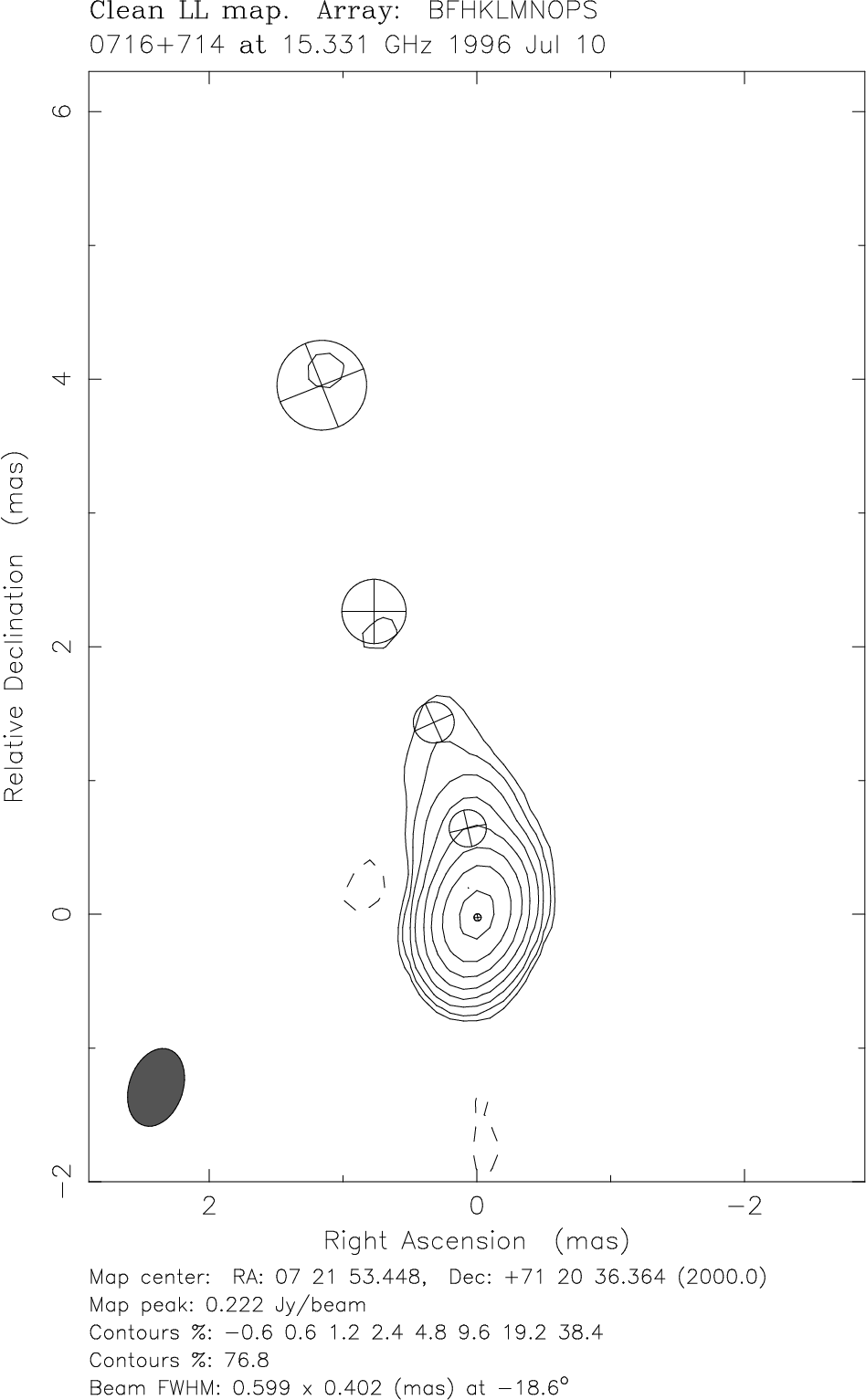}}
\subfigure[]{\includegraphics[clip,width=4cm]{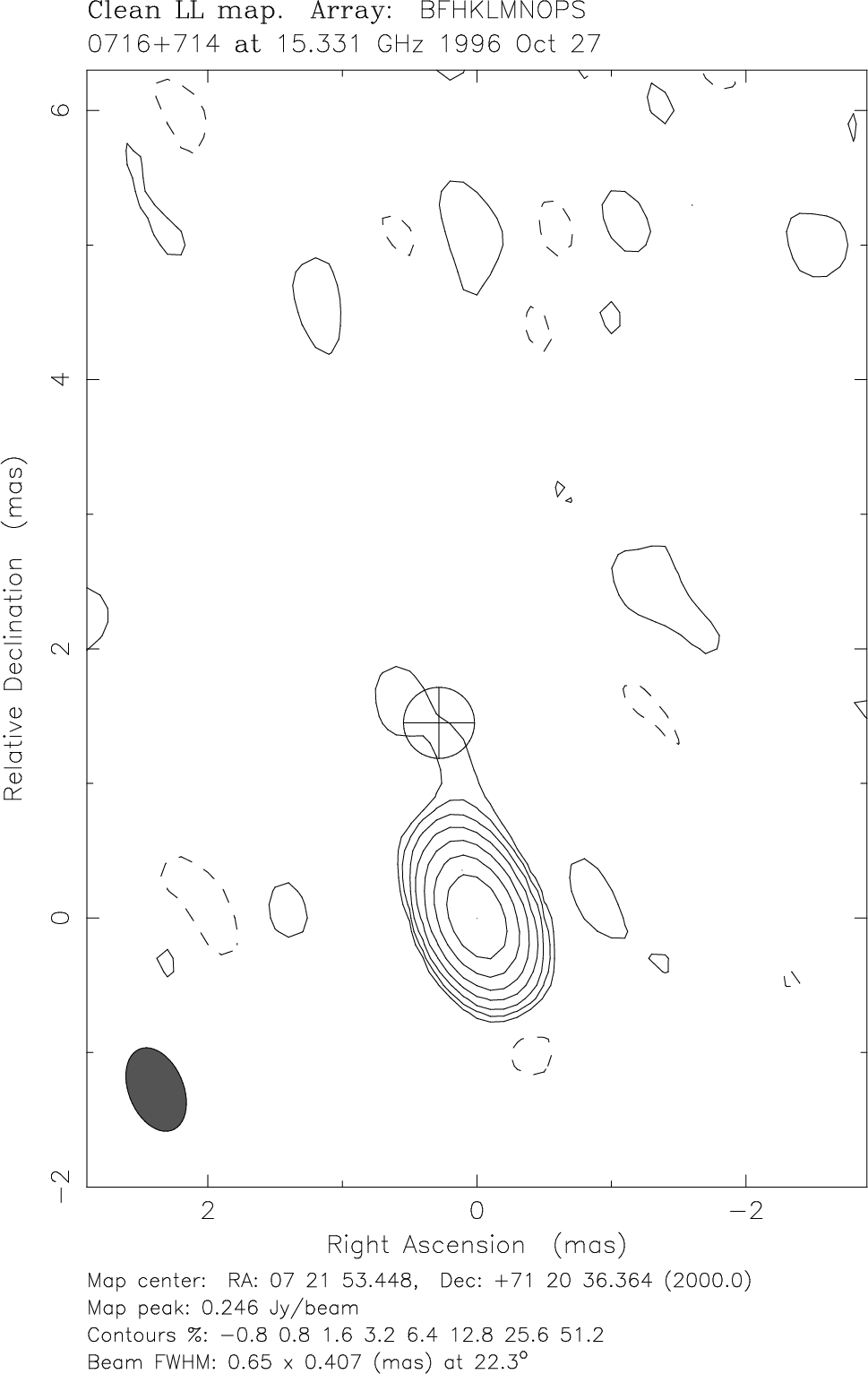}}\\
\subfigure[]{\includegraphics[clip,width=4cm]{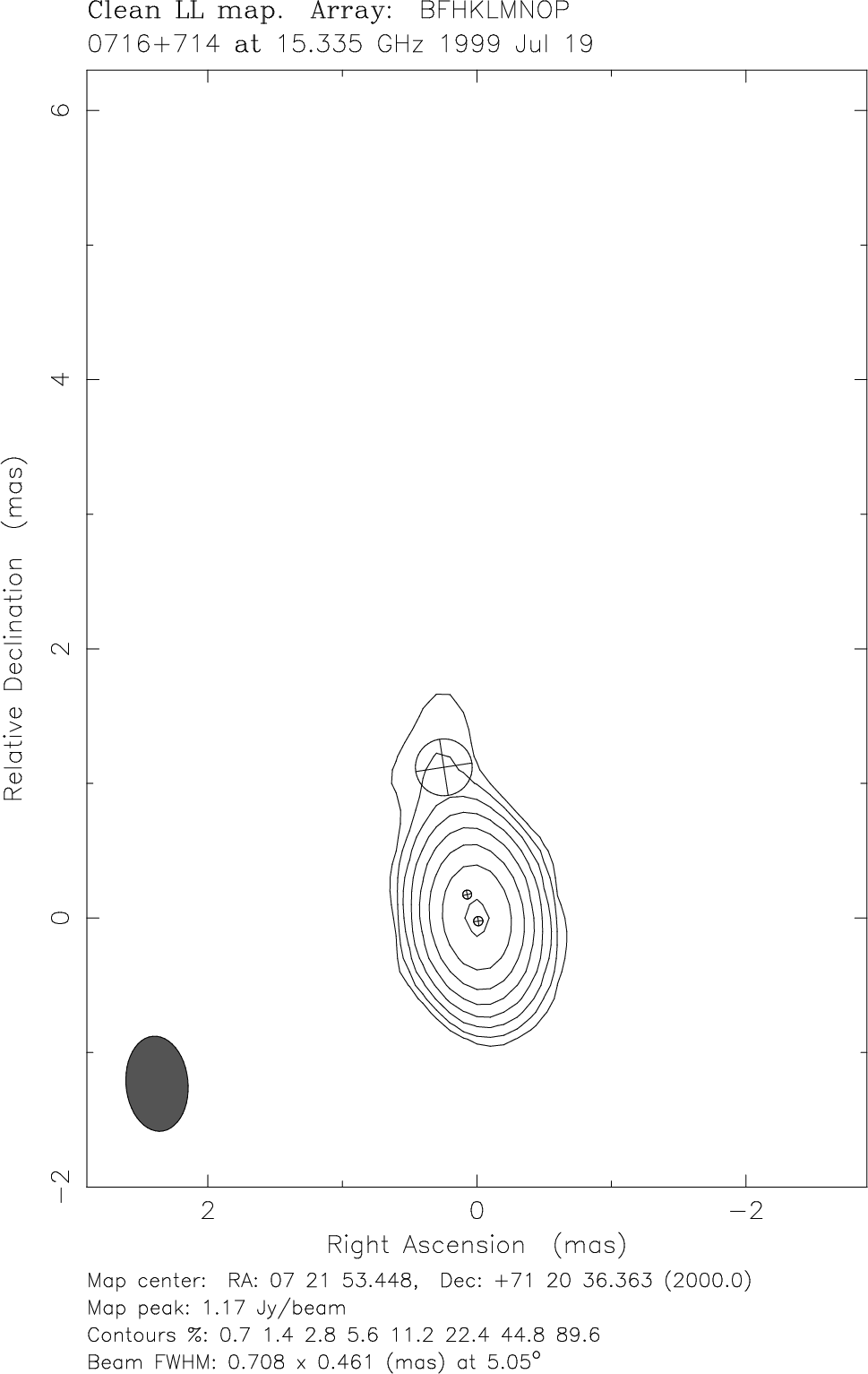}}
\subfigure[]{\includegraphics[clip,width=4cm]{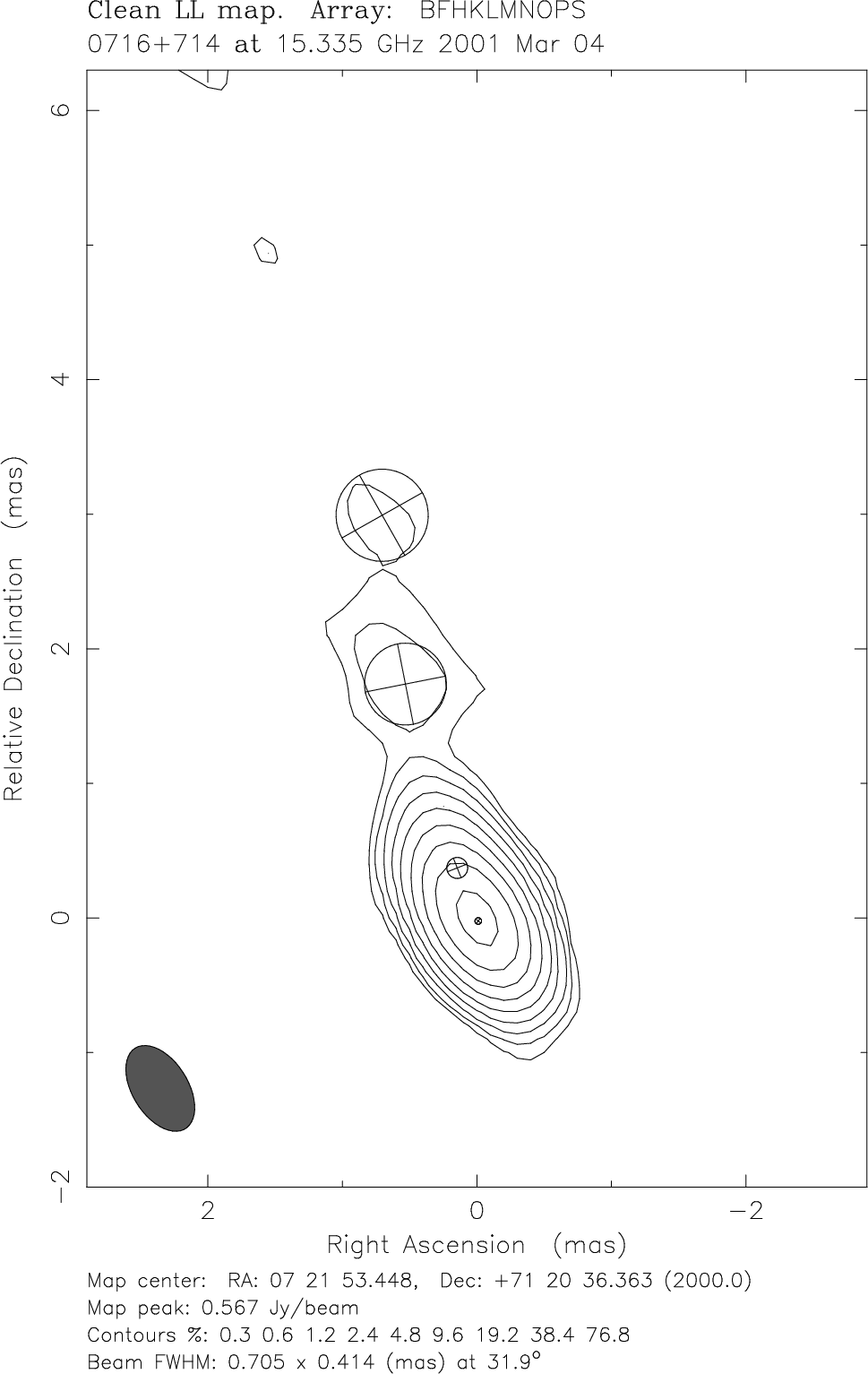}}
\subfigure[]{\includegraphics[clip,width=4cm]{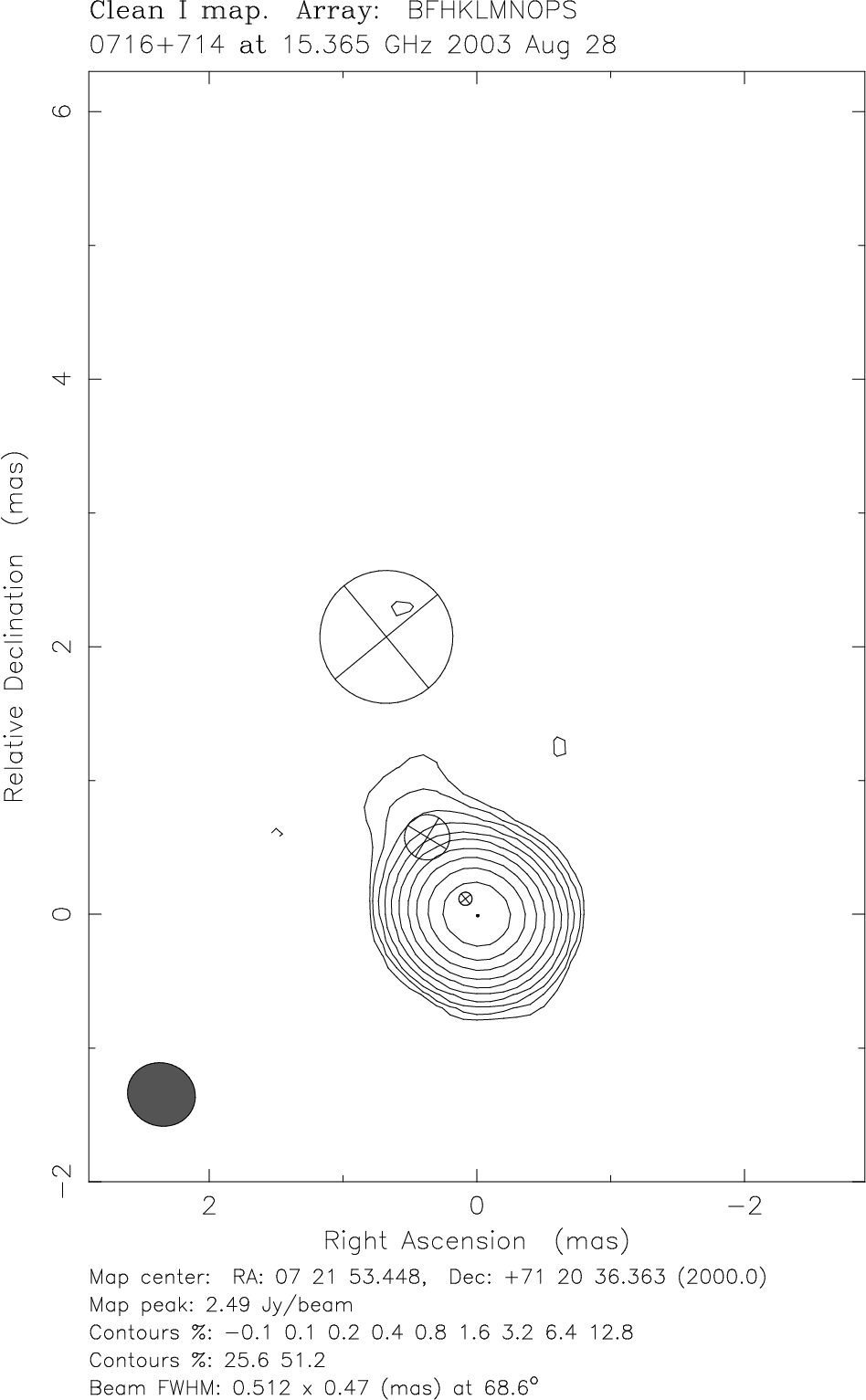}}\\
\subfigure[]{\includegraphics[clip,width=4cm]{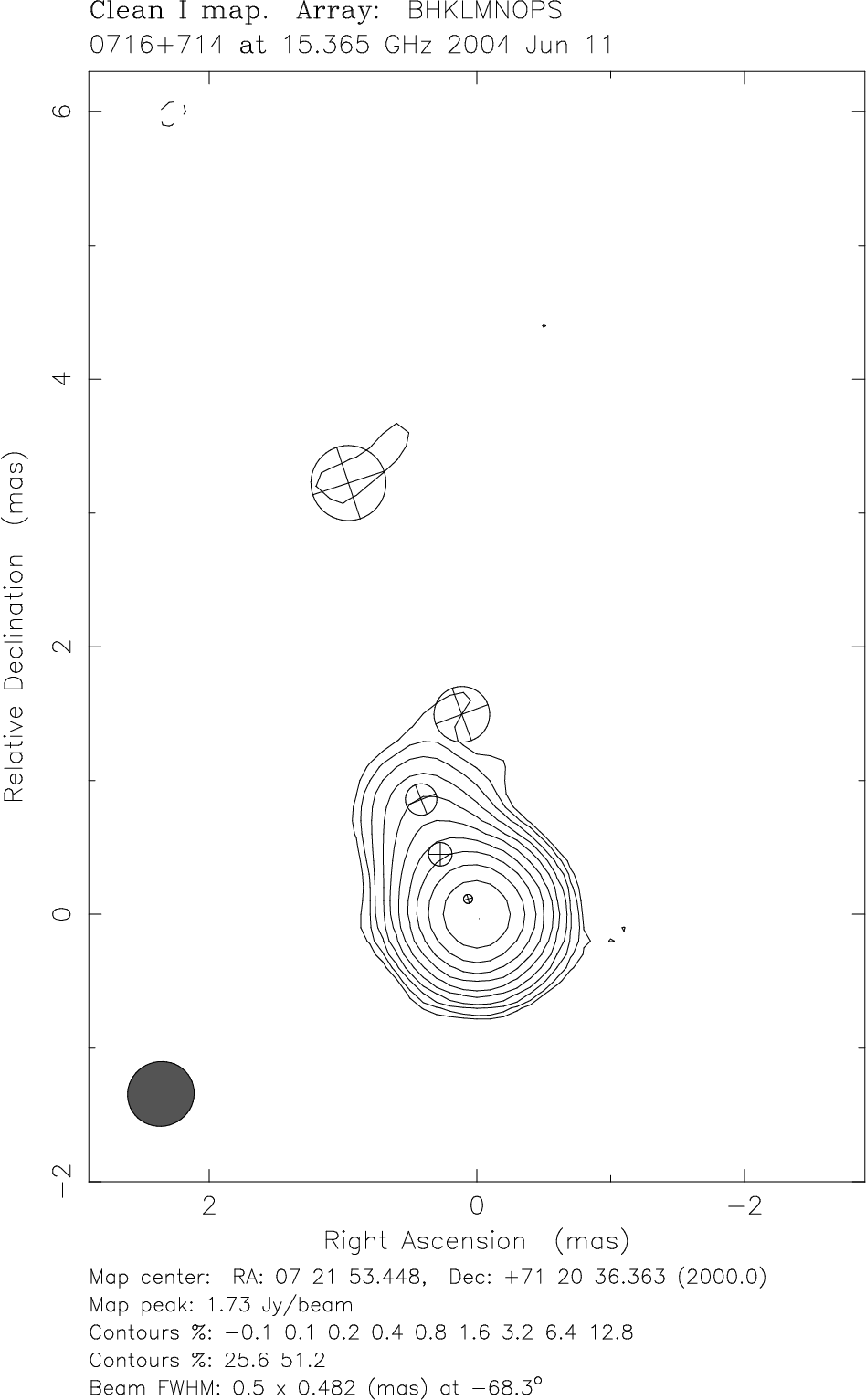}}
\subfigure[]{\includegraphics[clip,width=4cm]{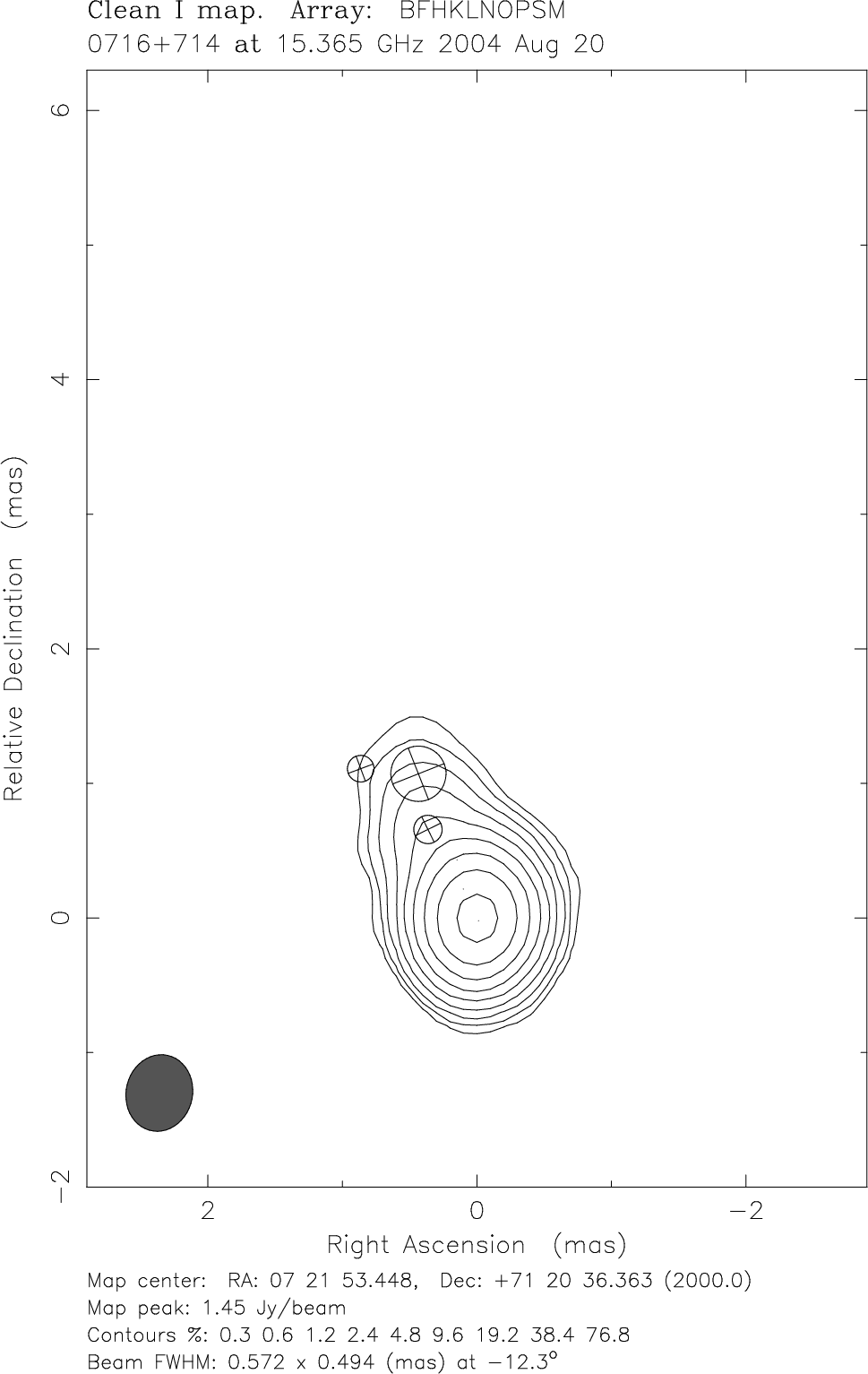}}
\subfigure[]{\includegraphics[clip,width=4cm]{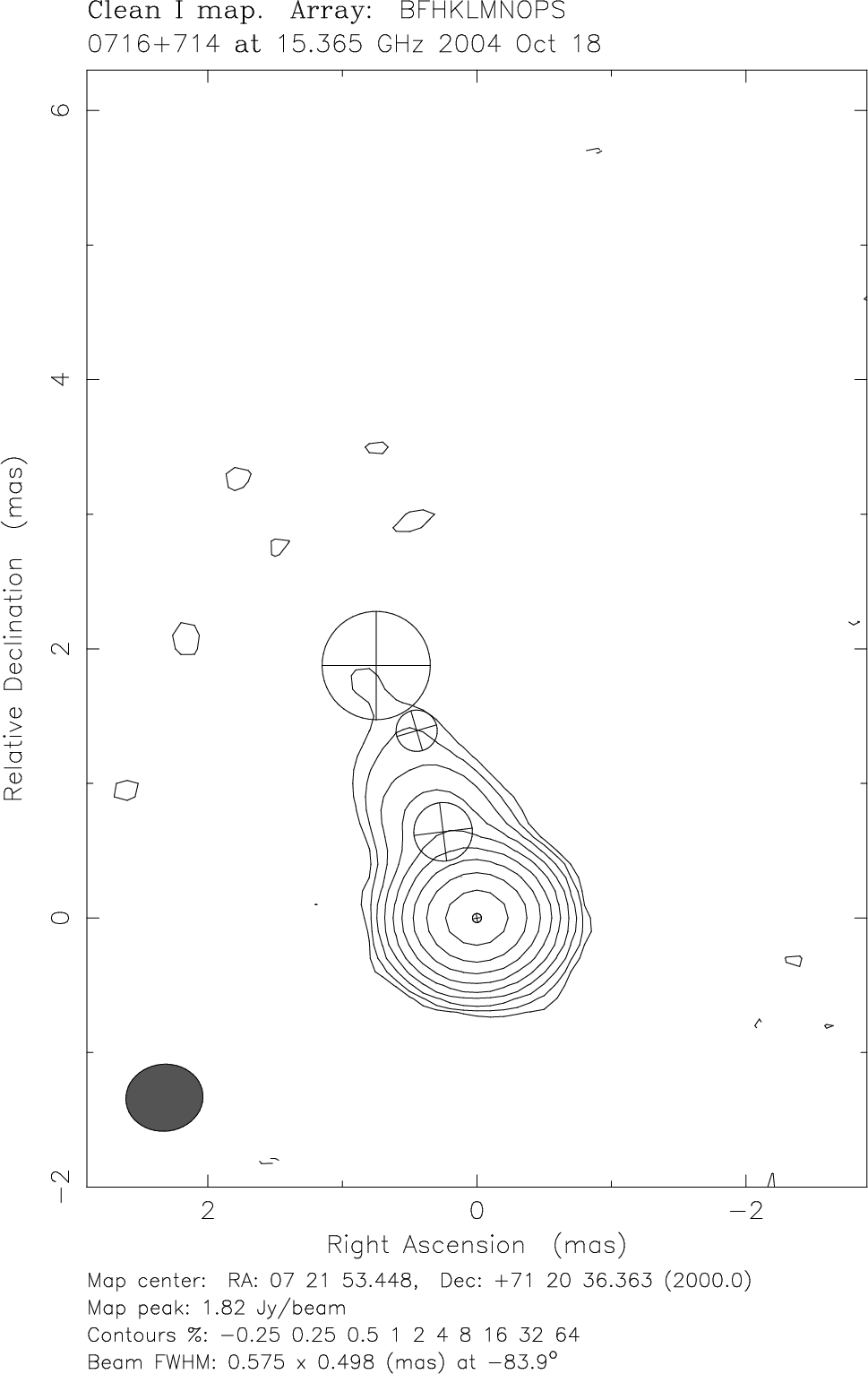}}\\
\end{center}
\caption{First part of images of S5 0716+714 at 15 GHz. }
\label{maps151}
\end{figure*}

\begin{figure*}[htb]
\begin{center}
\subfigure[]{\includegraphics[clip,width=4cm]{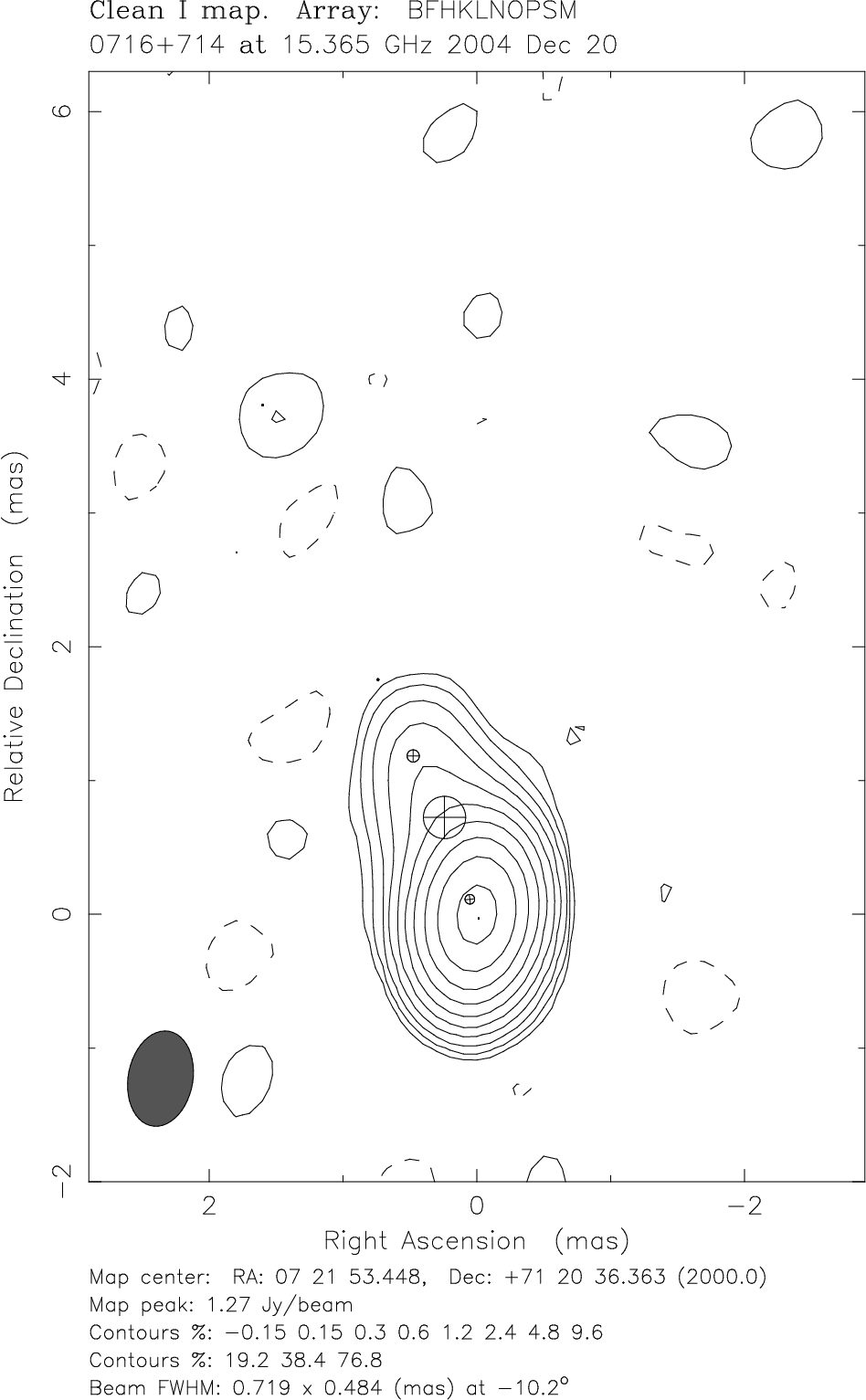}}
\subfigure[]{\includegraphics[clip,width=4cm]{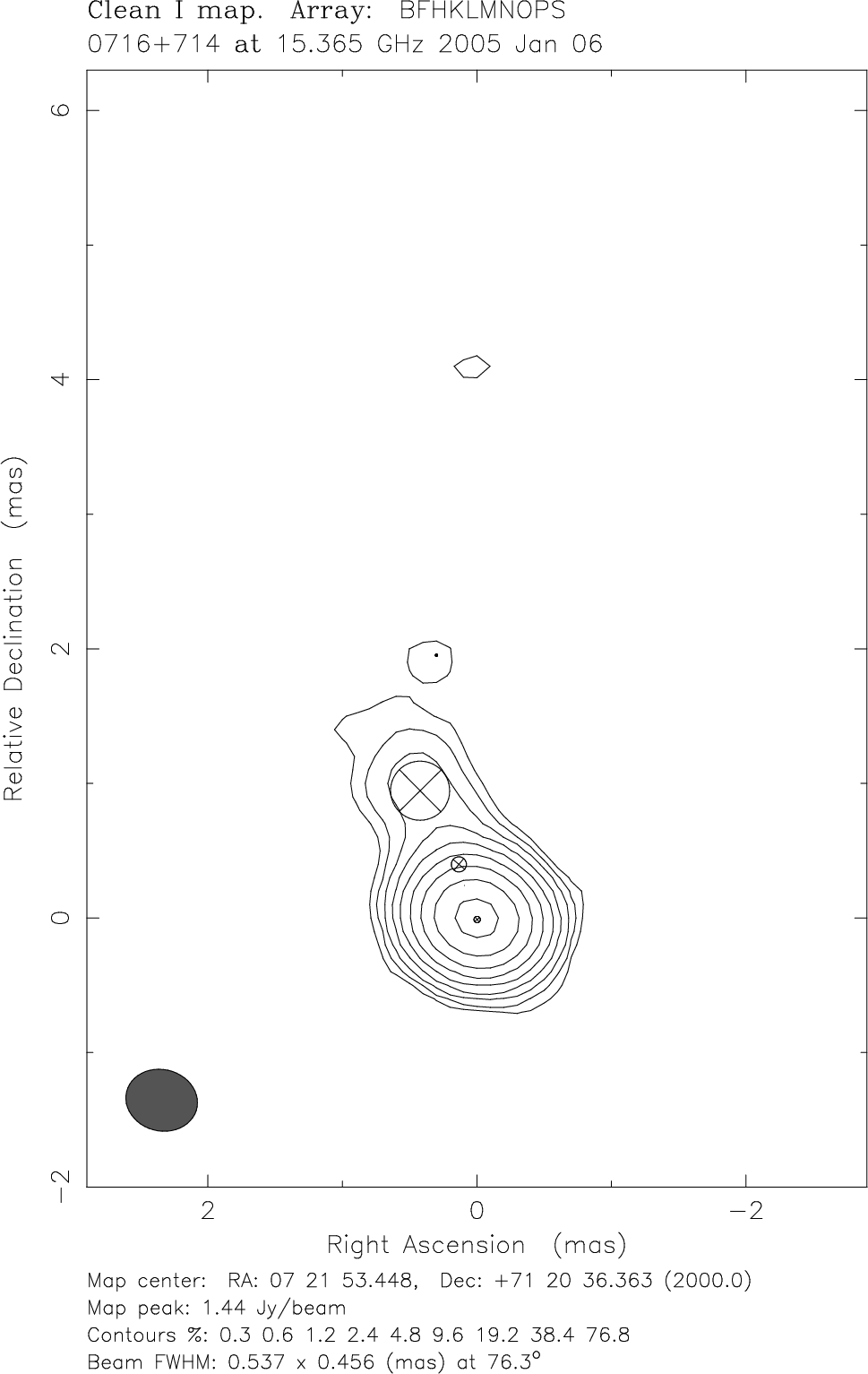}}
\subfigure[]{\includegraphics[clip,width=4cm]{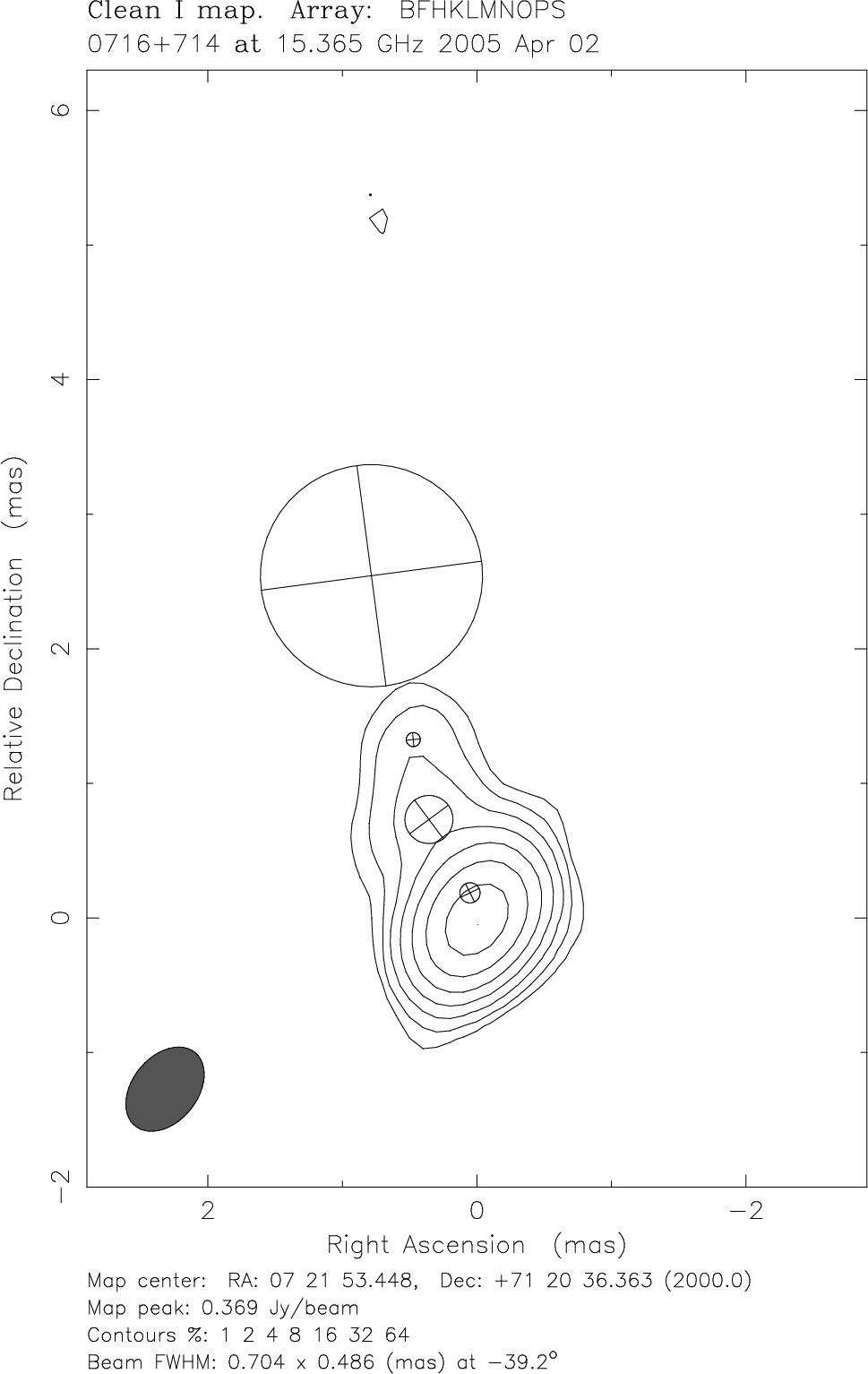}}\\
\subfigure[]{\includegraphics[clip,width=4cm]{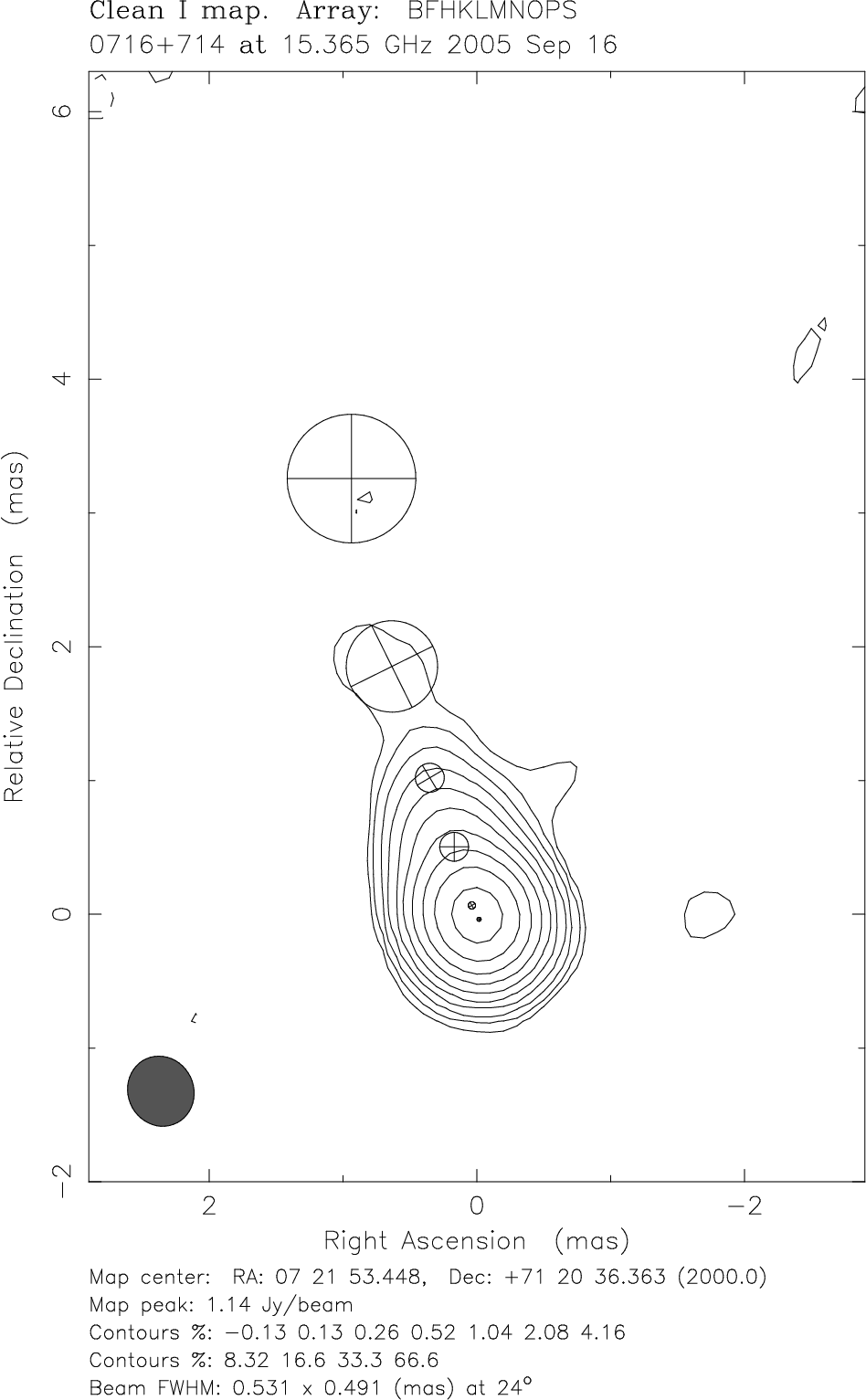}}
\subfigure[]{\includegraphics[clip,width=4cm]{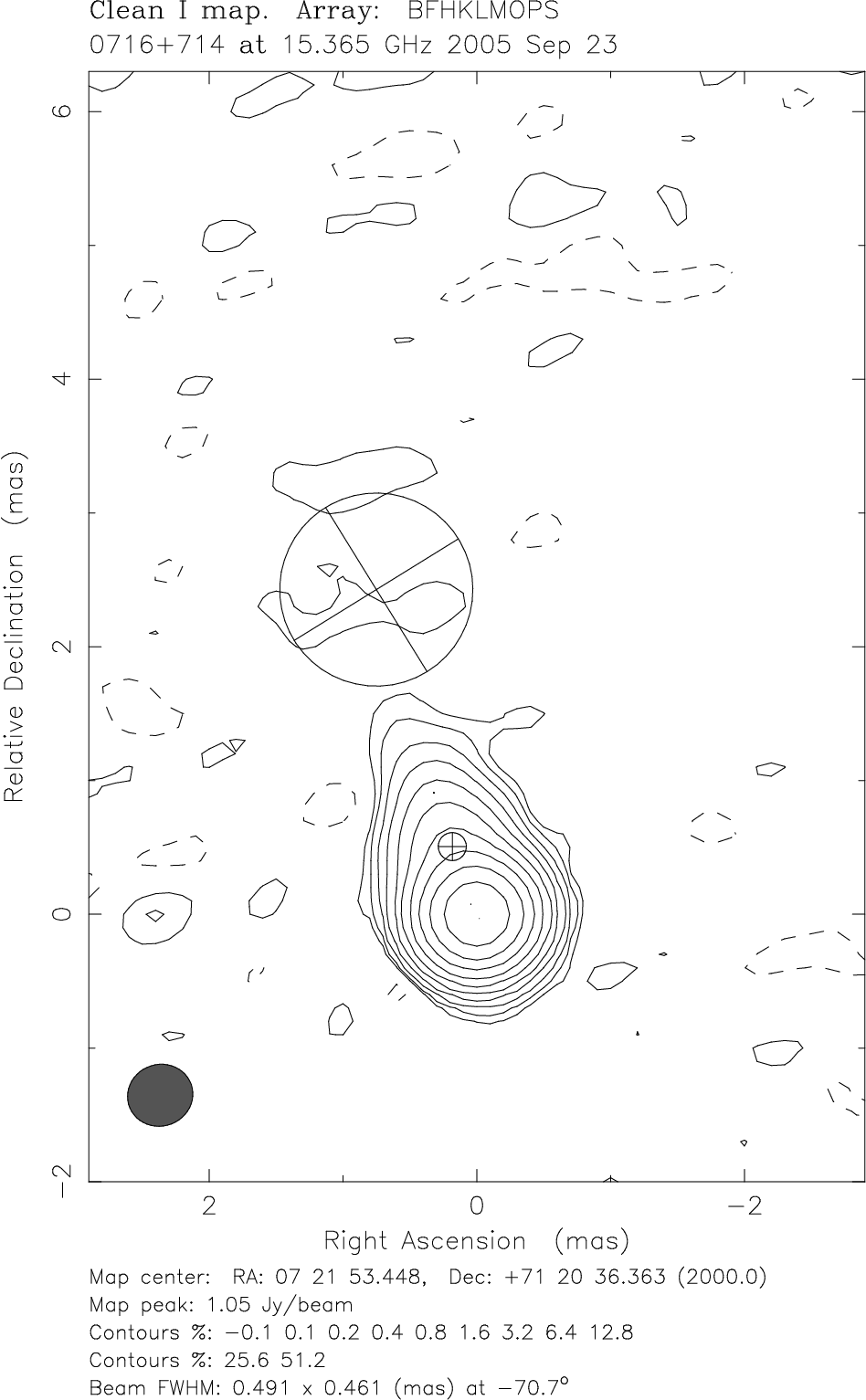}}
\subfigure[]{\includegraphics[clip,width=4cm]{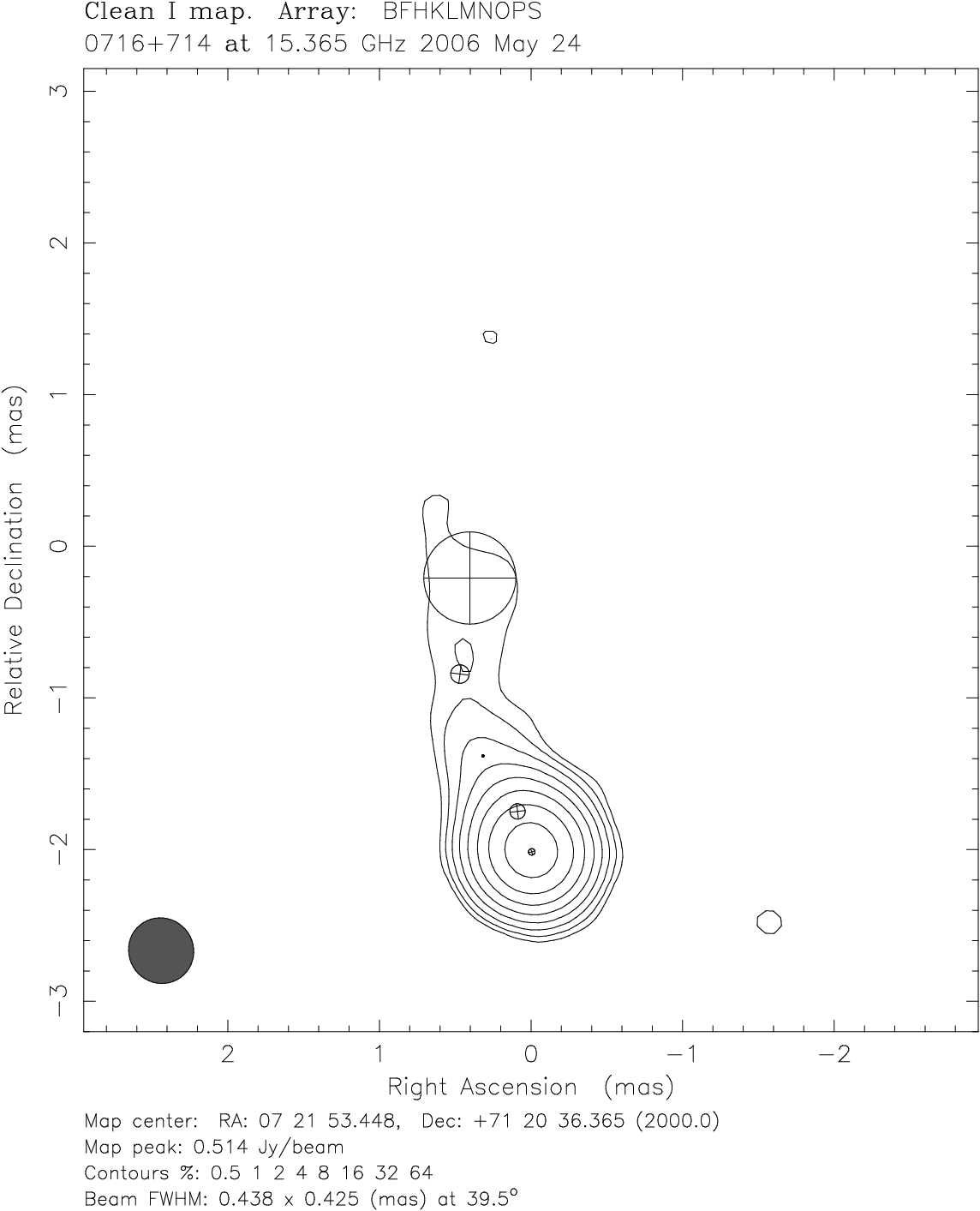}}
\end{center}
\caption{Second part of images of S5 0716+714 at 15 GHz.}
\label{maps152}
\end{figure*}

\begin{figure*}[htb]
\begin{center}
\subfigure[]{\includegraphics[clip,width=4cm]{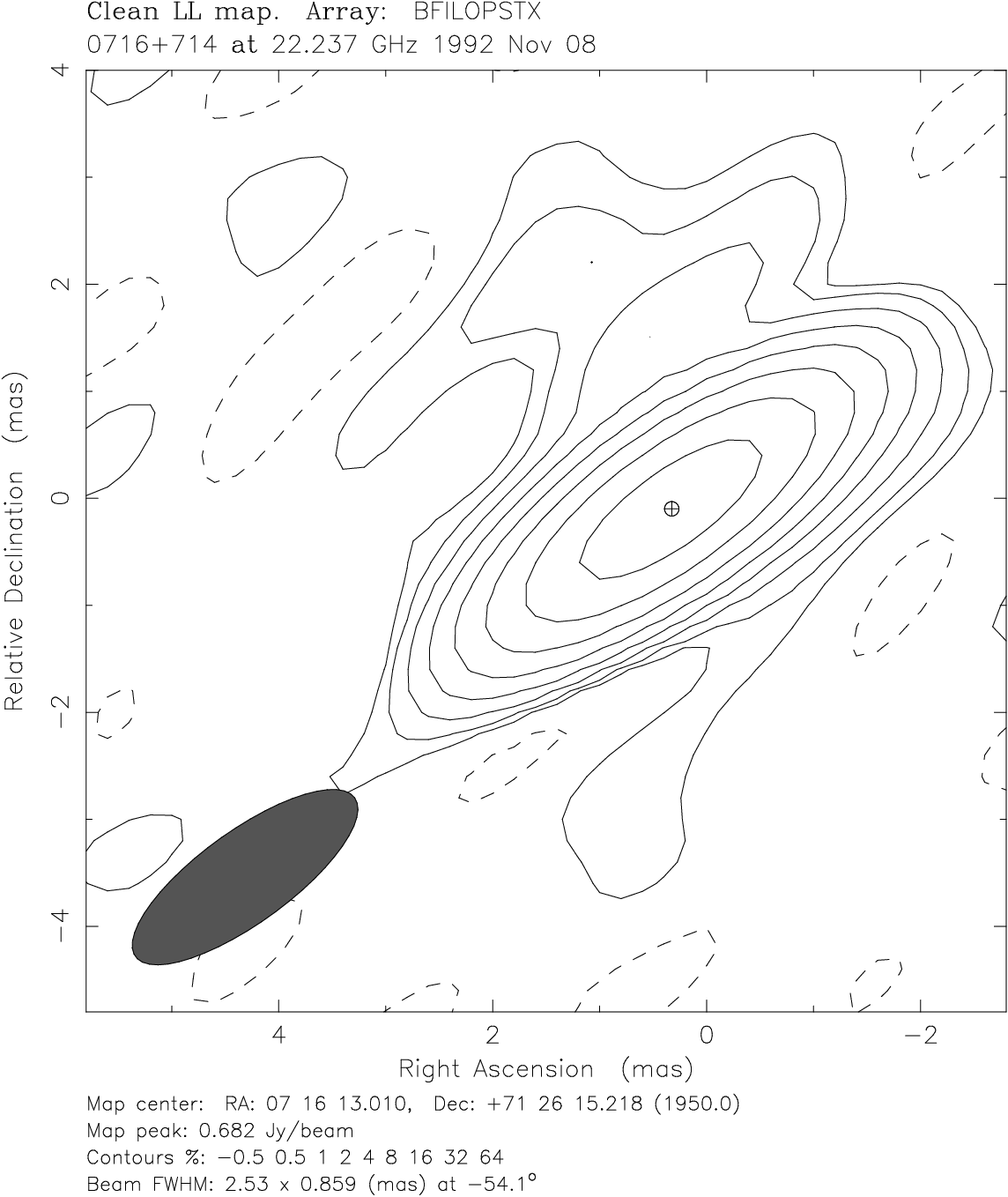}}
\subfigure[]{\includegraphics[clip,width=4cm]{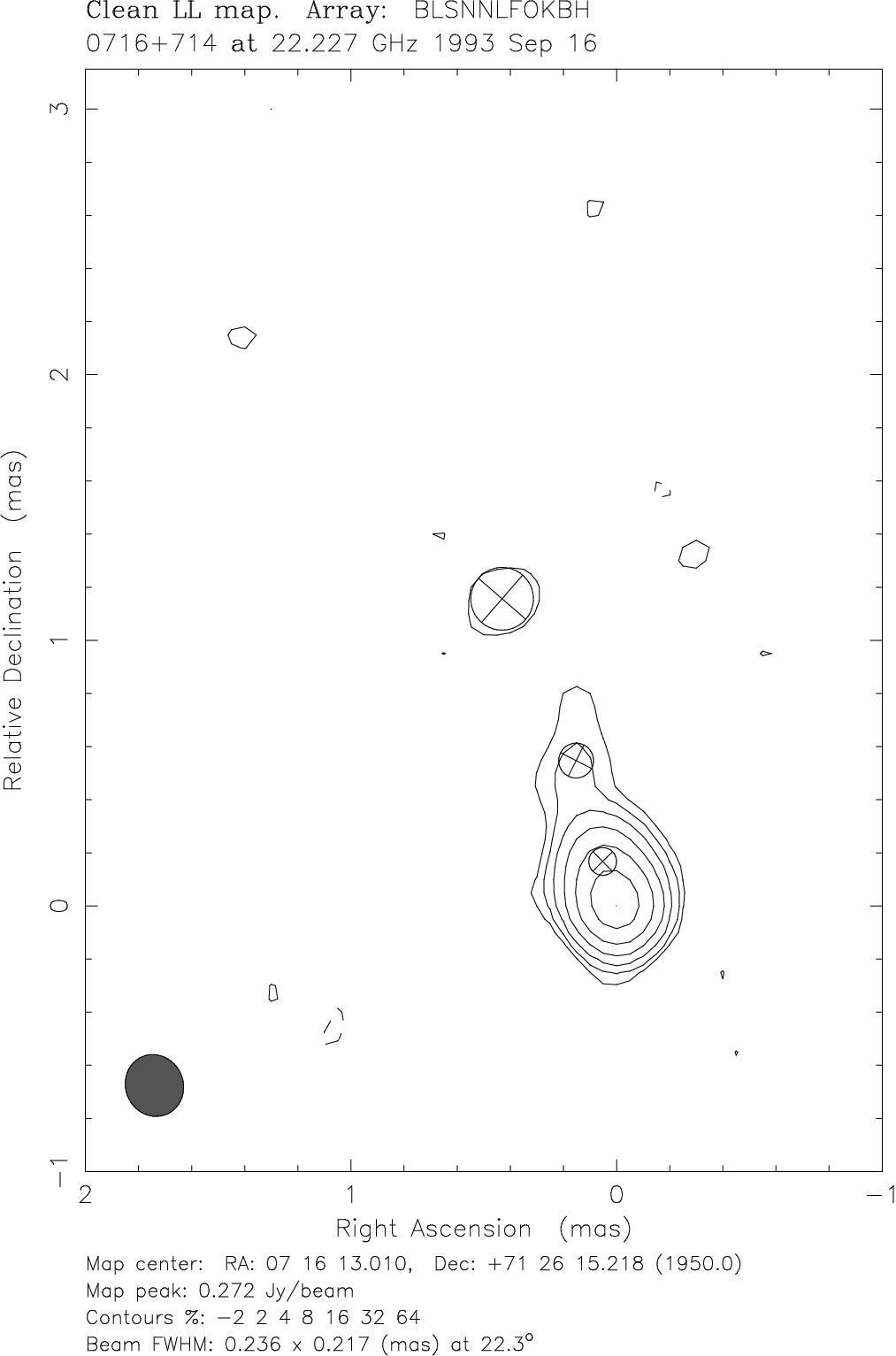}}
\subfigure[]{\includegraphics[clip,width=4cm]{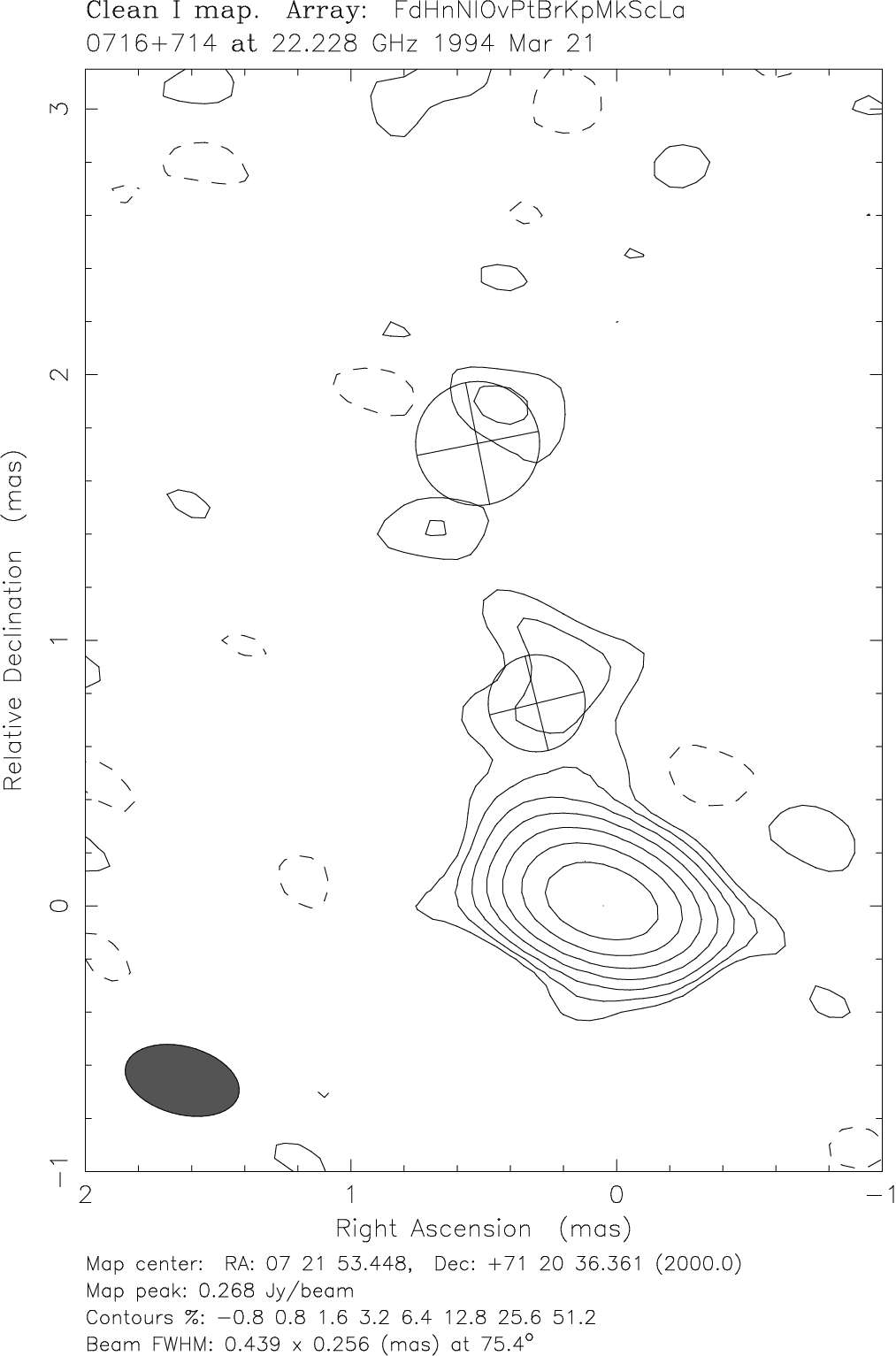}}\\
\subfigure[]{\includegraphics[clip,width=4cm]{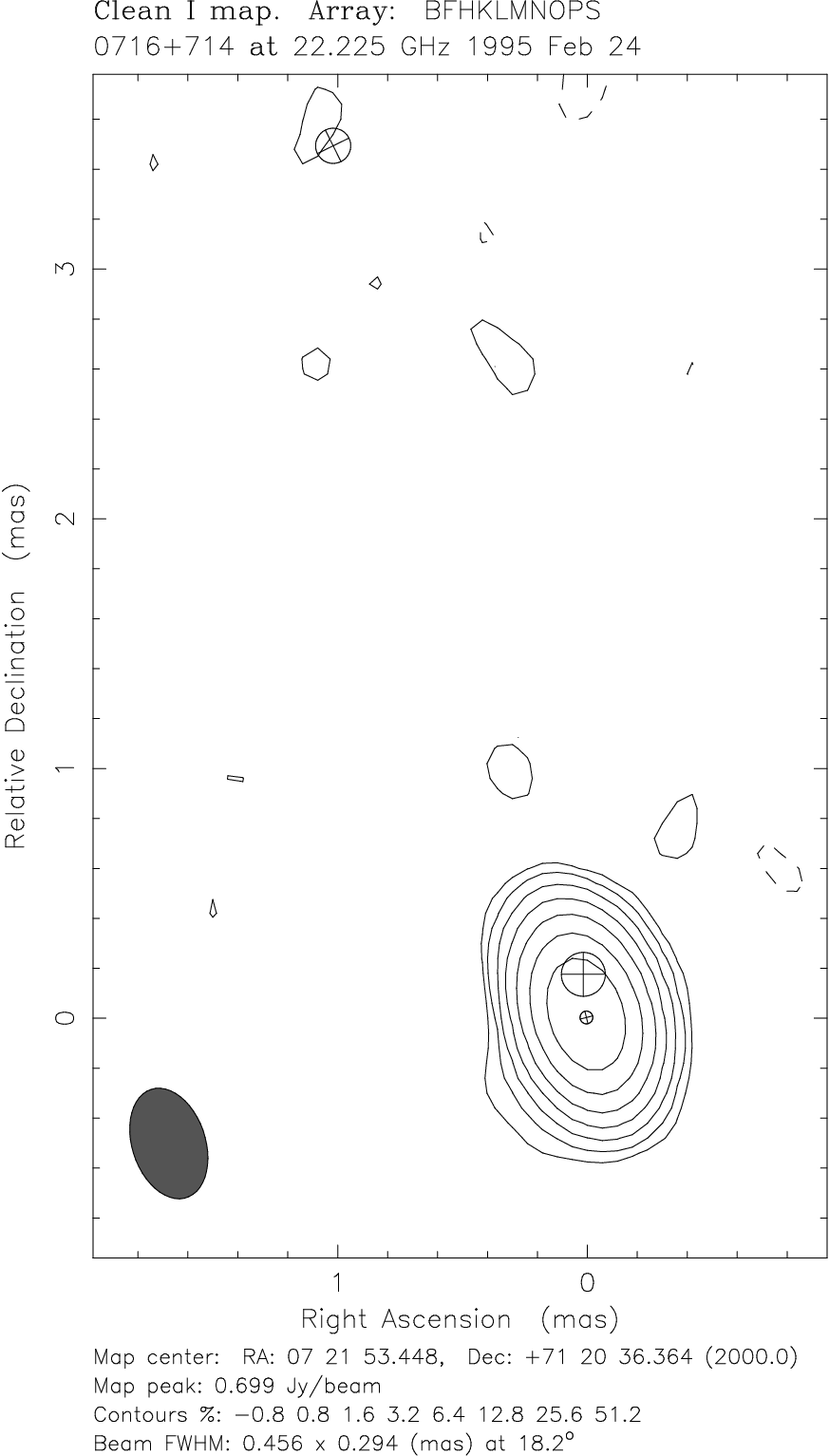}}
\subfigure[]{\includegraphics[clip,width=4cm]{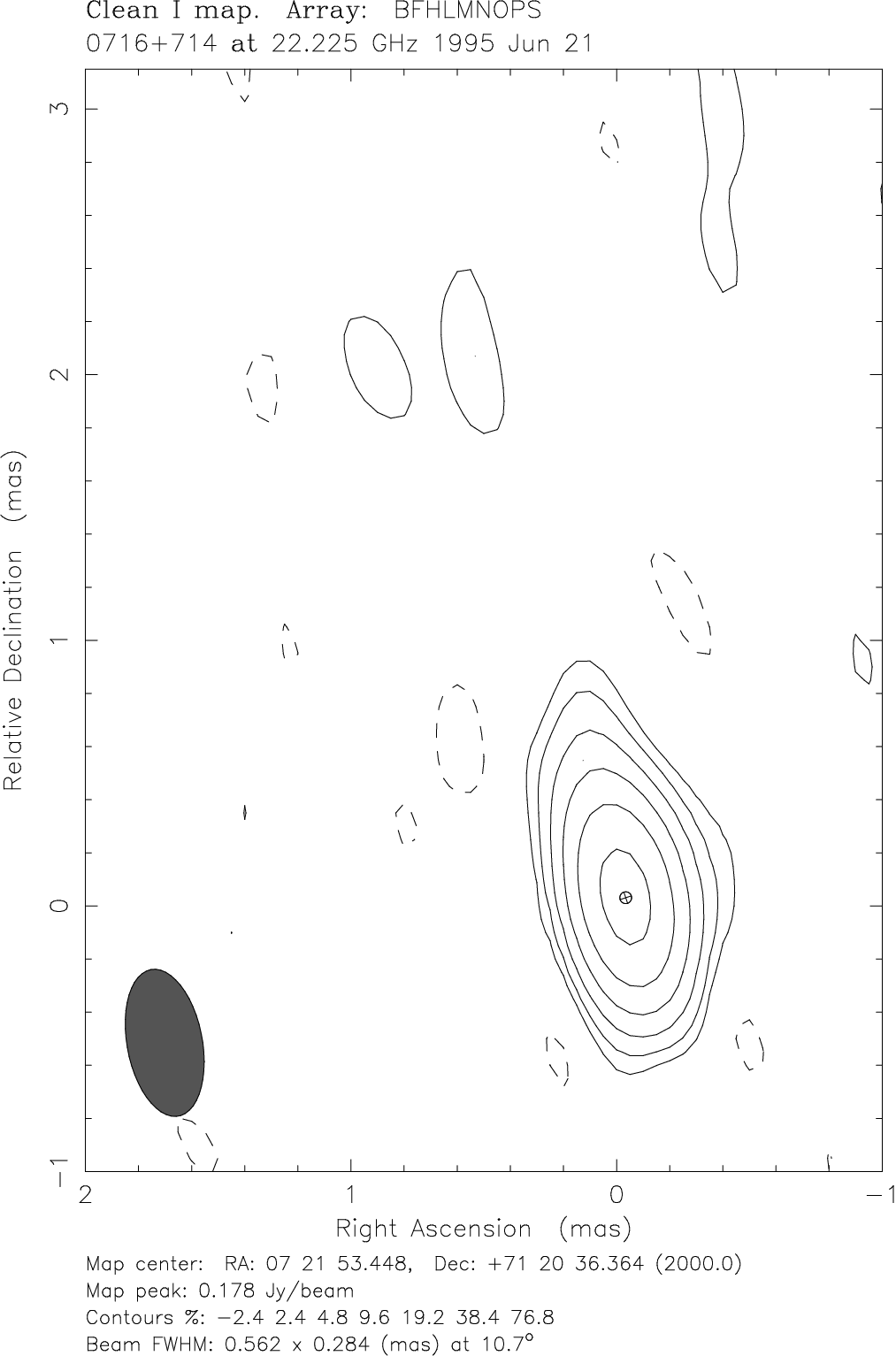}}
\subfigure[]{\includegraphics[clip,width=4cm]{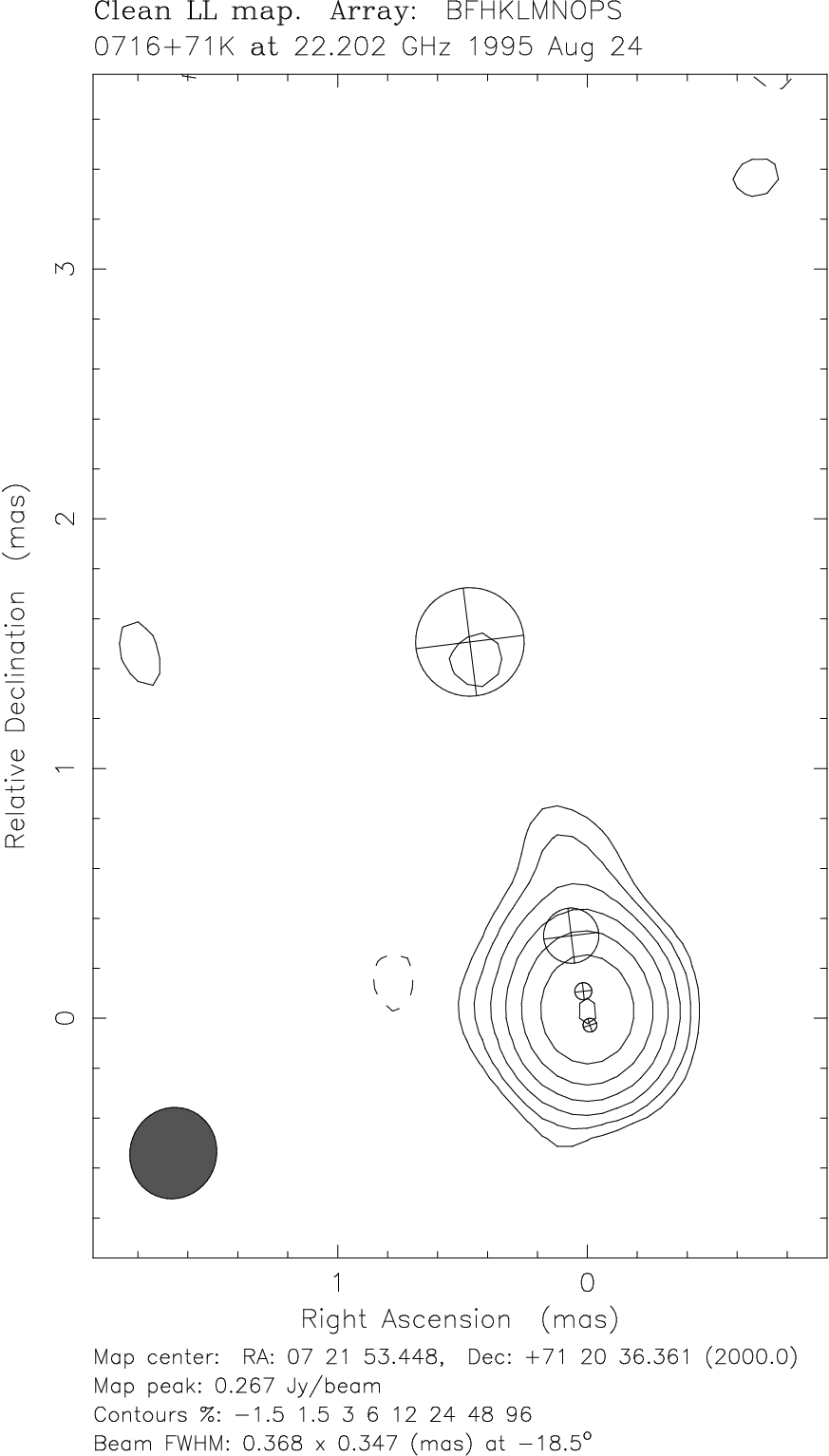}}\\
\subfigure[]{\includegraphics[clip,width=4cm]{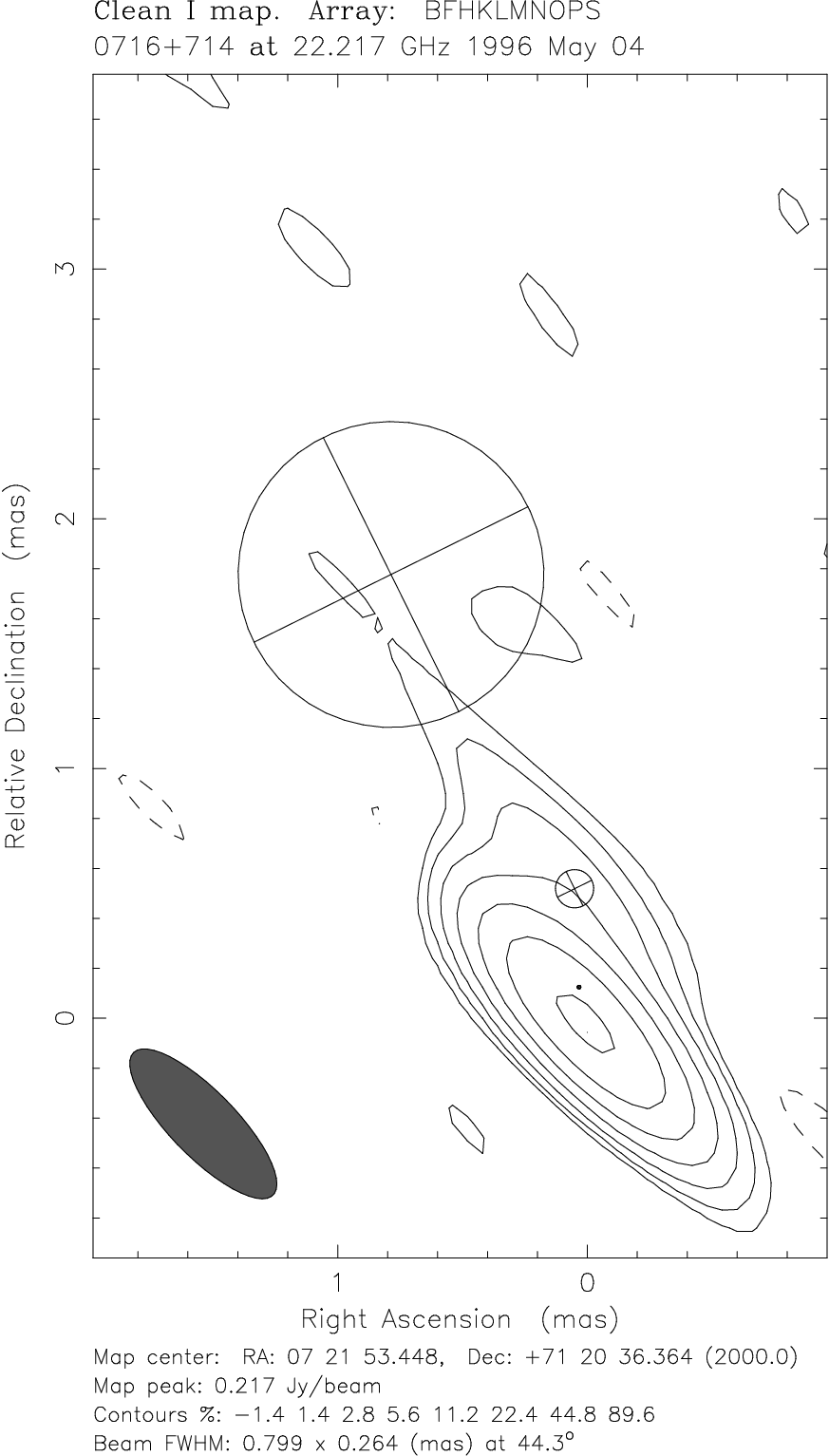}}
\subfigure[]{\includegraphics[clip,width=4cm]{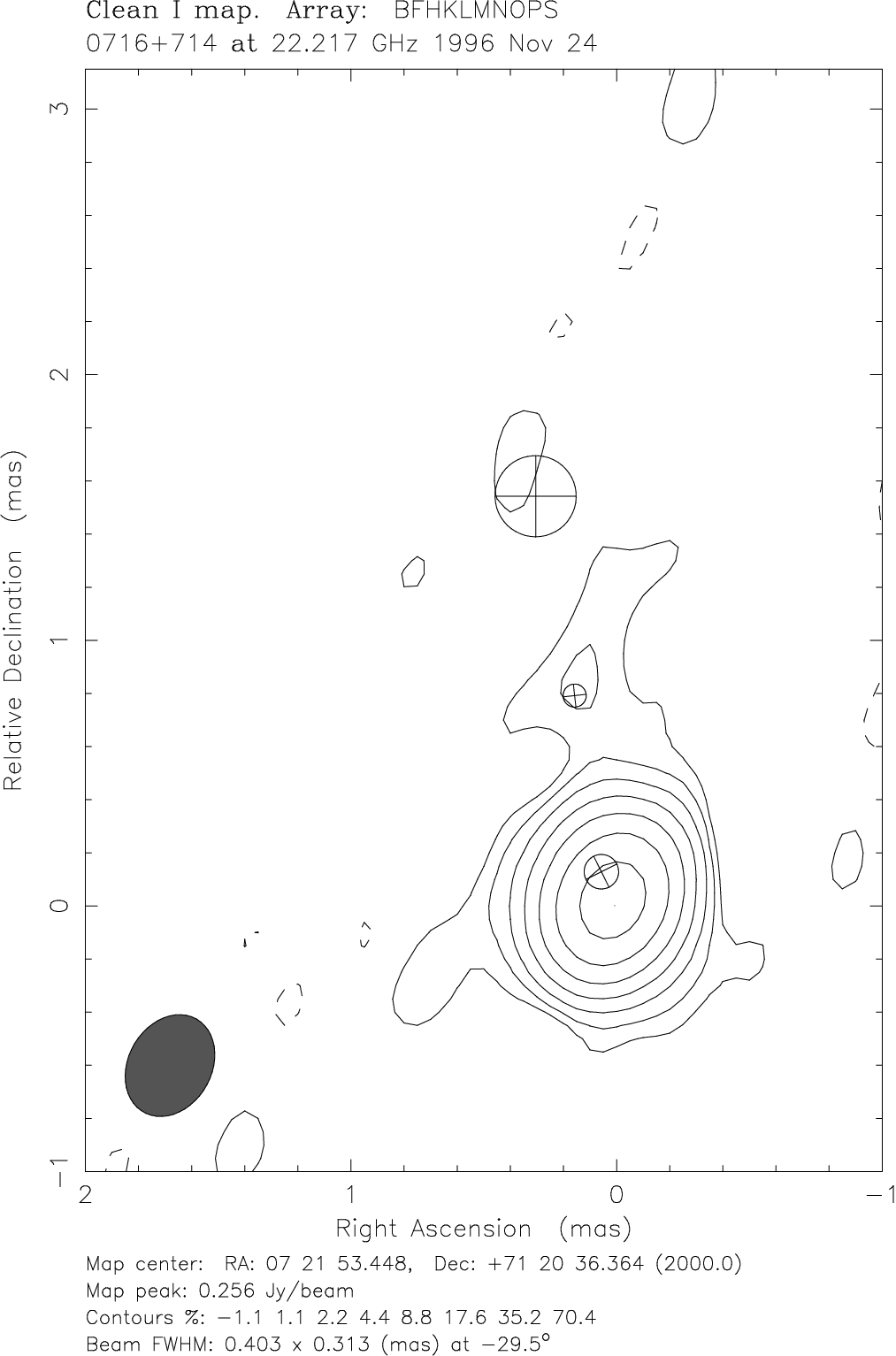}}
\subfigure[]{\includegraphics[clip,width=4cm]{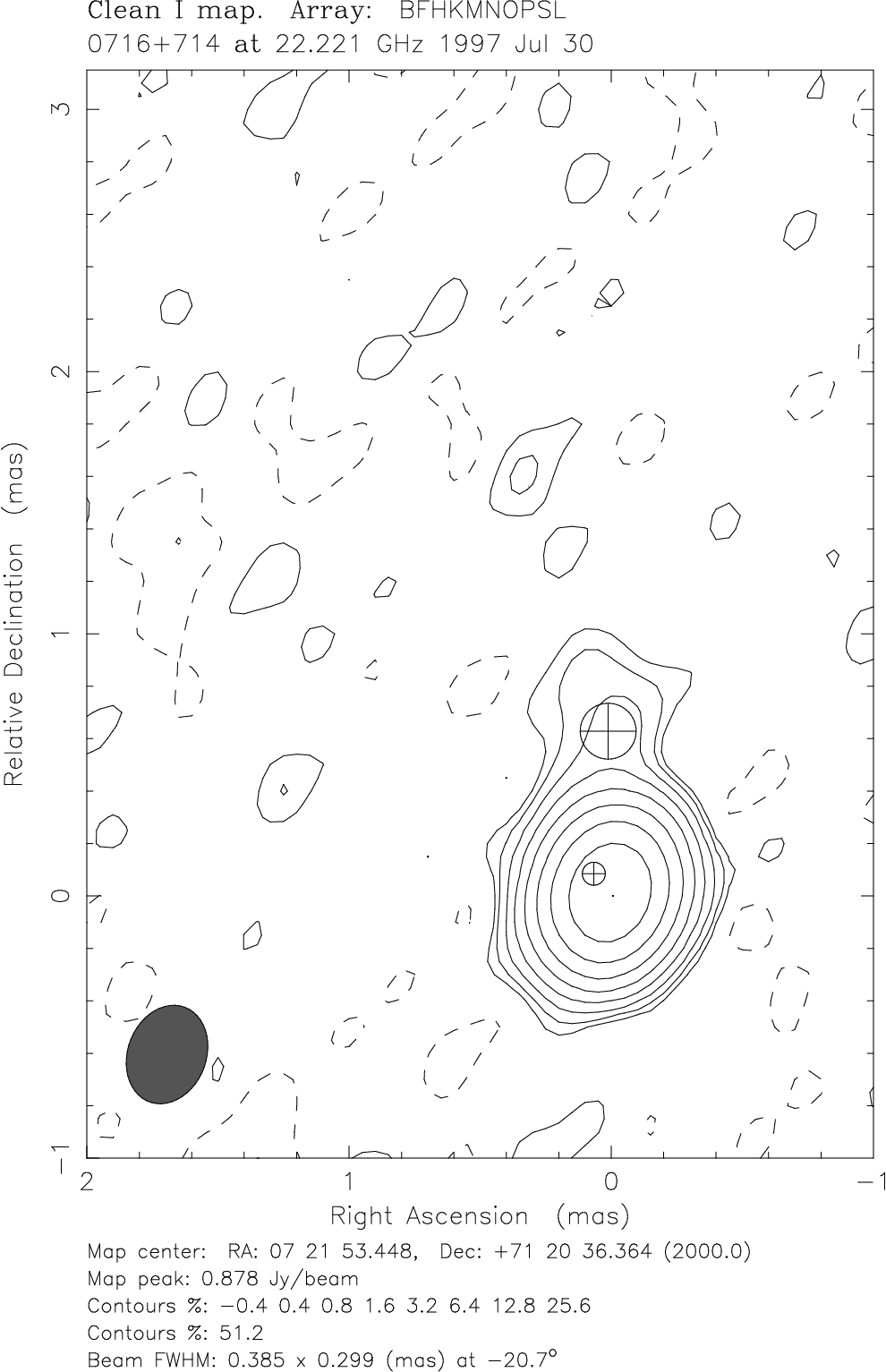}}\\
\end{center}
\caption{First part of images of S5 0716+714 at 22 GHz.}
\label{maps221}
\end{figure*}

\begin{figure*}[htb]
\begin{center}
\subfigure[]{\includegraphics[clip,width=4cm]{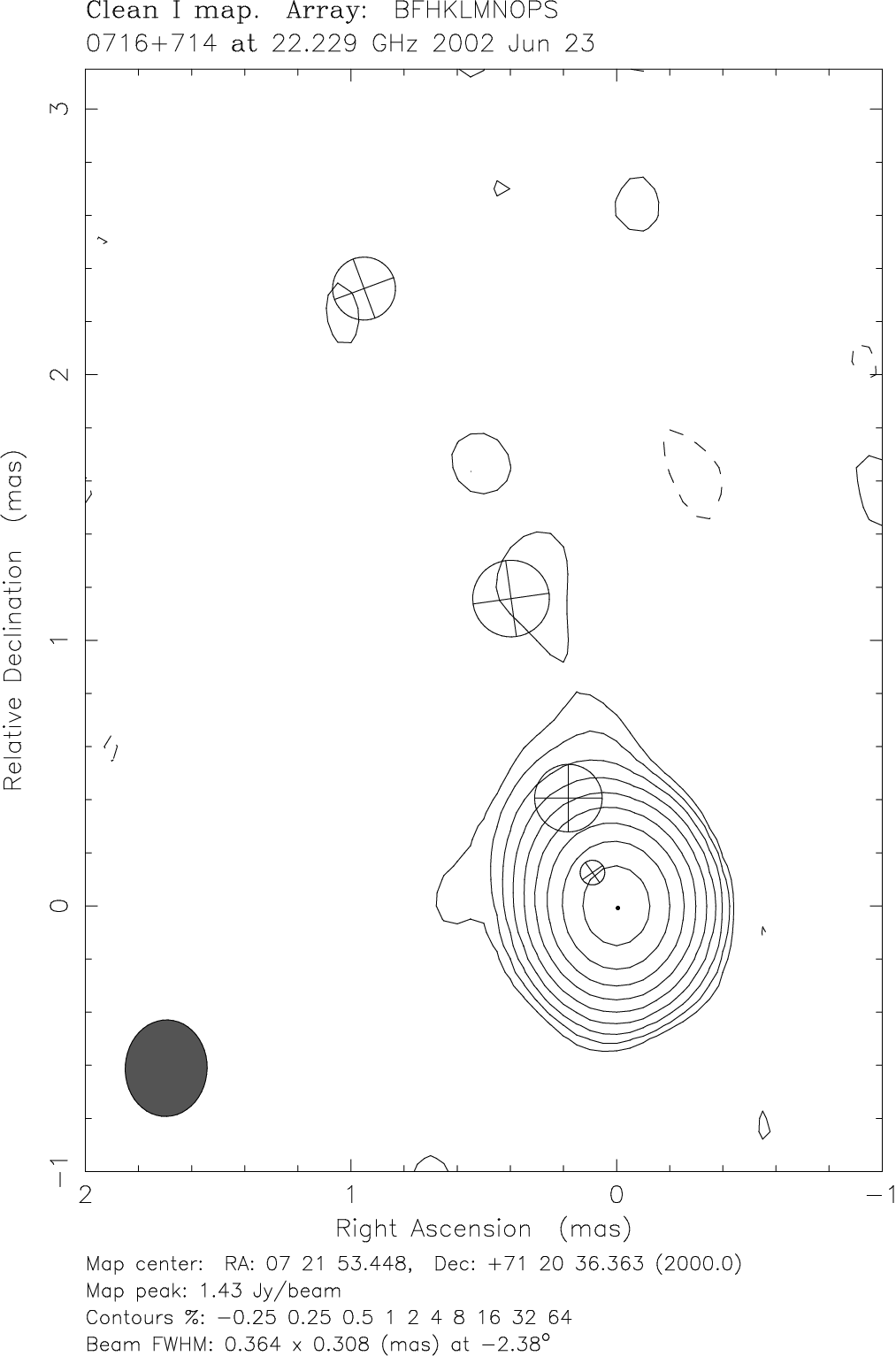}}
\subfigure[]{\includegraphics[clip,width=4cm]{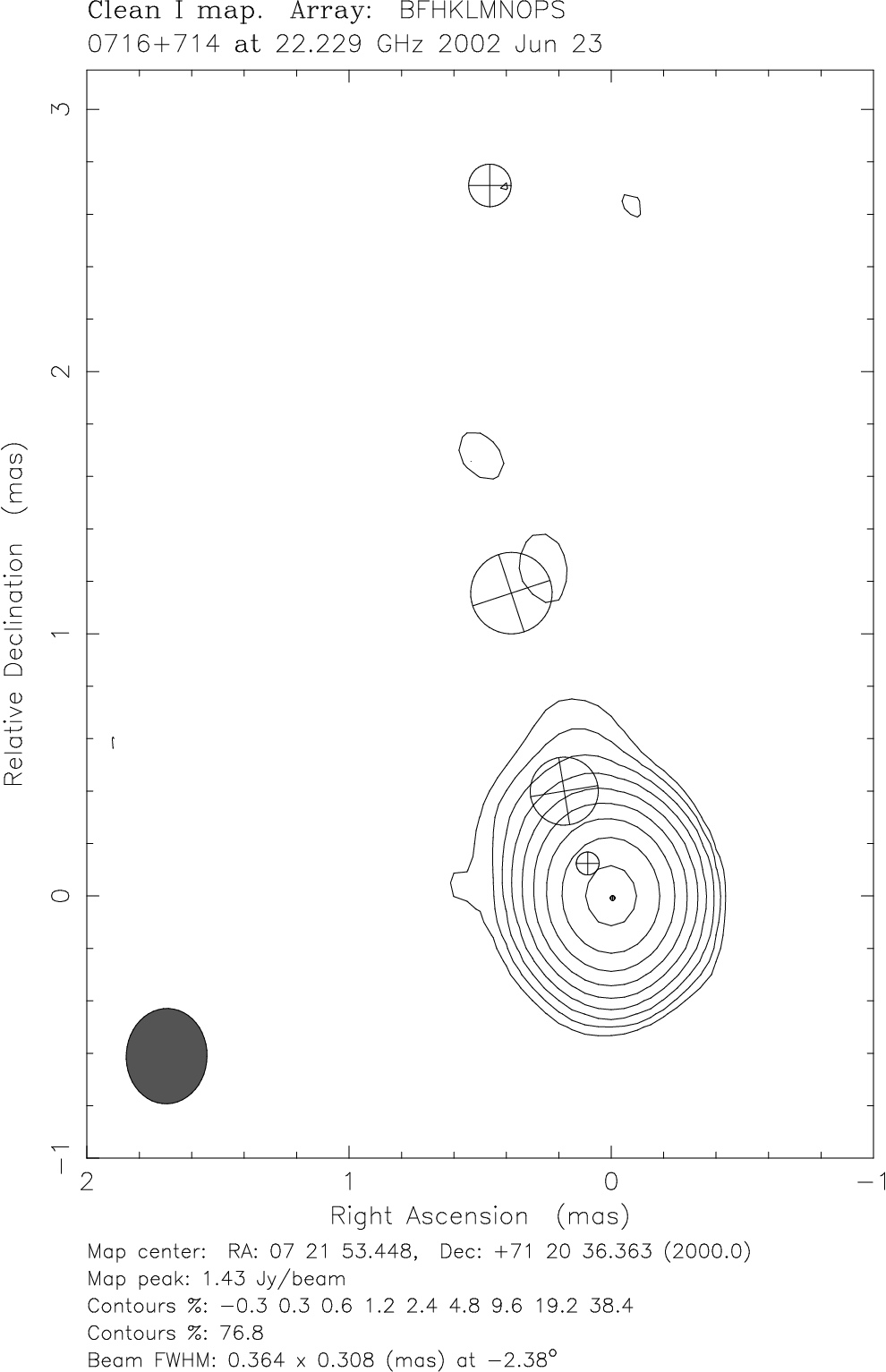}}
\subfigure[]{\includegraphics[clip,width=4cm]{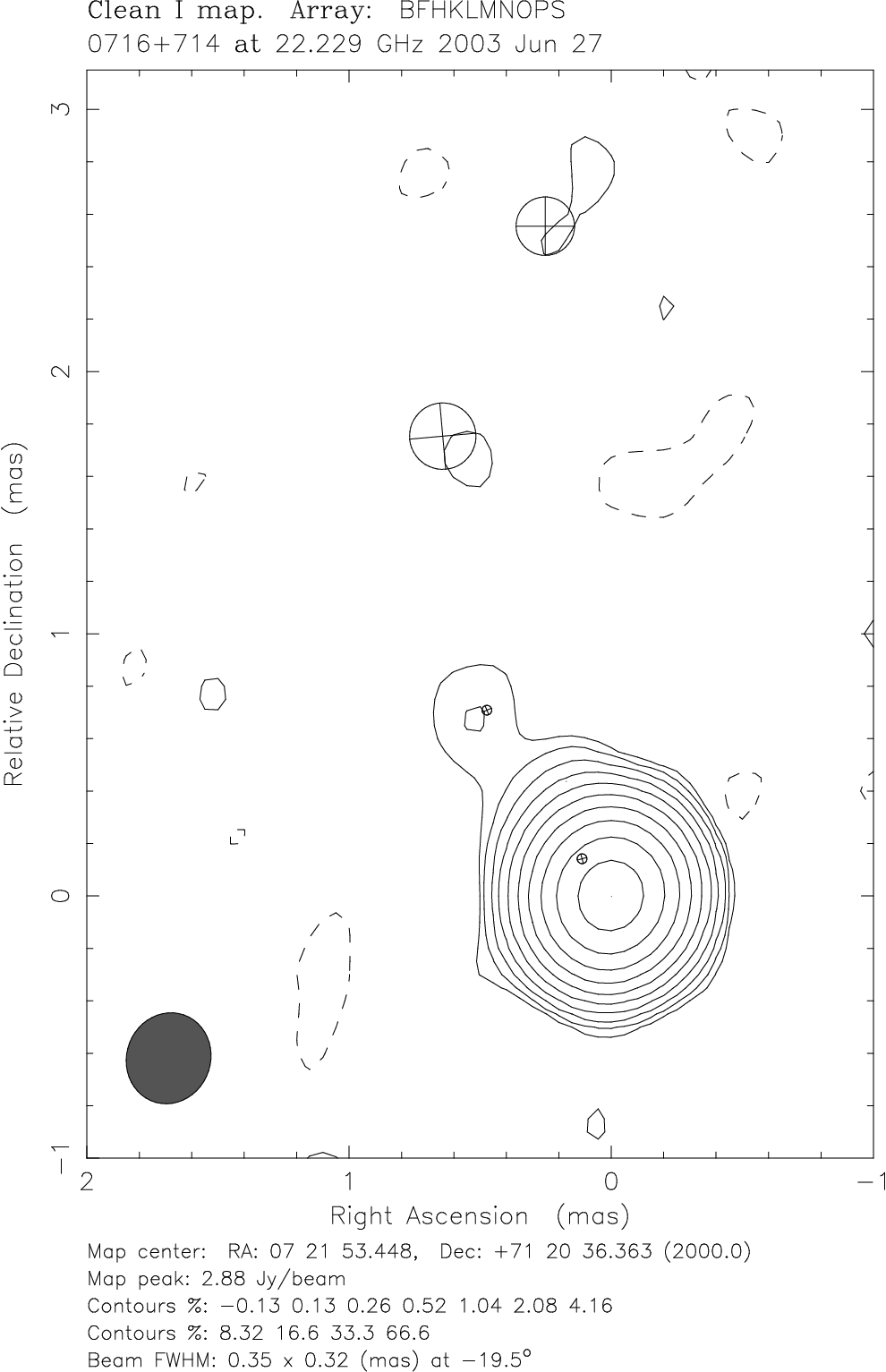}}\\
\subfigure[]{\includegraphics[clip,width=4cm]{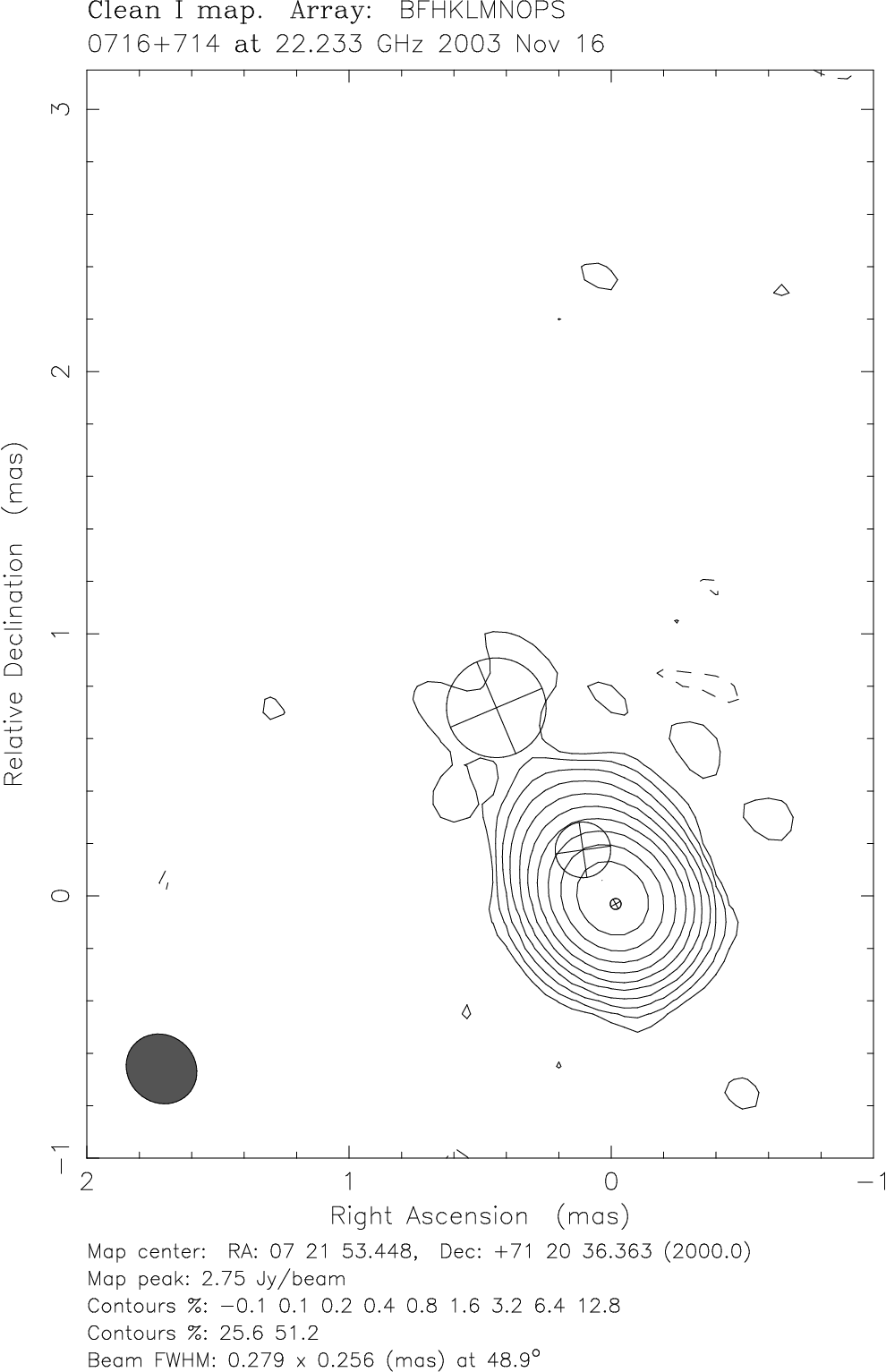}}
\subfigure[]{\includegraphics[clip,width=4cm]{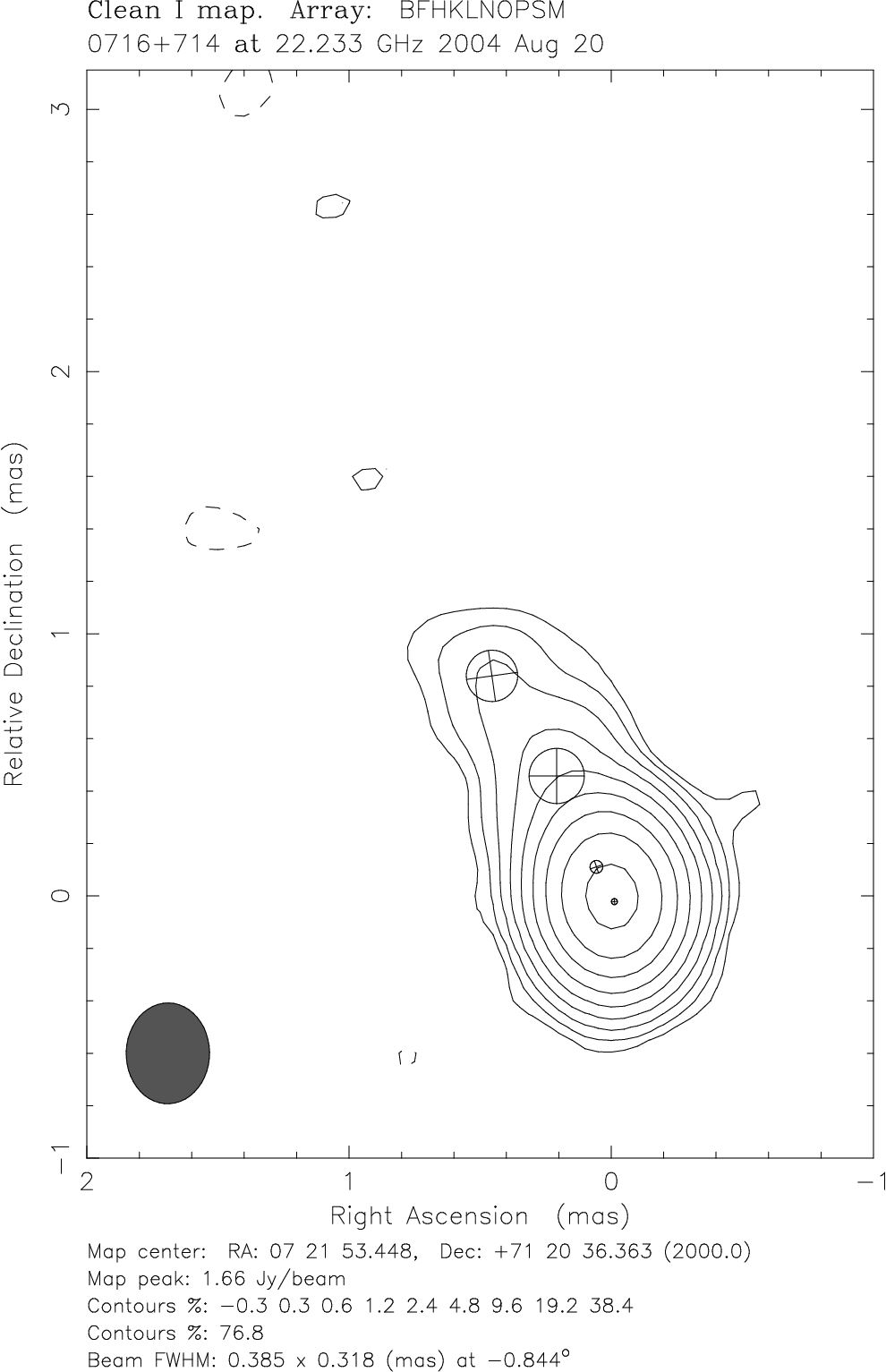}}
\subfigure[]{\includegraphics[clip,width=4cm]{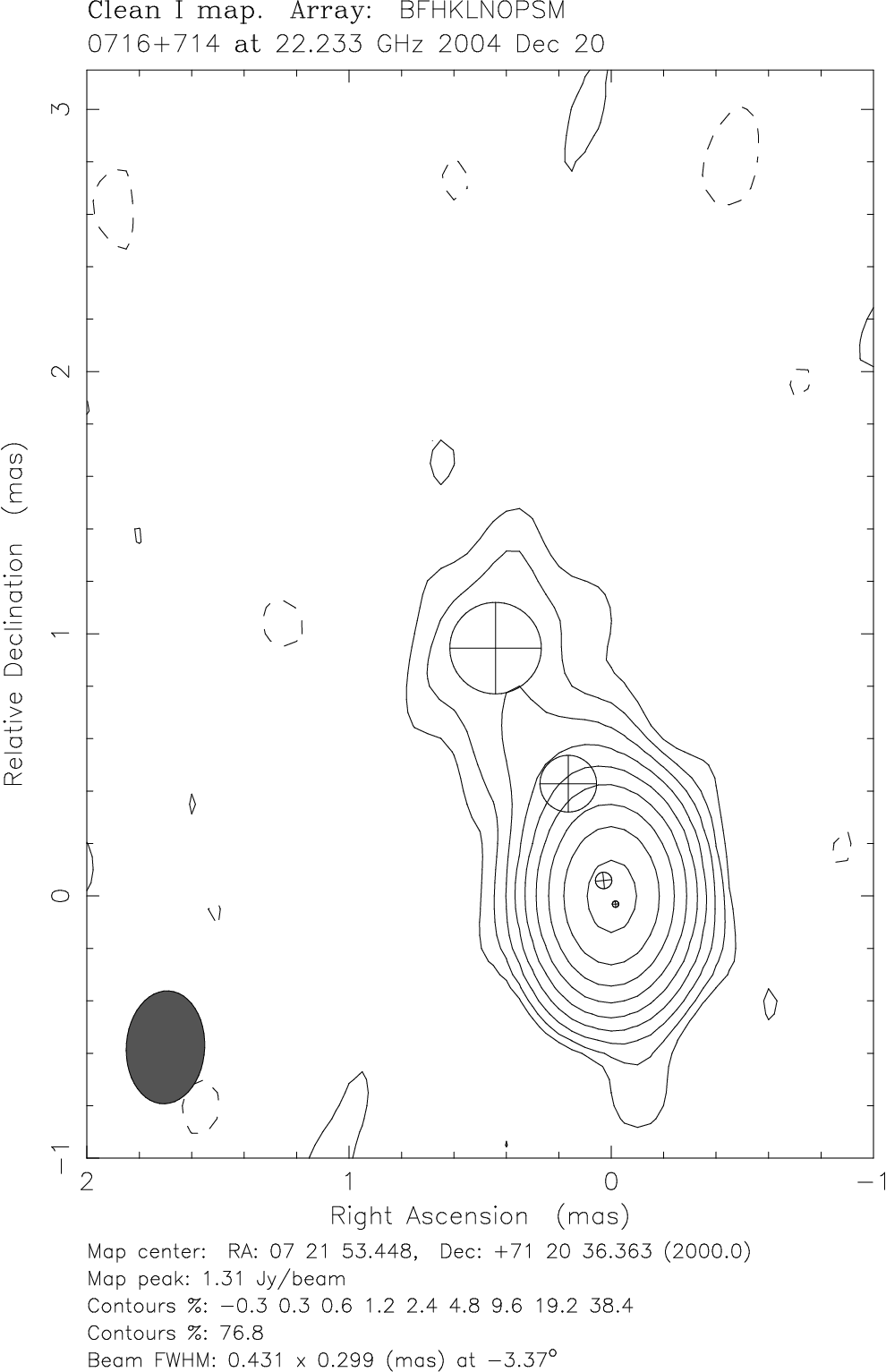}}\\
\subfigure[]{\includegraphics[clip,width=4cm]{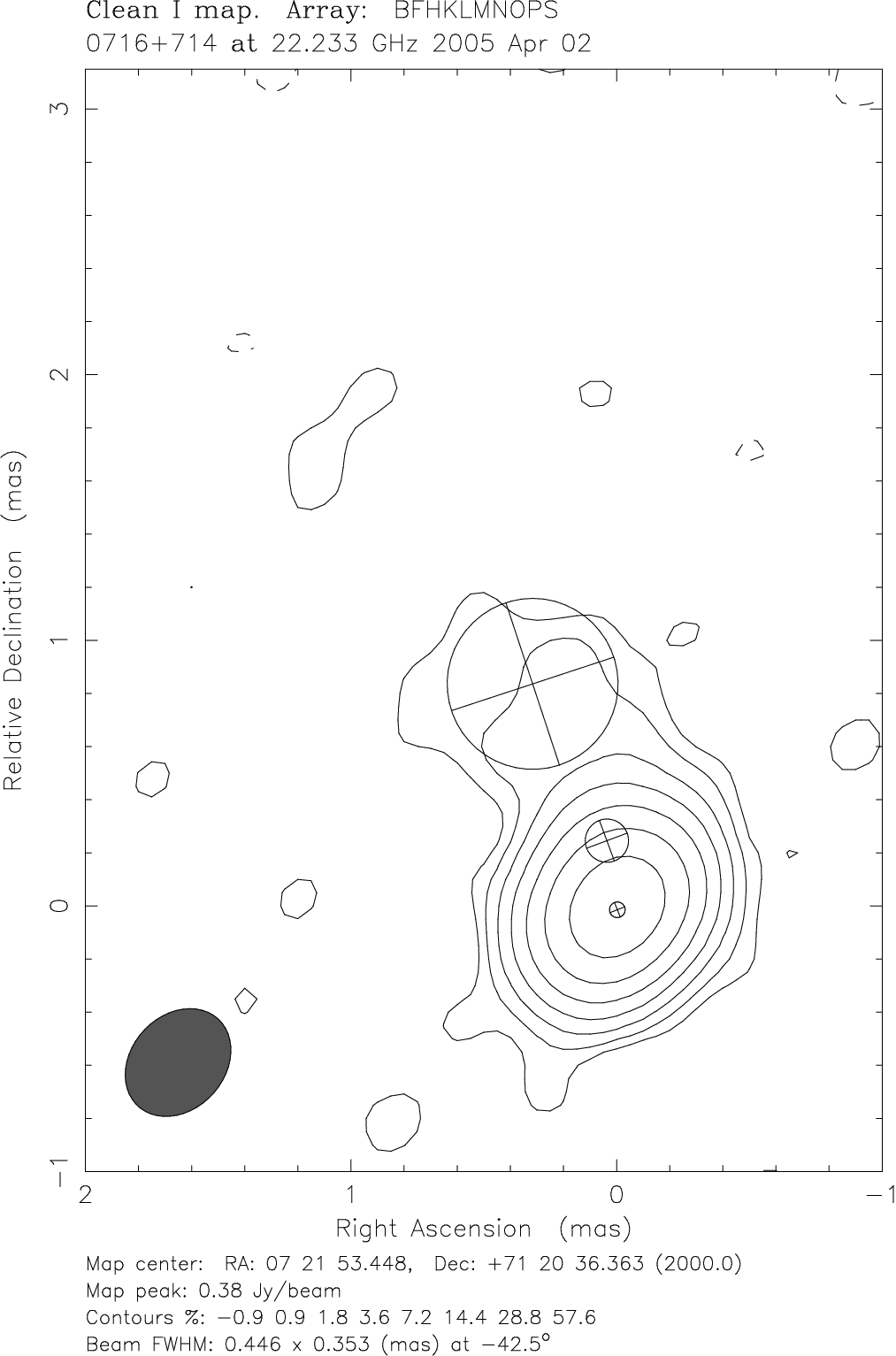}}
\end{center}
\caption{Second part of images of S5 0716+714 at 22 GHz.}
\label{maps222}
\end{figure*}

\begin{figure*}[htb]
\begin{center}
\subfigure[]{\includegraphics[clip,width=4cm]{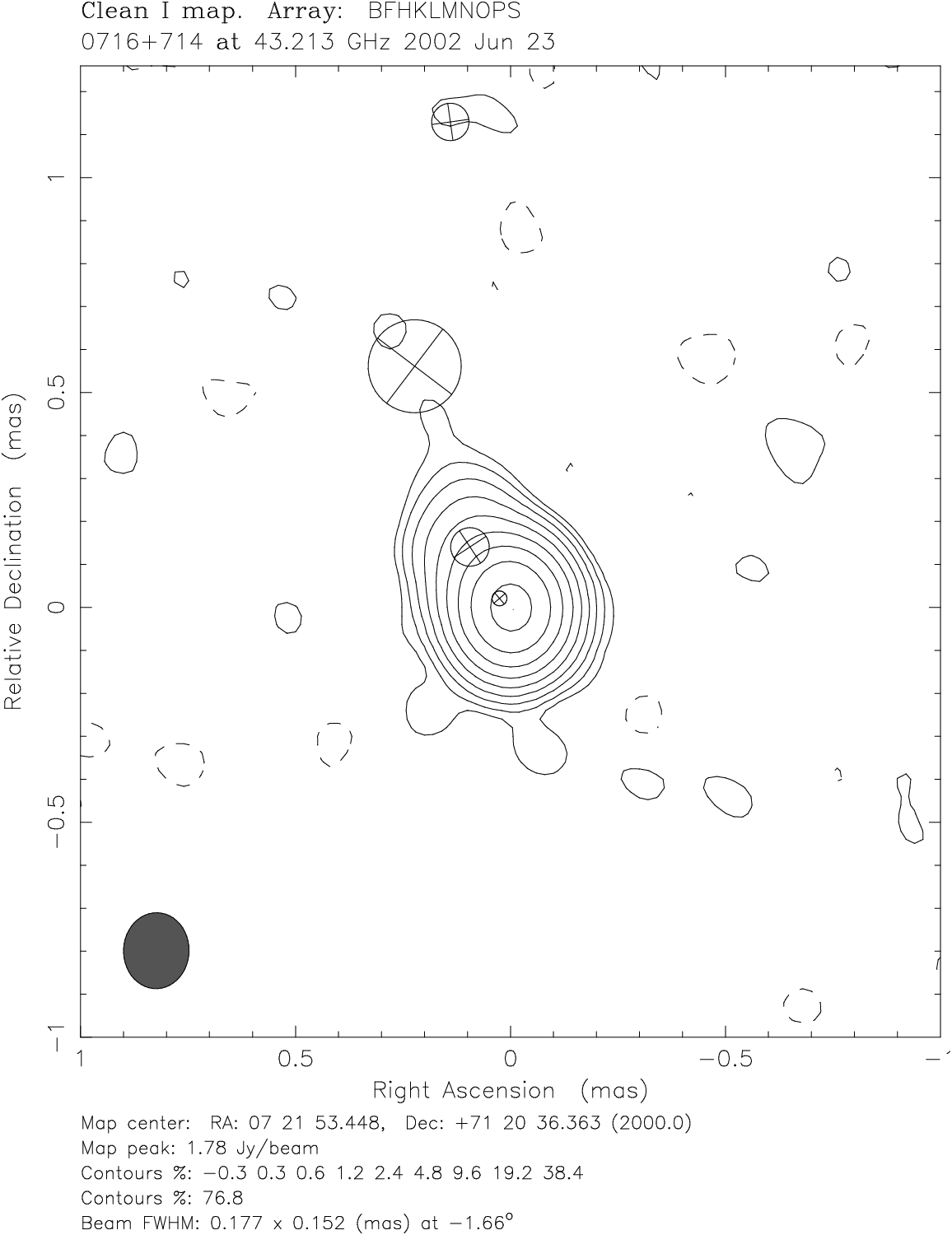}}
\subfigure[]{\includegraphics[clip,width=4cm]{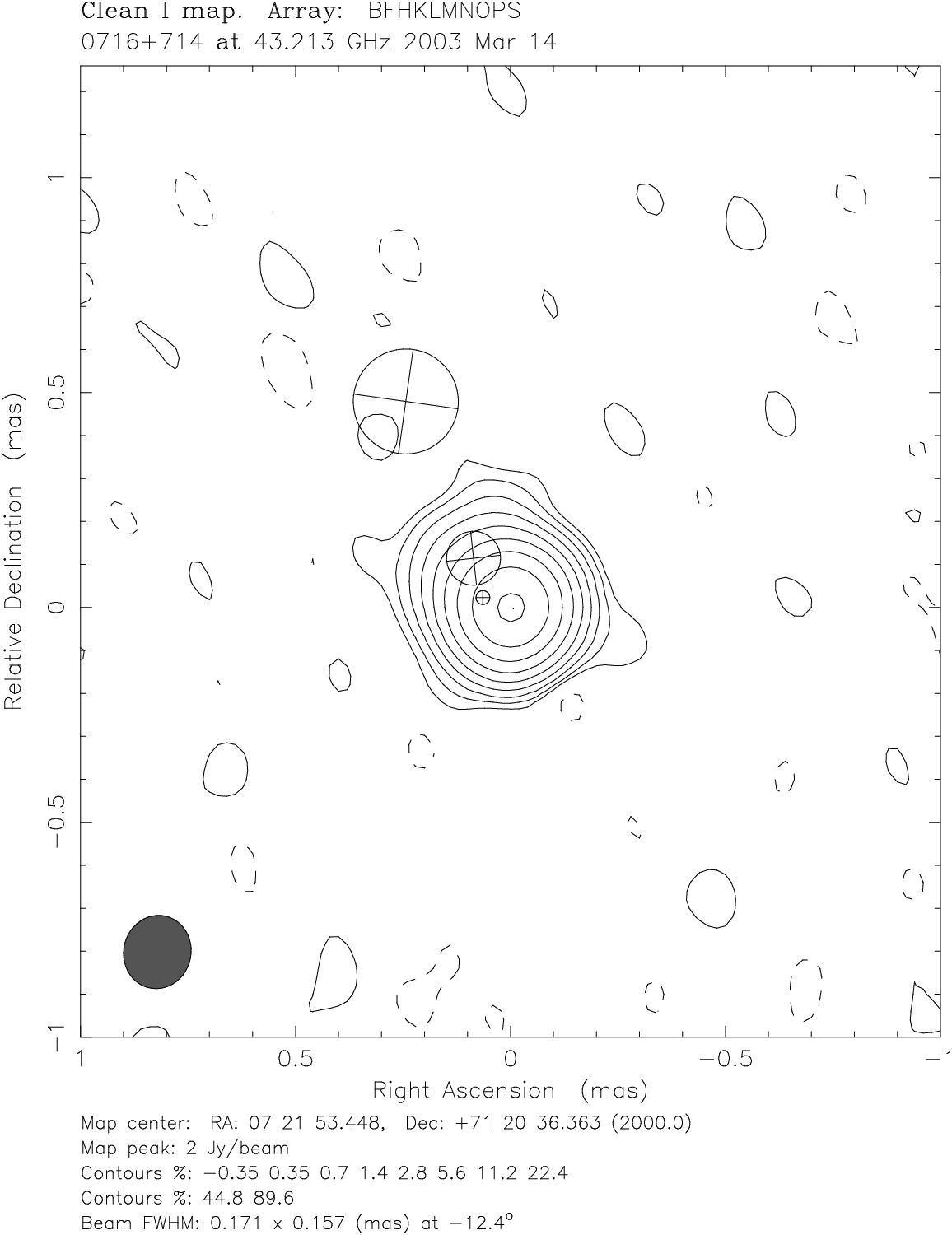}}
\subfigure[]{\includegraphics[clip,width=4cm]{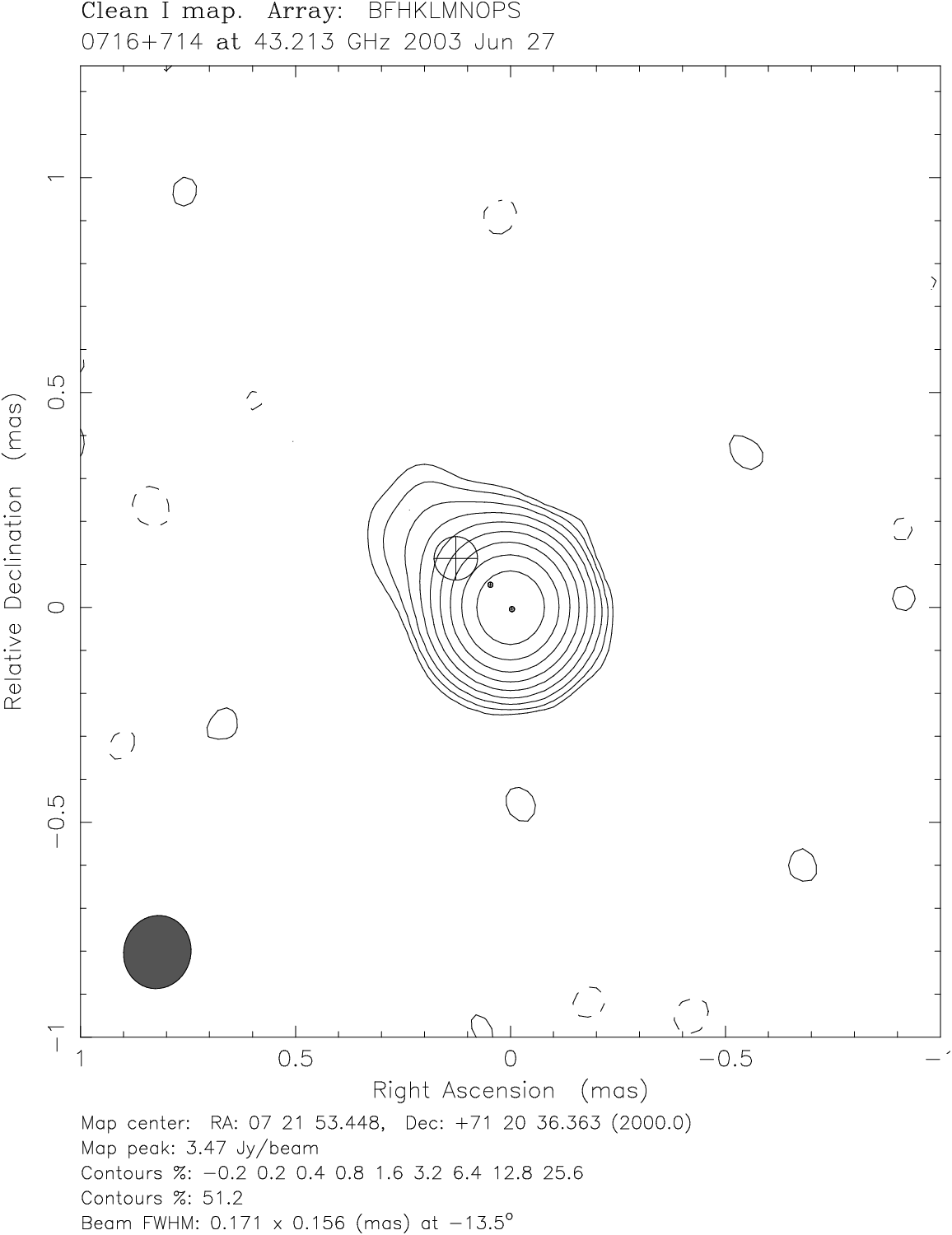}}\\
\subfigure[]{\includegraphics[clip,width=4cm]{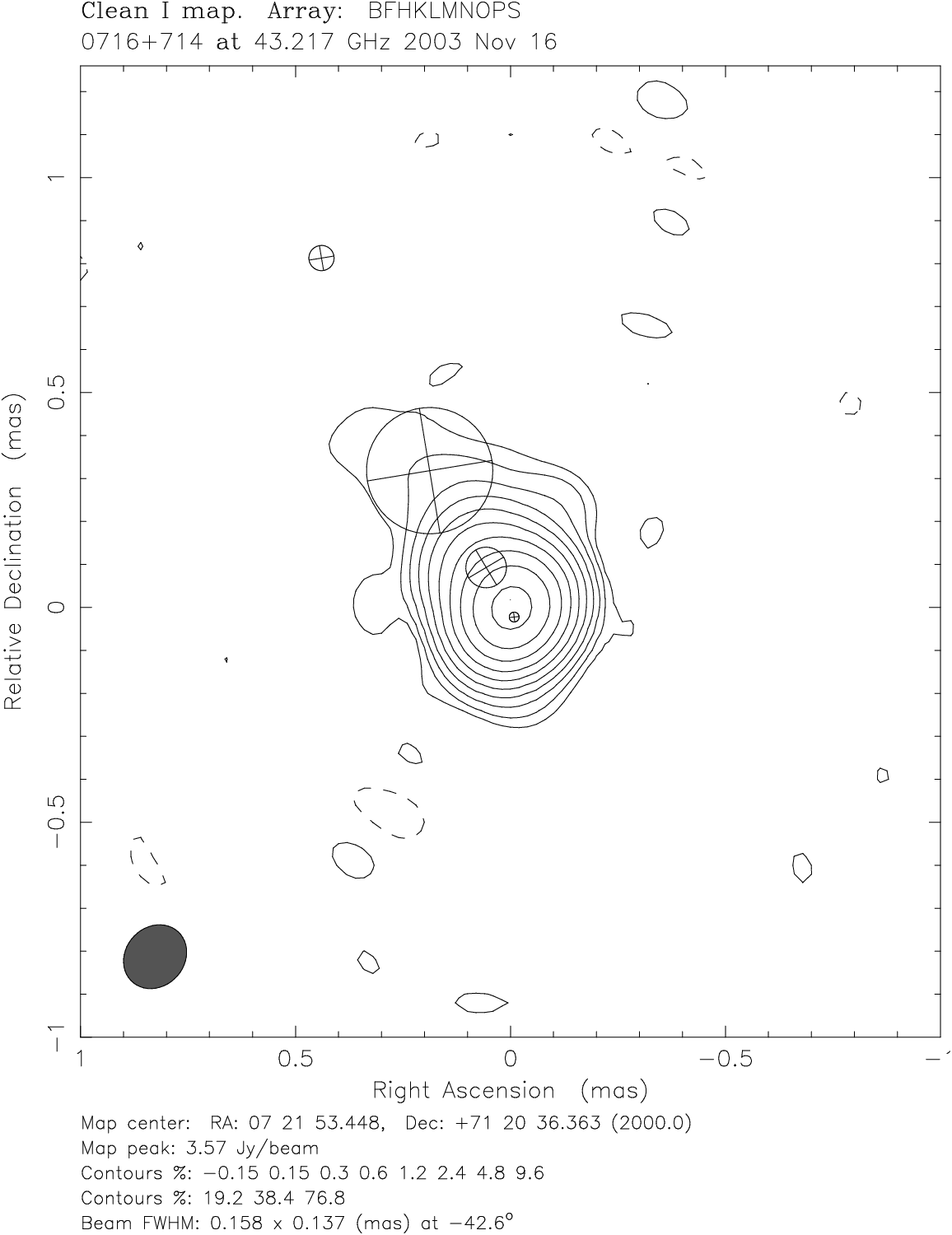}}
\subfigure[]{\includegraphics[clip,width=4cm]{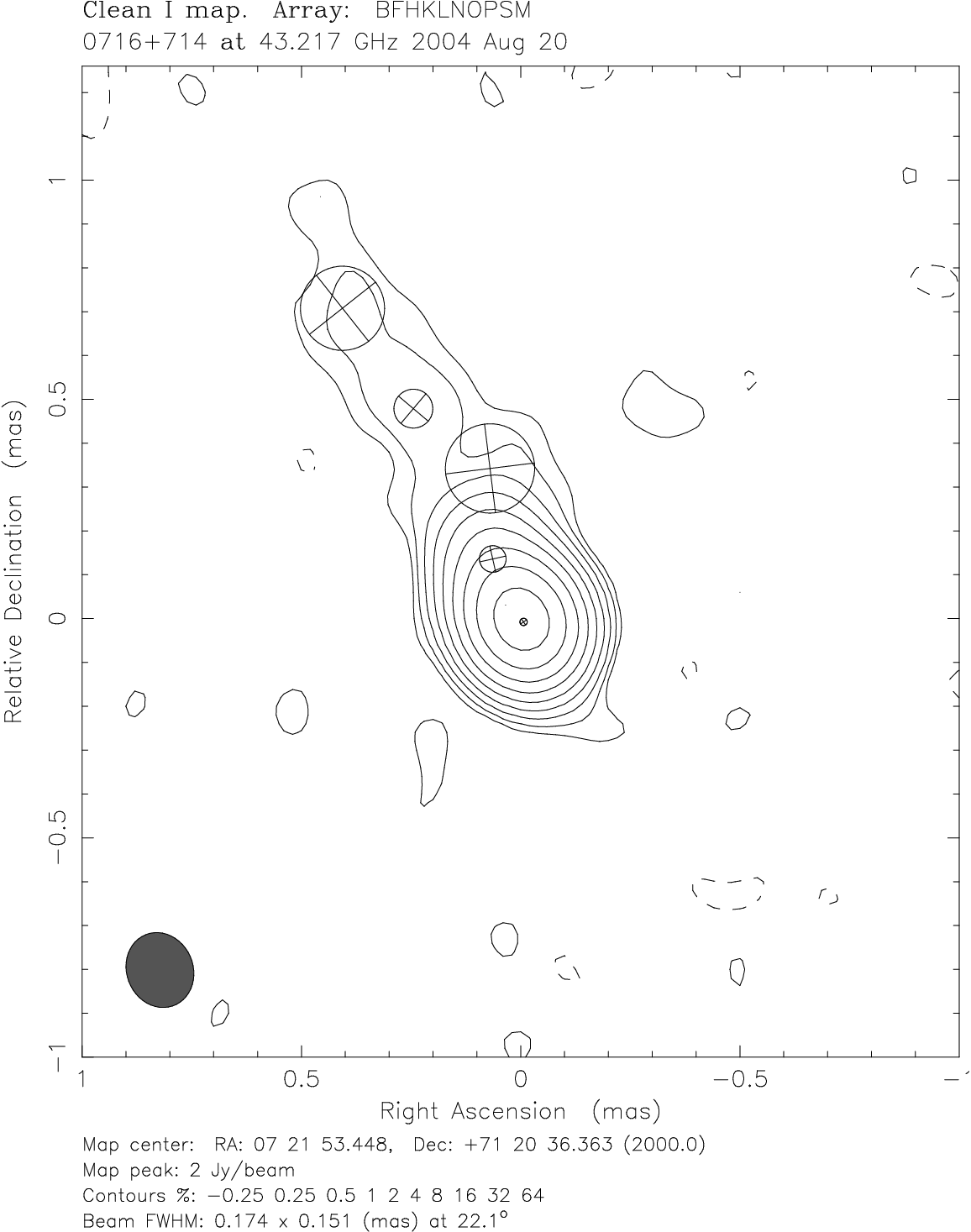}}
\subfigure[]{\includegraphics[clip,width=4cm]{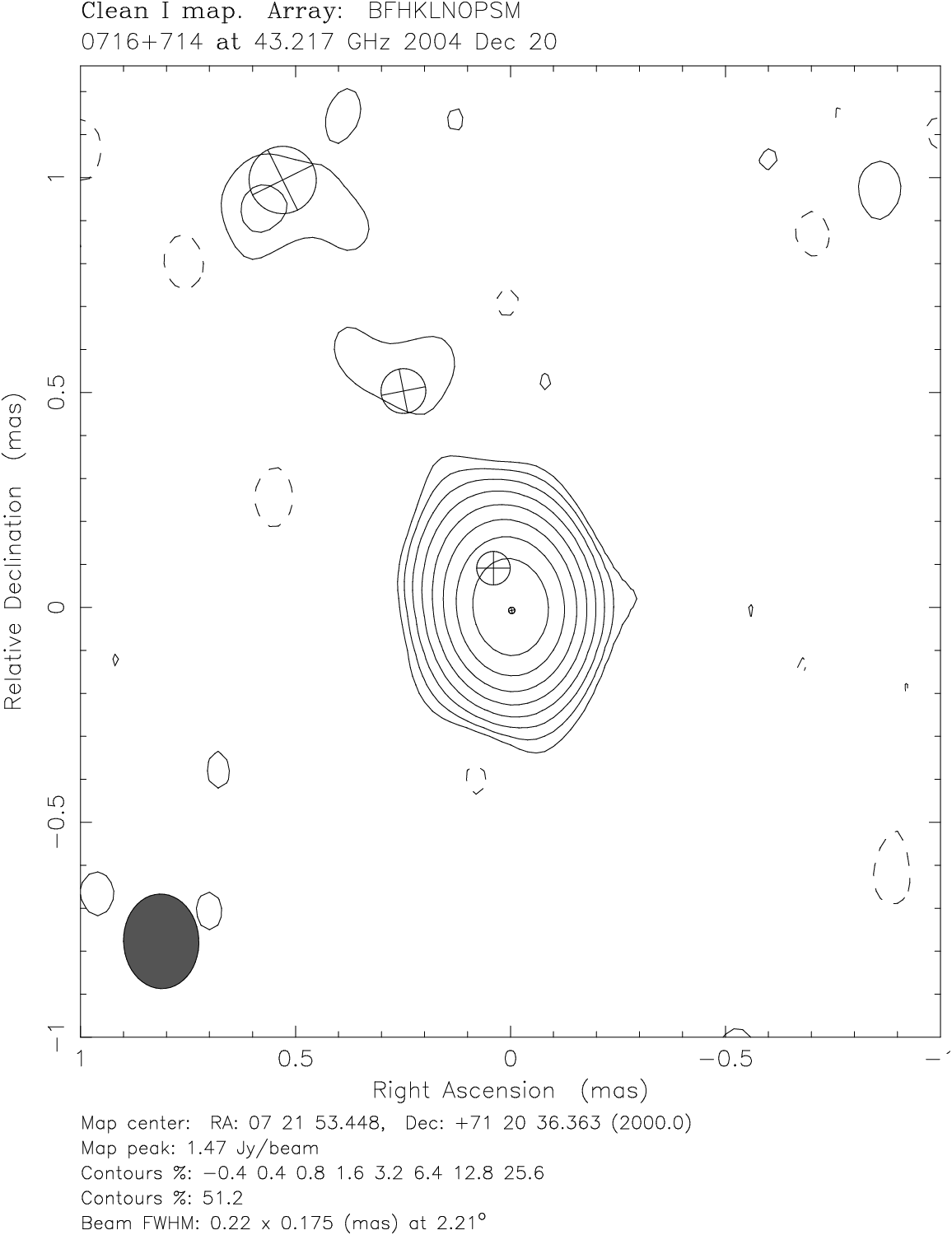}}\\
\subfigure[]{\includegraphics[clip,width=4cm]{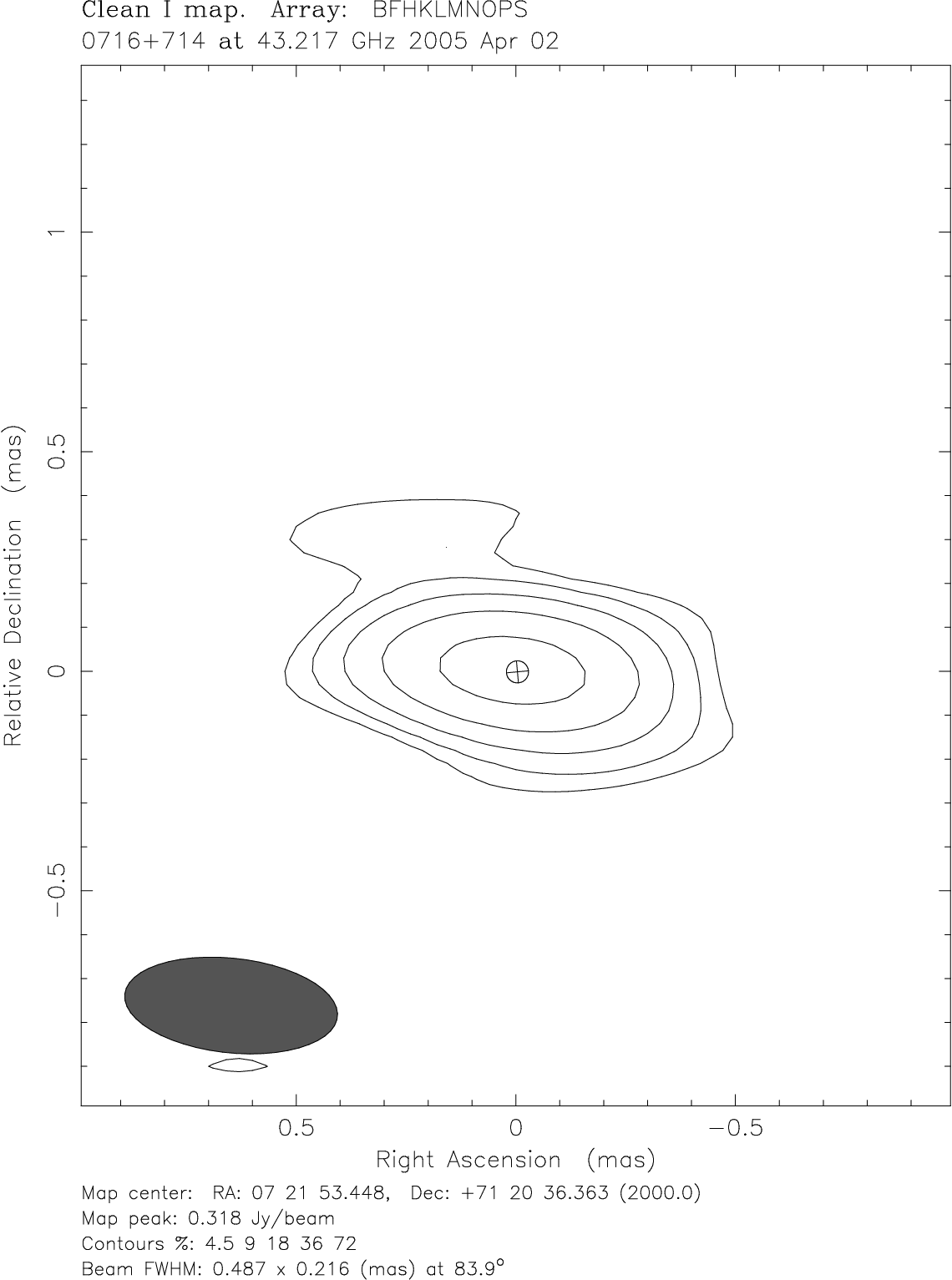}}
\end{center}
\caption{Images of S5 0716+714 at 43 GHz.}
\label{maps43}
\end{figure*}

Model-fits  have been carried out using the {\it difmap}-package (e.g., Lovell 2000) and starting with a point-like model in all of the observations. The final number of jet components necessary to fit the data adequately was reached when adding another jet component did not lead to a significant improvement in the value of chi-squared.
We aimed at obtaining the optimum number of components for each individual data set. The quality of the observations can differ from epoch to epoch and, in addition, the flux-density of the source - in particular in this case of 0716+714 - is known to vary between epochs. A brighter core component can obscure part of the jet structure, and not all jet components might be visible in this particular epoch. Vice versa, in the case of a faint core component when more jet components might be detectable. Thus, the optimum number of Gaussian components that fits the data best, differs between epochs.
The uncertainties of the model component parameters have been determined by comparing the parameter ranges obtained by performing model-fits with different numbers of model components ($\pm$1 component).
Where this was not possible, we used for the core separation r and the position angle an uncertainty of 5 \%. The uncertainties of the flux-density increase with increasing distance from the core. We accounted for this by adjusting the uncertainties accordingly: r$\leq$ 1mas: 5 \%, 1$\leq$ r$\>$ 2mas: 10 -- 20 \%, and for r$>$2mas : 20\%.
We list all parameters of all re-analyzed epochs in Table~\ref{modelfit}. We list the frequency, epoch of observation, flux-density, core separation, position angle, axis size, and component identification. In Figs.~\ref{maps5},~\ref{maps8},~\ref{maps151},~\ref{maps152},~\ref{maps221},~\ref{maps222}, and ~\ref{maps43} we show the maps with the Gaussian components superimposed.
\subsection{Component identification}
\begin{figure}[htb]
\hspace*{-4cm}{\includegraphics[clip,width=11cm,angle=-90]{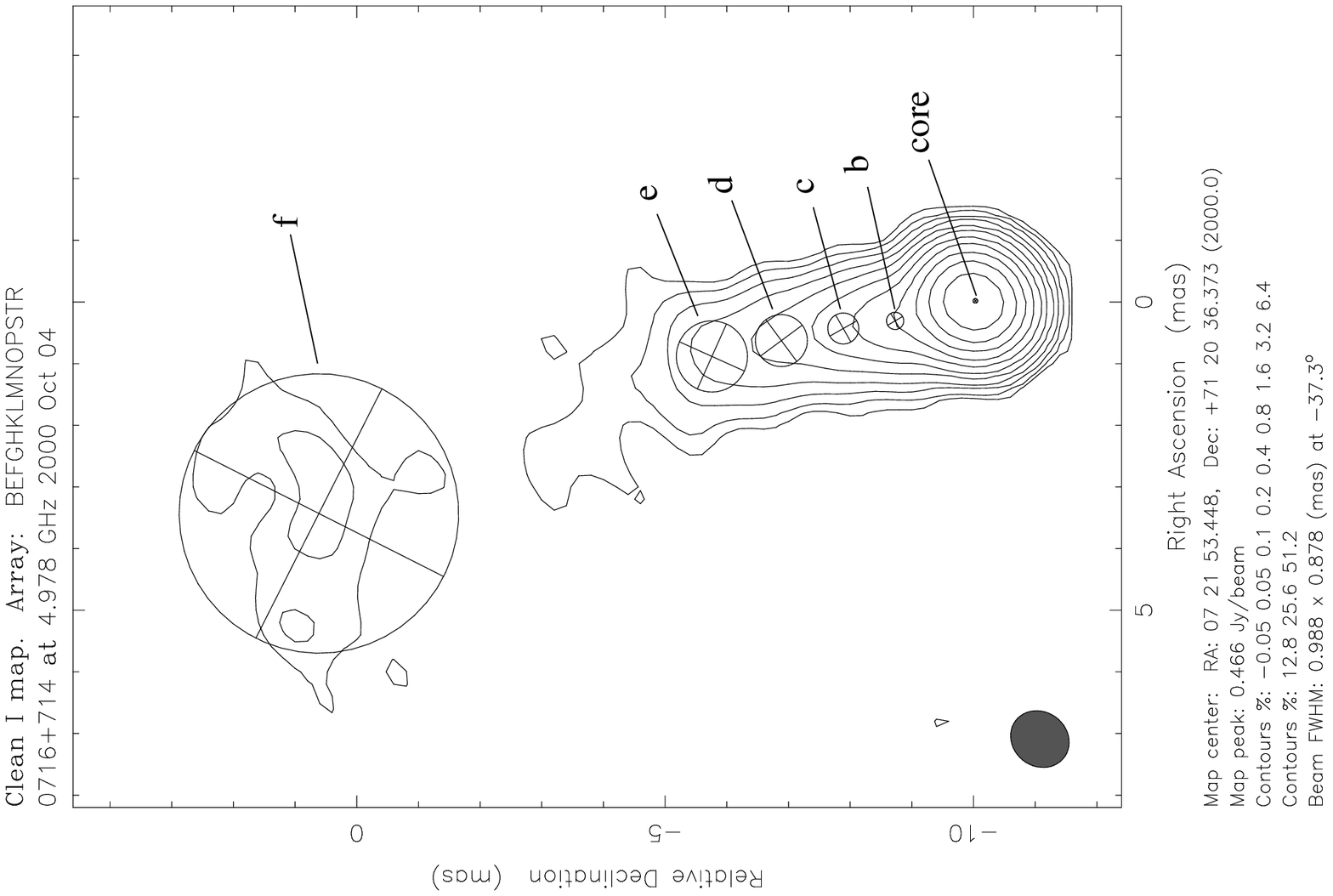}}
\caption{A map of S5 0716+714 with Gaussian model-fit components superimposed. Letters denote individual components.}
\label{identification}
\end{figure}

The component identification was carried out based on the assumption that changes of the flux density, core separation, and position angle of the modeled jet components should be small on time scales between adjacent epochs. We identified the components across the epochs by taking {\it all} model-fit parameters into account. We show an example for the component identification in Fig.~\ref{identification}. The identified model components are marked by letters which are used throughout the paper for the identification scenario presented here.
We are aware of the problems involved when tracing and identifying components through the epochs. A detailed discussion of the problems and uncertainties involved has been published by us in Britzen et al. (2008) with regard to the analysis of the kinematics of the CJF-survey (for 293 AGN). Whereas in most VLBI data analyses the small number of observations hampers a proper component identification the situation is much better in the case of 0716+714 with a comparably large number of data sets. However, 0716+714 is a rather faint AGN in the cm-regime, and the jet is even fainter. Thus the modeling of jet components is difficult, and the identification of components is even more difficult. Although a large number of VLBA observations (50) is available for the analysis, the data are -as usual in VLB analysis - unevenly sampled. In addition to this fact, VLBI data are usually of non-uniform quality. The identification across the epochs has to take differences in the number of data points, uv-coverage, data quality and resolution of the data sets into account. To prevent  a  potentially large systematic error from arising due to the incorrect cross-identification of moving features from epoch to epoch, we chose to adopt the simplest scheme when identifying the jet-features in the jet of 0716+714. We used {\it all} available model-fit parameters (flux-density, core separation, position angle, size of axis) to find the best suited set of cross-identifications. The result presented here is meant to provide the most robust and simplest identification based on the largest VLB data set currently available for this source. Stroboscopic effects can occurr in very fast sources where too few data are available. For this source we present here the analysis of the largest data set so far available and hope we can exclude stroboscopic effects. 
\begin{figure*}[htb]                                                                                              
\begin{center}                                                                                                    
\subfigure[]{\includegraphics[clip,width=5.8cm,angle=-90]{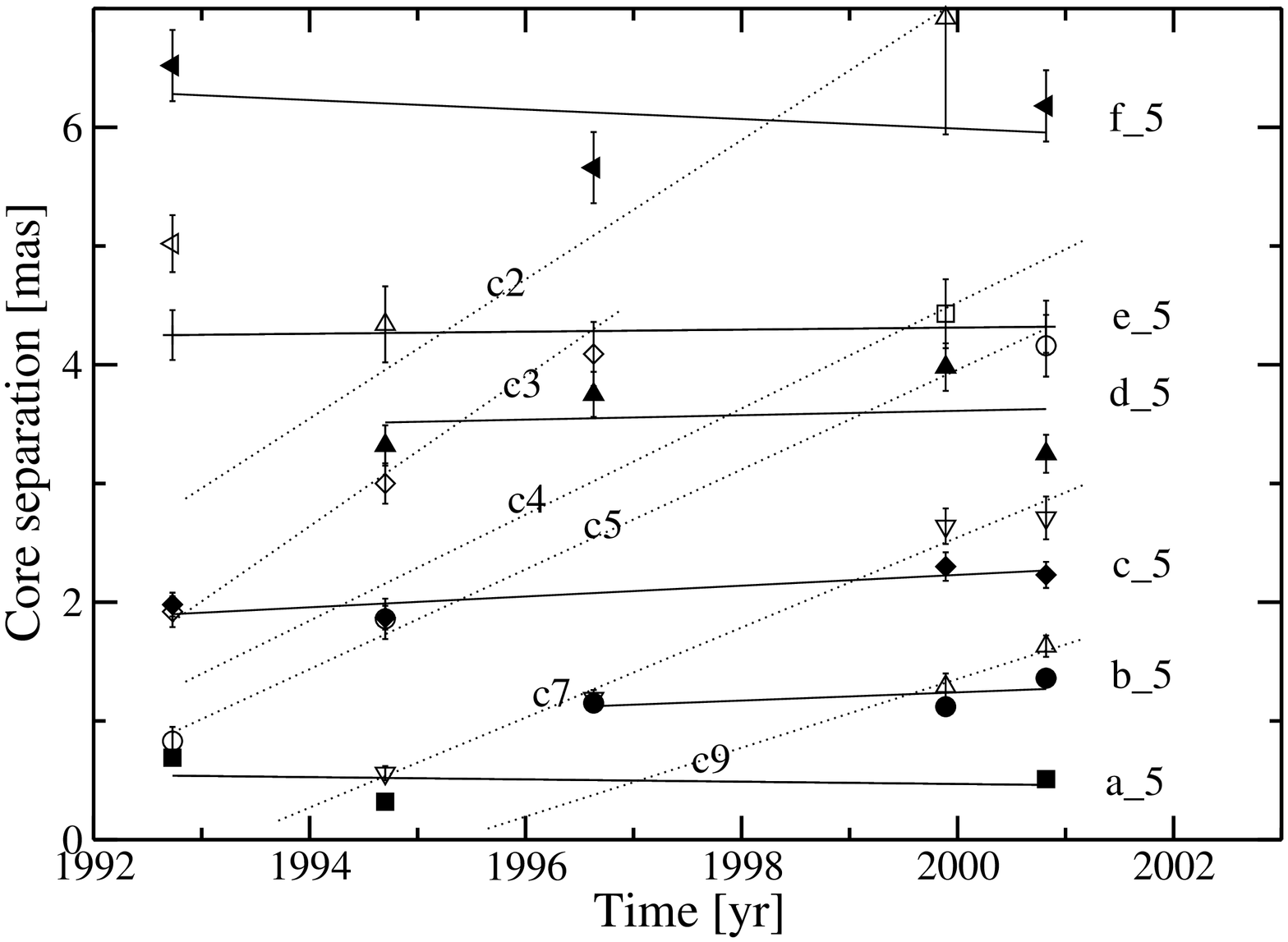}}                                
\subfigure[]{\includegraphics[clip,width=5.8cm,angle=-90]{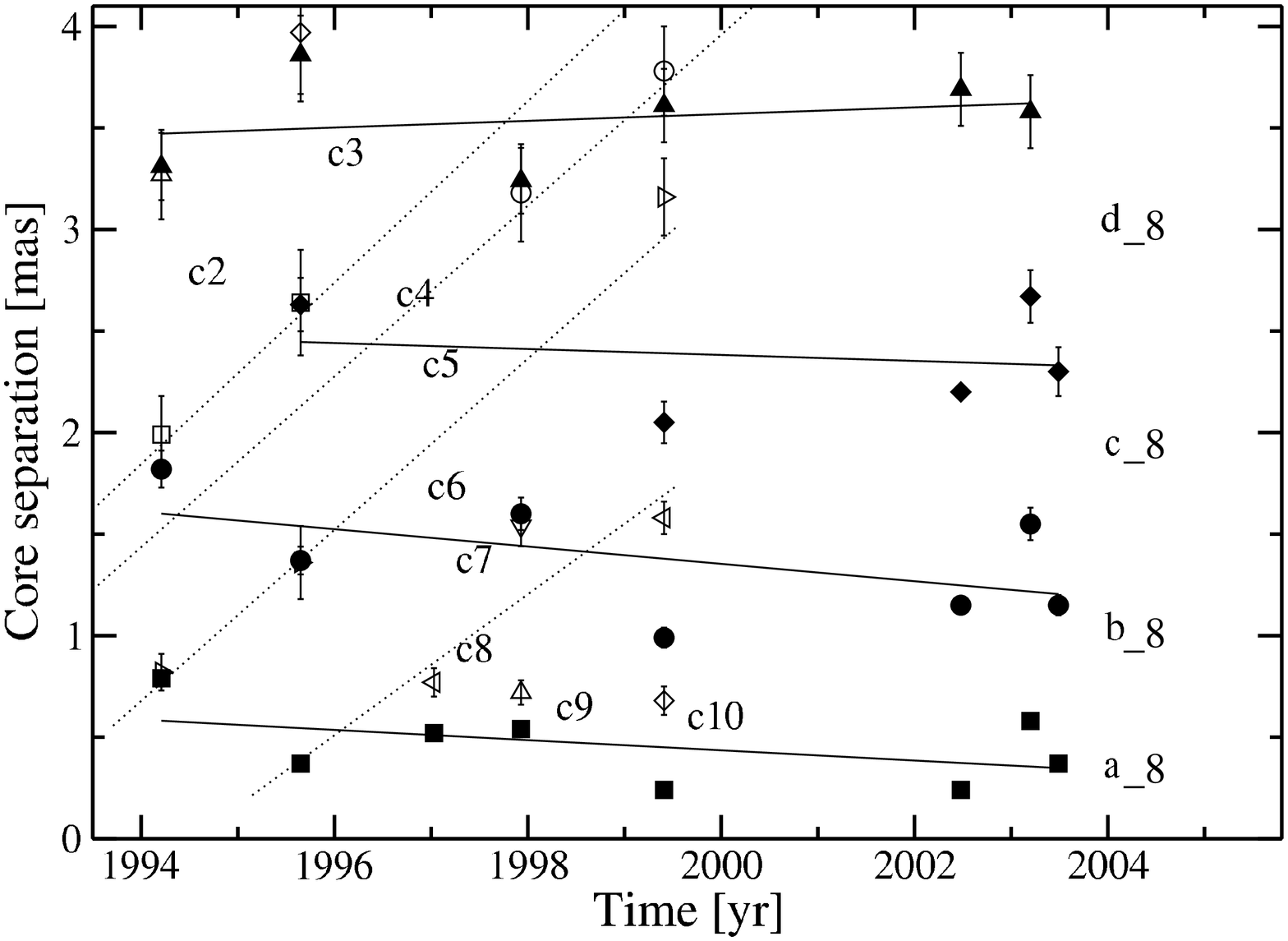}}\\                                             
\subfigure[]{\includegraphics[clip,width=5.8cm,angle=-90]{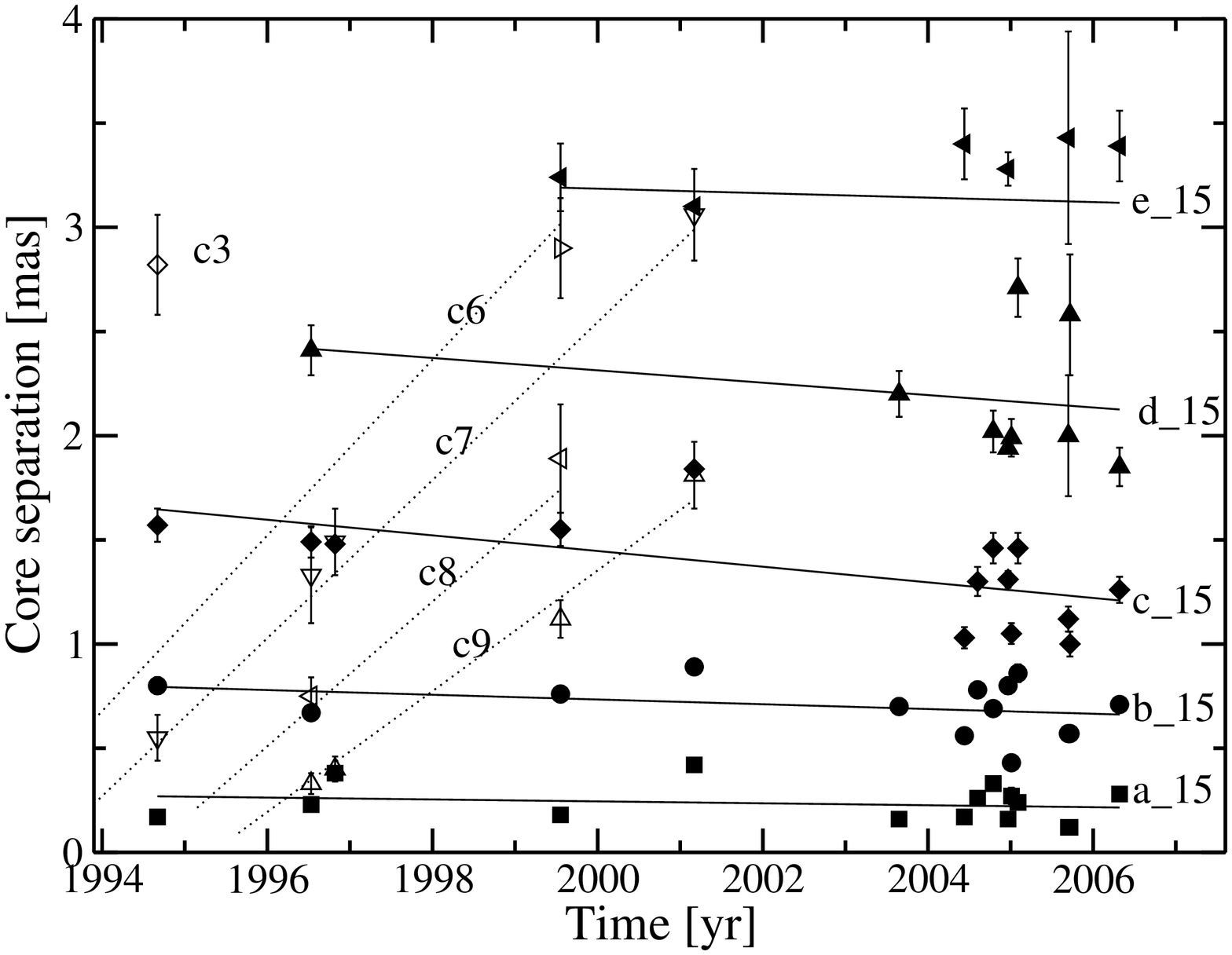}}                                            
\subfigure[]{\includegraphics[clip,width=5.8cm,angle=-90]{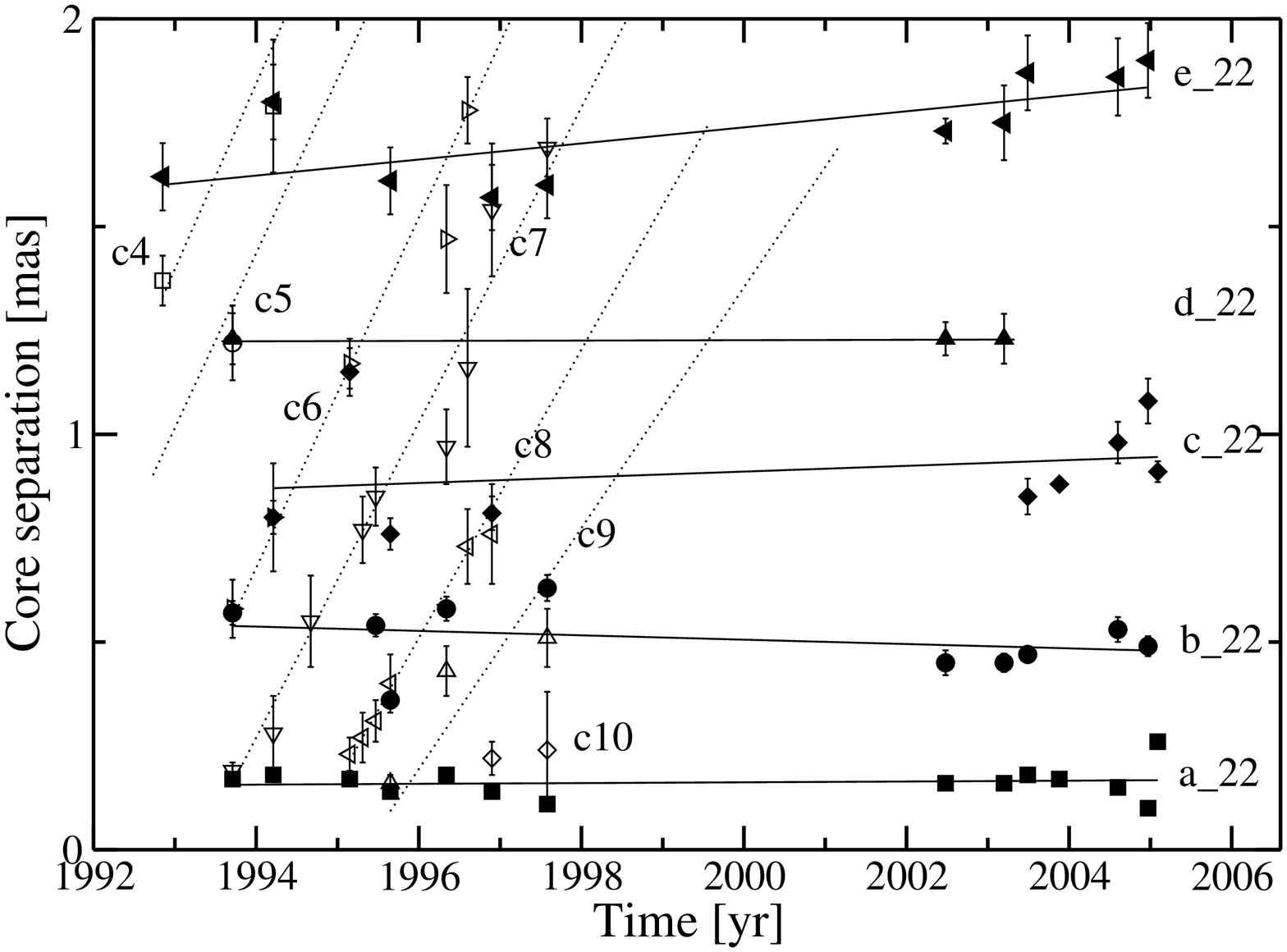}}\\                           
\subfigure[]{\includegraphics[clip,width=5.8cm,angle=-90]{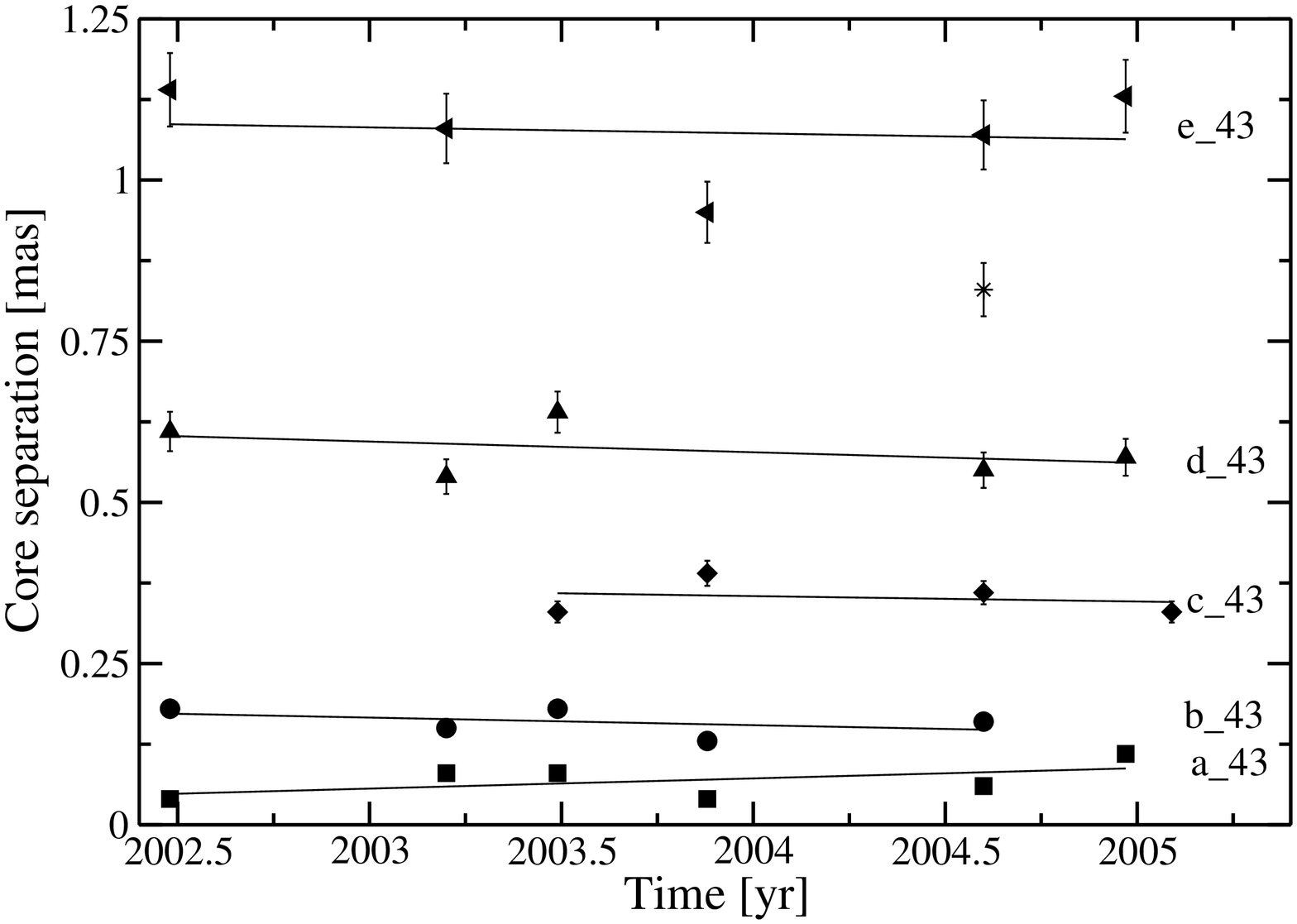}}                               
\end{center}                                                                                                      
\caption{The core separation as a function of time is shown for all frequencies investigated: (a) at 5 GHz, (b) at 8 GHz, (c) at 15 GHz, (d) at 22 GHz, and (e) at 43 GHz. We show the new identification scenario described in detail in this paper (filled symbols, linear regression is shown as solid line). In addition we show -- where possible -- the previous jet component identifications adopted by Bach et al. (2005) (open symbols, linear regression as dotted line). }                                                                                                       \label{coreseparation}                                                                                            \end{figure*}               
\begin{table*}
\caption{The table lists the average values of the reduced $\chi^{2}$-values.}
\label{chiagain}
\begin{center}
\begin{tabular}{|l|c|c|c|c|c|}
\hline
linear regression    & data &$\nu$ [GHz] &time span& $\chi^{2}_{\rm red}$&number of data points\\
\hline
`stationary scenario' & Meyer 2007 & 22  & 1995.15--1996.90&0.015&10\\
Bach et al. (2005)    & Bach et al. (2005) & &1995.15--1996.90&0.016&29\\
\hline
\end{tabular}
\end{center}
\end{table*}

\section{Results}
In the following sections, we present the results of our model-fitting analysis. S5 0716+714 reveals a core-jet structure with a rather faint jet. Tracing the jet components across the epochs allows us to investigate the kinematics in the pc-scale jet. In addition to tracing the components, we investigate the jet ridge line as a whole. This does not require any component identification and is thus independent of any identification scenarios. We first describe this evolution of the jet ridge line, and then we concentrate on the core separation behavior of the individual components as a function of time. In the next subsection we compare our identification scenario - based on the observed core separation/time relation - with results presented by other authors. We then discuss the position angle changes, calculate the apparent velocities in the standard manner (core separation/time) and taking all the observed motions into account (XY). We compare the derived values with the results presented by other authors. We then study the frequency dependence of the individual components and finally search for correlations between the kinematics and the flux-density evolution. 
\subsection{Evolution of the jet ridge line}
In Fig.~\ref{jetridgeline} we show the evolution of the jet ridge line (defined by the positions in core separation and position angle for all the components per epoch) with time at 15 GHz (a--d) and at 22 GHz (e--h). Obviously the components define rather different paths between adjacent epochs. The difference in position angle is larger than the uncertainties of the individual components. Based on these plots, it seems that the individual components do not follow a well-defined path, instead, the ridge line changes between adjacent epochs. We are currently investigating much denser sampled data (one epoch per day over one week of observations) for S5 0716+714 to investigate this effect in more detail and to determine whether this evolution is periodic.

\subsection{Core separation as function of time}
In Fig.~\ref{coreseparation}(a)--(e) we show the core separation as a function of time for all model-fitted components for all data sets from 5 -- 43 GHz. The components are shown with filled symbols. Superimposed are linear regressions (solid line) based on our component identification. In addition to our own results, the results published by Bach et al. (2005) for a subset of the data presented here are shown as well. The component identification across the observing epochs by Bach et al. is based on a combined data set composed of all data at all frequencies (5, 8, 15, and 22 GHz). The data by Bach et al. are indicated by open symbols, the linear regression by Bach et al. is shown with dotted lines. \\
The larger amount of data analyzed in this paper leads to a different component identification compared to Bach et al. We find that components tend to remain at similar core separations over the entire observing period. This behavior is found at all five frequencies. 

\begin{table*} 
\begin{center}
\caption[]{Proper motions and apparent speeds for individual components (for further explanation see section 3.4). }
\label{speeds}
\begin{tabular}{|c|c||c|c||c|c|}
\hline
$\nu$ & Comp. & $\mu$ & $\beta_{\rm app}$ (coresep.) &$\mu$ & $\beta_{\rm app}$ (rect. coord.)  \\
  $\rm [GHz]$ &  & [mas/year]& [$c$] &[mas/year] & [$c$]    \\
\hline
5.0 & a & -0.009$\pm$0.042 & -0.19$\pm$0.84 & 0.117$\pm$0.058 & 2.19$\pm$1.09  \\
      & b &  0.034$\pm$0.048 &  0.68$\pm$0.95 & 0.110$\pm$0.055 & 2.06$\pm$1.03  \\
      & c &  0.045$\pm$0.017 &  0.89$\pm$0.34 & 0.121$\pm$0.060 & 2.27$\pm$1.13  \\
      & d &  0.018$\pm$0.085 &  0.36$\pm$1.68 & 0.424$\pm$0.212 & 7.96$\pm$3.98  \\
 \hline
  8.4 & a & -0.025$\pm$0.018 & -0.49$\pm$0.37 & 0.169$\pm$0.084 & 3.17$\pm$1.58  \\
      & b & -0.042$\pm$0.030 & -0.84$\pm$0.59 & 0.205$\pm$0.102 & 3.85$\pm$1.92  \\
      & c & -0.014$\pm$0.046 & -0.29$\pm$0.91 & 0.164$\pm$0.082 & 3.08$\pm$1.54  \\
      & d &  0.016$\pm$0.031 &  0.32$\pm$0.61 & 0.276$\pm$0.138 & 5.18$\pm$2.59\\  
 \hline
 15.3 & a & -0.004$\pm$0.006 & -0.09$\pm$0.13 & 0.24$\pm$0.12 & 4.5$\pm$2.3  \\
      & b & -0.011$\pm$0.009 & -0.22$\pm$0.19 & 0.32$\pm$0.16 & 6.0$\pm$3.0  \\
      & c & -0.037$\pm$0.013 & -0.74$\pm$0.27 & 0.54$\pm$0.27 & 10.0$\pm$5.0  \\
      & d & -0.029$\pm$0.037 & -0.58$\pm$0.73 & 1.60$\pm$0.80 & 29.8$\pm$15.0  \\
 \hline
 22.2 & a &  0.001$\pm$0.002 &  0.02$\pm$0.05 & 0.043$\pm$0.021 & 0.80$\pm$0.40  \\
      & b & -0.005$\pm$0.006 & -0.10$\pm$0.12 & 0.088$\pm$0.044 & 1.65$\pm$0.82  \\
      & c &  0.006$\pm$0.010 &  0.14$\pm$0.20 & 0.269$\pm$0.134 & 0.50$\pm$0.25  \\
      & e &  0.019$\pm$0.006 &  0.38$\pm$0.13 & 0.005$\pm$0.002 & 0.09$\pm$0.04  \\
 \hline
 43.2 & a &  0.015$\pm$0.012 &  0.31$\pm$0.24 & 0.031$\pm$0.015 & 0.58$\pm$0.29  \\
      & b & -0.011$\pm$0.014 & -0.23$\pm$0.27 & 0.044$\pm$0.022 & 0.82$\pm$0.41  \\
      & c & -0.008$\pm$0.027 & -0.17$\pm$0.54 & 0.179$\pm$0.089 & 3.36$\pm$1.68  \\
      & d & -0.016$\pm$0.021 & -0.32$\pm$0.43 & 0.235$\pm$0.117 & 4.41$\pm$2.20\\  
      & e & -0.009$\pm$0.042 & -0.18$\pm$0.84 & 0.775$\pm$0.387 & 
 14.55$\pm$7.27 \\

\hline
\end{tabular}    
\end{center}
\end{table*}
\begin{table}
\begin{center}
\caption[]{Apparent velocities taken from Bach et al. (2005)}
\label{bach}
\begin{tabular}{|c|c|c|}
\hline
Id. & $\mu[mas/yr]$  & $\beta_{\rm app}$ [$c$] \\
\hline
C1  & 0.86$\pm$0.13  & 16.13$\pm$2.36        \\
C2  & 0.63$\pm$0.10  & 11.87$\pm$1.82        \\
C3  & 0.62$\pm$0.04  & 11.59$\pm$0.76        \\
C4  & 0.44$\pm$0.01  &  8.27$\pm$0.24        \\
C5  & 0.43$\pm$0.02  &  8.16$\pm$0.29        \\
C6  & 0.41$\pm$0.02  &  7.80$\pm$0.33        \\
C7  & 0.37$\pm$0.01  &  6.89$\pm$0.19        \\
C8  & 0.32$\pm$0.01  &  6.04$\pm$0.20        \\
C9  & 0.26$\pm$0.01  &  4.98$\pm$0.27        \\
C10 & 0.24$\pm$0.01  &  4.52$\pm$0.45        \\
c11 & 0.29$\pm$0.02  &  5.48$\pm$0.55        \\
\hline
\end{tabular}
\end{center}
\end{table}

\subsubsection{Comparison of goodness of fit}
In the previous section we introduced a new scenario of jet component identification for S5 0716+714. Based on this core separation/time relation we obtain different apparent velocities compared to Bach et al. These apparent velocities will be discussed in section 3.4. Here we compare how good the assumed identification (outwards moving components in the case of Bach et al. (2005) and rather stationary components as assumed in this paper) fits the distribution of the observed data. We therefore compare the reduced chi-squared values obtained when fitting these two scenarios to the data sets with the most intense sampling in order to have the most reliable data sets for a meaningful analysis:\\
- the ``stationary scenario'' for the 22 GHz data by Meyer et al. (2007) between 1995.15 and 1996.90\\ 
- the ``fast scenario'' by Bach et al. (2005) for the data from Bach et al. (2005) between 1995.15 and 1996.90\\
In Table \ref{chiagain} we list in the first column the kind of scenario we apply, the second column identifies the data reduction, the third column lists the frequency, the fourth column lists the time span covered by these observations, the fifth column gives the reduced chi-squared value and the last column contains the number of data points that entered this study. We do not list the frequency for the data of the Bach et al. paper, since data from all frequencies were taken together to identify individual components; thus, all frequencies enter the reduced chi-squared value test. This explains why the number of data points is significantly higher compared to the data set from Meyer (2007).  
Based on the assumed scenario (``fast'' or ``stationary'') we apply a linear regression to the data and measure the quality of the fit by means of the chi-squared values. The $\chi^{2}$ values were calculated as follows

\begin{equation}
\chi^{2} = \Sigma\frac{\left({\mbox{observed values - expected values}}\right)^{2}} {\mbox{expected values}}
\end{equation}

\begin{equation}
\chi^{2}_{\rm red} = \frac{\chi^{2}}{\mbox{number of data} - 1}
\end{equation}
$\chi^{2}_{\rm red}$ is the reduced $\chi^{2}$ value.

The resulting values are listed in Table \ref{chiagain}. 
We obtain a comparable (insignificantly better) chi-squared value for the ``stationary'' scenario-based linear regression for the 22 GHz data compared to the Bach et al. ``fast scenario'' for the same time range. 
We conclude that our scenario of rather stationary components fits the observed data well, and the test using the time spans with the most intensive sampling
demonstrates quantitatively that the fit is at least comparable to, if not better than, the ``fast motion scenario''.

A general comparison of the two scenarios is complicated by several factors:\\
\begin{itemize}
\item
Bach et al. (2005) determined the linear regression for individual components based on the {\it multi-frequency} dataset. All component parameters obtained at all investigated frequencies have been taken together to determine individual component motions.\\
We investigated each frequency individually.\\
\item
The time span covered by Bach et al. is significantly shorter compared to the time span that the data presented by us cover.\\
\item
The number of components in Bach et al. is much larger compared to the number of components which our identification requires.\\
\end{itemize}
\subsection{Position angle as function of time}
\begin{figure}[htb]
\begin{center}
{\includegraphics[clip,width=5.8cm,angle=-90]{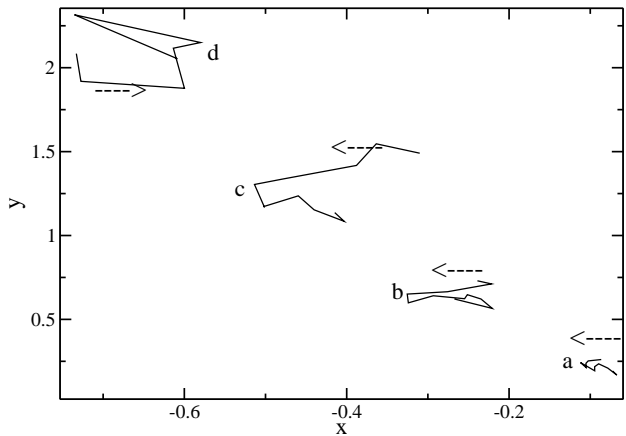}}
\end{center}
\caption{Motion of the inner four components at 15 GHz in rectangular coordinates (5-point running average for {\bf a}, 4-point running average for {\bf b} and {\bf c}, and 3-point running average for {\bf d}). The arrows indicate the direction of the motion.}
\label{xy}
\end{figure}

All jet components show significant changes with regard to the position angle. We show the motion of the inner four jet components at 15 GHz in rectangular coordinates in Fig.~\ref{xy}. We used a running average for the data of {\bf a}, {\bf b}, and {\bf c}. The arrows in Fig.~\ref{xy} indicate the direction as the paths evolve with time. To follow the position angle evolution with time completely, more densely sampled data in time are required.    

\begin{figure}[htb]
\begin{center}
\subfigure[]{\includegraphics[clip,width=4.8cm,angle=-90]{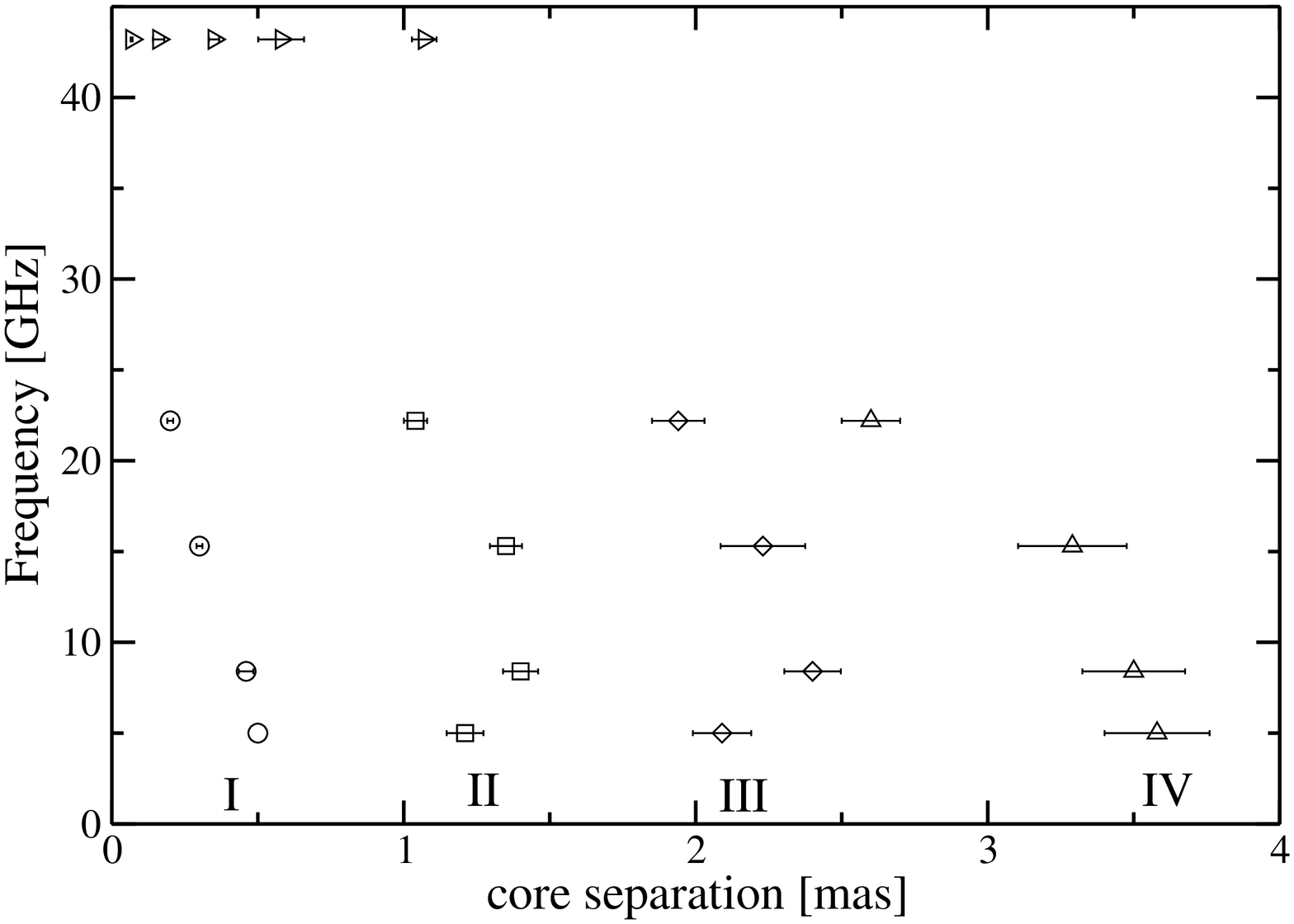}}\\
\subfigure[]{\includegraphics[clip,width=4.8cm,angle=-90]{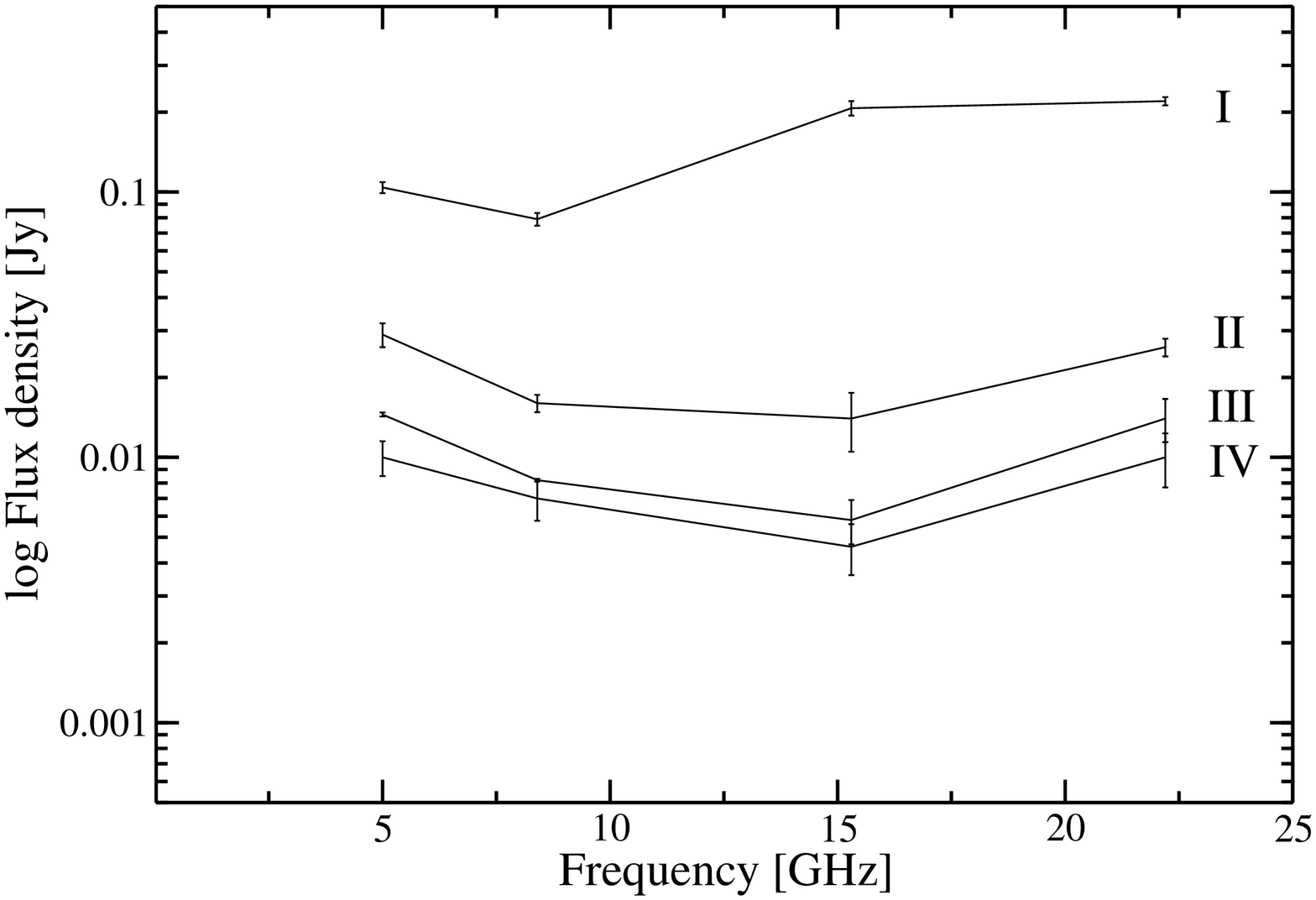}}\\
\subfigure[]{\includegraphics[clip,width=4.8cm,angle=-90]{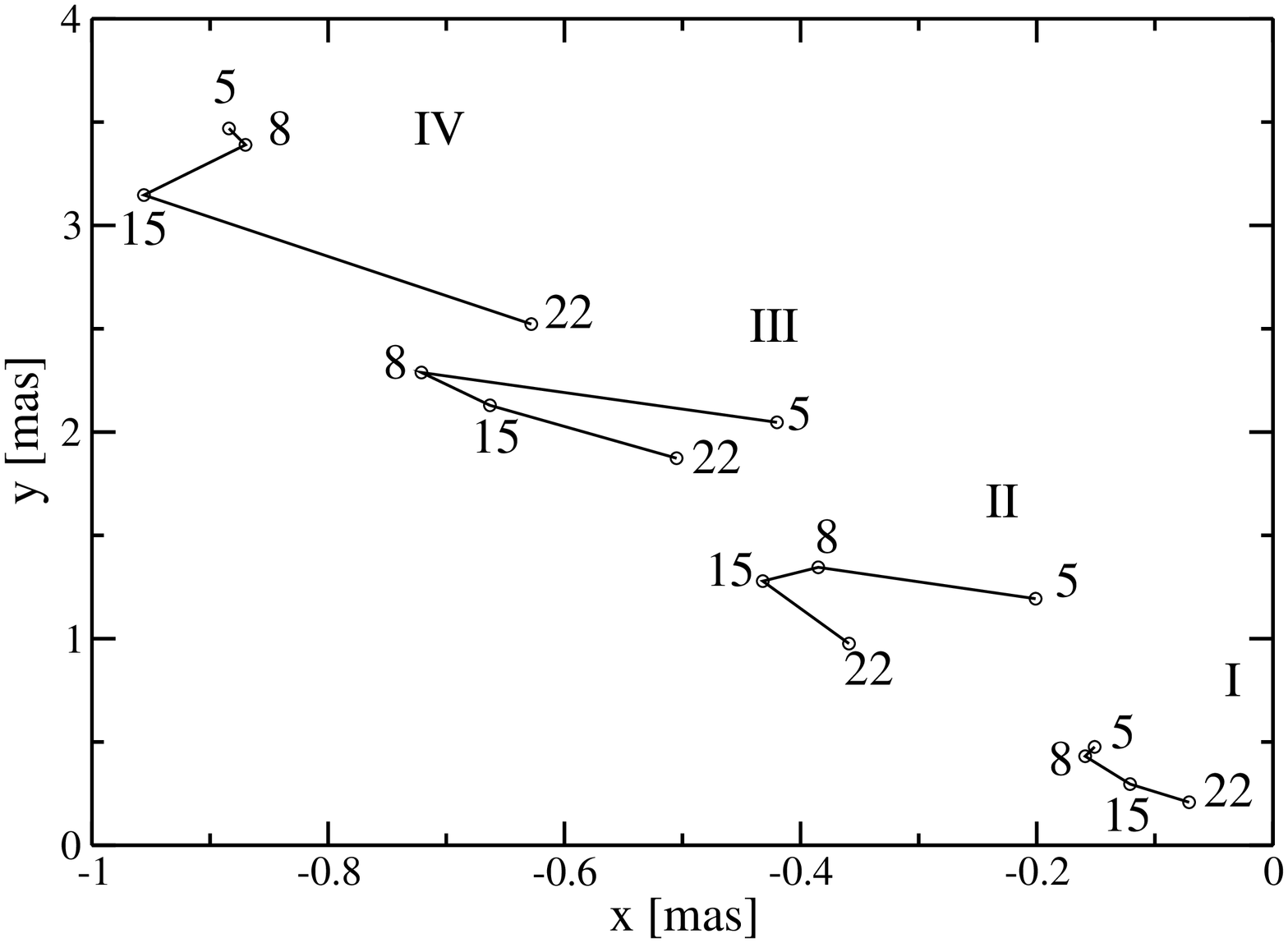}}\\
\end{center}
\caption{Identification of the components across the frequencies. (a) shows the average core separation of each component at each frequency. (b) shows the logarithm of the flux-density as function of frequency. (c) shows the position of the components according to the frequency at which they were observed in rectangular coordinates along the jet.}
 \label{spectrum}
 \end{figure}

\subsection{Apparent velocities}
In the subsections above we showed that a ``stationary'' scenario - components showing no long-term outward motion - describes our results best. 
We calculate the apparent speeds based on the observed ``stationarity'' and list them in Table~\ref{speeds}. The apparent speeds have been calculated for the inner 2.5 mas at 5.0 GHz, for the inner 3.5 mas at 8.4 GHz, for the inner 2.5 mas at 15.3 GHz, for the inner 2 mas at 22.2 GHz, and for the inner 1.3 mas at 43.3 GHz.\\
Some of the model components reveal apparent inward motion, which is not uncommon in AGN. We discuss this in the Discussion section. However, none of these values is of $3\sigma$ significance. For the apparent velocities derived for the motion based on the core separation evolution with time only, we thus exclusively find subluminal values. We list these values in Table~\ref{speeds} on the left side. The values have been calculated depending on the assumed angle to the line of sight of 4.9$^\circ$ or $\simeq$0$^\circ$, respectively. Angles to the line of sight are taken from Bach et al. (2005). \\
The fast apparent velocities found by e.g., Bach et al. (see Table~\ref{bach}, taken from Bach et al. 2005) are not confirmed by the apparent speeds we derive for the ``stationary'' components.\\
The MOJAVE 2cm survey webpage\\ (http://www.physics.purdue.edu/astro/MOJAVE/sourcepages/) quotes a number of 1.199$\pm$ 0.116 mas/y for the motion in S5 0716+714 (Lister et al., in prep.). This value is significantly higher than the values we find based on the core separation/time relation for individual jet components. \\
In columns (5) and (6) of Table~\ref{speeds}, we calculate lower limits for the proper motion and apparent speed based on the motion perpendicular to the radial motion - taking the motion with regard to the position angle into account. We estimated the distance traveled by each component in rectangular coordinates. \\ 



\subsection{Frequency dependence} 
After having identified all components independently according to core separation, position angle, flux-density, and major axis - as described in previous sections - we then traced the individual components across the frequencies. We did not {\it a priori} make any assumptions concerning frequency dependent properties. We averaged the component positions (weighted by their flux-density) and found that they can easily be grouped into the four components shown in Fig.~\ref{spectrum}(a). We there show the average core separation of each component at each frequency. The frequency identification is hampered by the different resolution at different frequencies. Thus, at 22 GHz, components close to the core can be seen, which are not visible at lower frequencies. We want to mention that - for the afore mentioned reason - at 15 GHz we grouped the components {\bf a} and {\bf b} into one component and at 22 GHz the components {\bf c} and {\bf d} into one component and {\bf e} and {\bf {\rm x}$_{1}$} and {\bf x$_{2}$} and {\bf x$_{3}$} into one component. We find that we can easily identify four components (labeled I--IV) between 5 and 15 GHz. In Fig.~\ref{spectrum}(b) we show the logarithm of the flux-density as a function of frequency for these 4 components. Fig.~\ref{spectrum}(c) shows the position of the components according to the frequency at which they were observed in rectangular coordinates along the jet.   
We can trace components at different frequencies along the jet and find that at lower frequencies we see consistently and preferrentially the right side of the jet and towards higher frequencies the left side of the jet becomes visible. 
\section{Correlation: total flux-density evolution/position-angle evolution}
\begin{figure*}[htb]                                                                                                         \begin{center}                                                                                                               \subfigure[]{\includegraphics[clip,width=7.8cm,angle=-90]{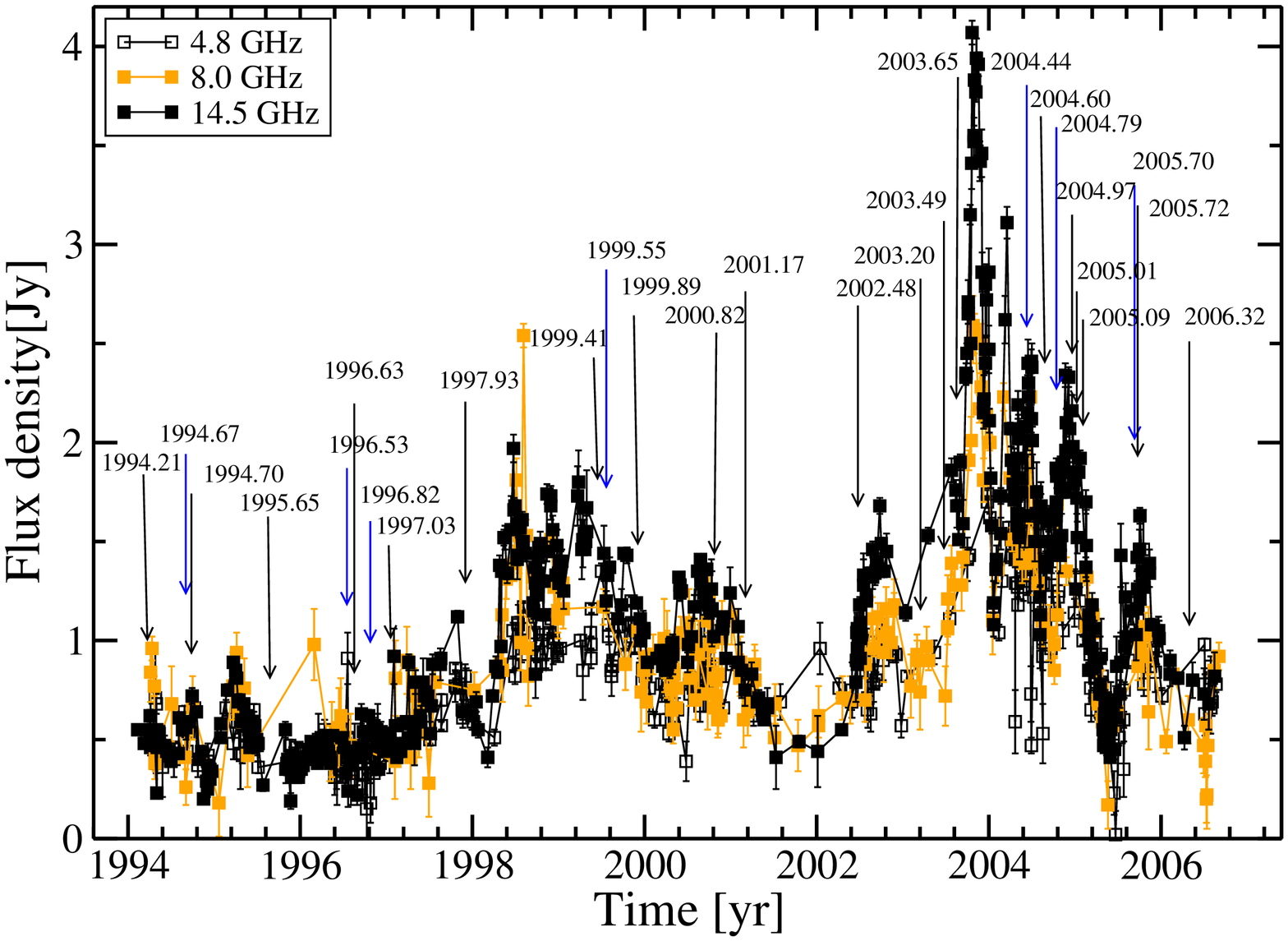}}                                                      \subfigure[]{\includegraphics[clip,width=7.5cm,angle=-90]{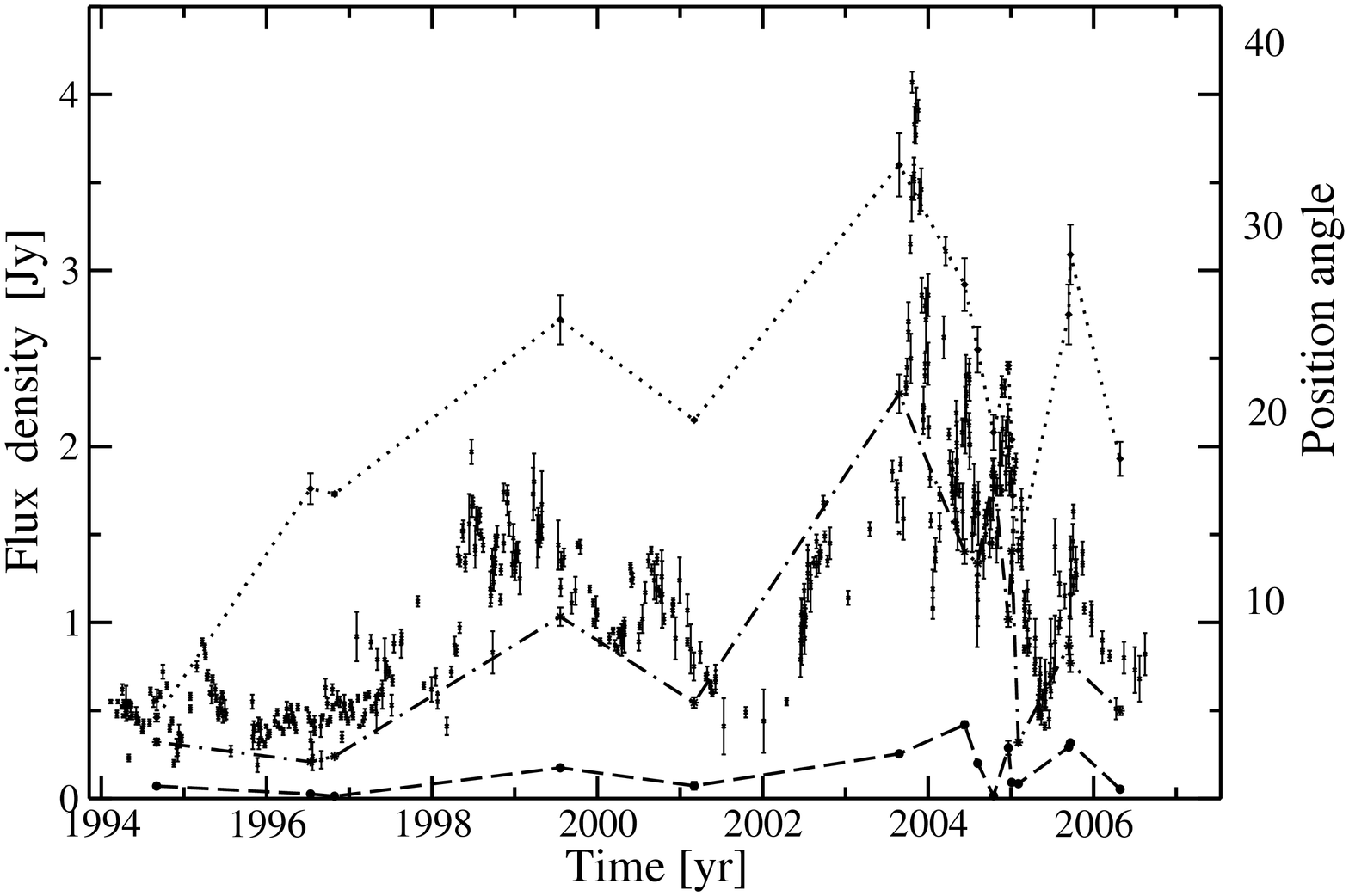}}                                                      \end{center}                                                                                                                 \caption{(a) shows the flux-density light curves obtained by the Michigan monitoring programme at three different frequencies. Superimposed are part of the epochs of the available VLBI epochs investigated in this paper. In (b) a compilation of several datasets is displayed: the total flux-density at 14.5 GHz (black dots with error bars), the flux density of the core (dot-dashed), the flux density of component a (dashed) as function of time and the position angle of component a (dotted). Most of the total flux-density is contained in the core component, as expected. }                                                    \label{a_zeit_flux.15}                                                                                                       \end{figure*}

In Fig.~\ref{a_zeit_flux.15}(a) we show the flux-density light-curves at 4.8, 8.0, and 14.5 GHz as obtained within the Michigan monitoring programme. Superimposed we show part of the epochs of the VLBI observations. This figure illustrates that morphological information is available to investigate a possible correlation between the longterm flux-density evolution and the kinematics. Although much more flux-density compared to VLBI information is available, the VLBI information covers the general trend well. In particular, maxima as well as minima in the flux-density light-curve can be accompanied by interferometric measurements.\\
In Fig.~\ref{a_zeit_flux.15}(b) we show a collection of several data sets (all at 15 GHz): the total flux density at 14.5 GHz (black dots, data), the flux density of the core (dot-dashed), the flux density of component {\bf a} (closest component to the core, dashed), and the position angle for component {\bf a} (dotted) as function of time. We find that most of the total flux density is contained in the core component - which is usual for this type of object. We also find evidence for a correlation between the total flux-density evolution with time and the position-angle evolution of component {\bf a} with time. Thus, we find a {\it correlation between flux-density variability and kinematic properties for the innermost jet component}. This means as well that the ``core'' is exhibiting   behavior similar to a jet-component. This might suggest that the longterm flux-density evolution in S5 0716+714 may have a non-negligible geometric contribution. The position angle information for {\bf a} is not sufficient to unambiguously prove a correlation, but the data consistently reproduce the same trend as seen in the total flux-density light-curve.
We plan to probe and investigate further this particular relation in more detail with more data and on shorter timescales.
\section{Discussion}
\subsection{``Fast'' versus ``Slow'' scenario}
In this paper we show that motion in the pc-scale jet of 0716+714 can be interpreted within a new kinematic scenario of non-radial motion. Components do not show the generally assumed fast apparent outward motion. Instead, stationary components with regard to the core separation are observed. In addition, significant motion perpendicular to the jet is observed.
In the following, we discuss our results and compare them to those from other studies of AGN pc-scale jet kinematics.
\subsection{Apparent stationarity and non-radial motion of jet components}
Our observations reveal a new picture of jet component motion in the BL Lac Object 0716+714: We do not find any longterm outward motion at any of the investigated frequencies. Instead, model components remain at similar core separations in the investigated time span. This is in contrast to previous studies of jet component motion in AGN, where components tend to show predominantly outward motion (e.g., Britzen et al. 2007, 2008). \\
In contrast to previous studies --- that relied mainly on investigating the core separation evolution --- we study the position-angle evolution with time as well. Although we do not find any evidence for long term outward motion, we instead find that all modelled jet components move with regard to the position angle. A similar motion phenomenon has already been observed and analyzed in the jet of 1803+784 (Kudryavtseva et al. 2007, Britzen et al. 2009).\\
In quite a number of AGN, jet components follow either the same path or a small number of different paths, which have been explained within precession scenarios (e.g., PKS 0528+134: Britzen et al. 1999). While components in general tend to follow the same jet ridge line in a number of AGN investigated in detail, the jet ridge line itself in 0716+714 shows strong evolution with time.\\ 
This can be explained as a result of precession of the base of the jet. This precession can be explained and described by a number of different models. Assuming that the core of the AGN hosts a single object (i.e. SMBH), precession can occur as a result of the Bardeen-Petterson effect (Caproni et al. 2006), magnetic torques (Lai 2003) or magnetohydrodynamic instabilities, prominently Kelvin-Helmholtz instabilities (e.g., Camenzind \& Krockenberger 1992, Hardee \& Norman 1988, Birkinshaw 1991, Zhao et al. 1992, Hardee et al. 1994, Hardee et al. 1997, Meier \& Nakamura 2006, Perucho et al. 2006).\\ 
If we assume that the AGN core hosts a binary system (presumably a binary black hole, BBH) then precession can occur as a result of the orbital motion of the binary system (e.g., Lobanov \& Roland 2005, Roos et al. 1993, Kaastra \& Roos 1992), tidal forces, gravitational torques (e.g., Katz 1997, Romero et al. 2000), as well as orbital motion of the system around the galactic gravity center (Roland et al. 2008).\\
Finally, it is possible that the bending of the jet ridge line is explained by external causes. Interaction with the surrounding intergalactic medium can cause a bending of the jet, it is, however, highly unlikely that this bending will show any periodic or quasi-periodic behaviour. Furthermore, bending of a jet ridge line can occur by gravitational interaction of the host galaxy with a nearby galaxy (Blandford \& Icke 1978, J\"agers \& de Grijp 1985, Lupton \& Gott 1982). Again, each and all the above phenomena, of external origin, can only partially but not fully explain the bending of the jet ridge line. For a more detailed discussion of the above models please see Britzen et al. (2009) and references therein.\\
We are currently working on a model to explain the quasi-stationarity within a model of non-ballistic motion (Gong 2008). 
\subsection{TeV blazars and subluminal motion}
In quite a large fraction of the TeV blazars, subluminal motion has been detected (e.g., Piner et al. 2008). The apparent pattern speeds measured in the TeV blazars are considerably slower than those measured in sources selected for their compact radio emission (e.g., CJF: Britzen et al. 2008), or for their GeV gamma-ray emission (Jorstad et al. 2001). S5 0716+714 has most recently been detected in the TeV-regime as well. The subluminal motion we find in this source thus fits nicely with the tendency for significantly slower apparent speeds in this class of objects. Piner et al. (2008) discuss the possible origin of these slow apparent motions and their connection to the TeV production mechanism. Since the rapid radio variability detected in S5 0716+714 requires quite high Doppler factors (e.g., Wagner \& Witzel 1995 and references therein), it is likely that the slow apparent speeds we see result from pattern speeds and do not reflect the speeds of the underlying bulk motion.\\ 
\subsection{Correlation between kinematics \& flux-density variability}
We find some evidence for a possible correlation between position angle changes of the innermost component and the long-term flux-density variability. This favors a geometric contribution to the origin of the longterm flux-density evolution. Further support for the geometric contribution has been discussed based on 
monitoring of the flux-density variability and studies of the light-curves:\\
The flux-density variability in the optical and in the radio band shows different timescales (Raiteri et al. 2003). Raiteri et al. find that the variation of the amplitude of the flux-density decreases with increasing wavelength and that the outbursts in the optical and radio regime are caused by the same mechanism. The flux-density in the radio regime seems to be the sum of two different spectral distributions. Raiteri et al. claim that the oscillations in both wavelength regimes have either an energetic or a geometric origin. Nesci et al. (2005) find in an optical light curve some evidence for long term variability on which short term variations
are superimposed. Indications for a precessing jet have also been found (Nesci et al.).\\
The tentative correlation that we find between the position angle changes of the innermost component and the long-term flux-density variability leads us to believe that a geometrical explanation for the long-term variability of S5 0716+714 is plausible. Camenzind \& Krochenberger (1992), first introduced the lighthouse effect, according to which the periodiocity (or quasi-periodicity) observed in the lightcurves of blazars can be attributed to the fast rotation of non-axisymmetric plasma bubbles near the central black hole. During the propagation of the bubble downstream in the jet, it experiences beaming effects, as the viewing angle changes, thus producing the outbursts in the light curves. The possible quasi-periodicity found from the light curves is characteristic of the rotation period of the plasma near the black hole, although it is significantly shorter due to relativistic effects. The above scenario requires a well collimated jet with a relatively small opening angle. In any different case the flux-density dies out quickly.\\
Schramm et al. (1993), find that the lighthouse effect is possibly responsible for the optical variability for the blazar 3C 345, although they consider that extrinsic effects e.g.
microlensing) could also contribute to the long-term variability observed. For OJ 287, Katz (1997) argues that a precessing disk consequently results in beaming effects for the jet and variability in the flux-density. For the same source, Villata et al. (1998), proposed a beaming model for which both black holes of a BBH system produce a jet. A combination of bending and beaming of the jets can produce the double-peaked lightcurve observed for that source, thus attributing the variability only to geometrical effects. Qian et al. (2007) make use of a similar model to explain the light curve of quazar 3C 345.3. Rieger (2004) also attributes variability of blazars to differential Doppler boosting effects, as the plasma follows helical trajectories, induced by different possible mechanisms (as orbital motion of a BBH system, intrinsic jet rotation and jet precession). A geometrical origin (relativistic beaming) for the variability of blazar AO 0235+16 is also considered by Raiteri et al. (2001).\\
Non geometrical phenomena can, however, also explain the long-term variability seen in the light curves of 0716+714. Most prominently, two different models have been proposed for OJ 287 which shows periodic variability in its light curves (in both the radio and the visual regime). Sillanp\"a\"a. et al. (1988) originally argued that tidal forces from the secondary black hole around the pericenter of its orbit will lead to an increased accretion rate for the primary black hole and thus result in an outburst. Lehto \& Valtonen (1996) proposed that the secondary black hole plunges into and through the accretion disk of the primary before and after the pericenter, resulting in the double outburst observed in the light curves (see also Valtonen et al. 2006). This procedure however, as argued by Valtaoja et al. (2000), is thermal and should not produce an outburst in the radio regime;  thus it seems unlikely for the case of 0716+714. In summary, we find that a geometrical explanation is a better match to the observational properties of  0716+714. 

\section{Conclusions}
In the previous sections we presented our results on the kinematics of jet component motion in S5 0716+714. We find a different scenario than previously described in the literature for this source. In particular, we find no long term outward motion for this source.  
We present a new model component motion scenario for S5 0716+714 with the following implications:\\
$\bullet$ The jet ridge line changes significantly from epoch to epoch. Components do {\it not} follow a well-defined path.\\ 
$\bullet$ In this scenario the jet components show no $3\sigma$ evidence for long term outward motion. Instead, they remain at similar core separations and seem to change significantly in position angle within the computed uncertainties. We calculate exclusively apparent subluminal speeds based on the observed core separation/time relation.\\
$\bullet$ S5 0716+714 has been detected as a TeV blazar and shows - as most of the other TeV blazars - subluminal motion. The observed apparent motions most likely result from pattern speeds.\\
$\bullet$ Taking the ``real'' motion in rectangular coordinates into account, we find significantly higher values for the apparent speeds. These values can only be limits.\\
$\bullet$ We find a possible correlation between the total flux density and the position angle of the component closest to the core ({\bf a}). This relation might suggest a non-negligible geometric contribution to the origin of the longterm flux-density variability in S5 0716+714.\\
The usually observed outward motion of jet components in AGN is --- according to our analysis --- not observable in the BL Lac object S5 0716+714. Instead, we find kinematic properties, e.g. non-radial motion, evolving jet ridge-line, etc. which have not been reported for this source so far and have only been observed in a couple of other AGN so far (e.g. Britzen et al. 2009). More observations and modelling are required to understand the physical origin of these phenomena. \\

\begin{acknowledgements}
We highly appreciate many comments and suggestions by T.P. Krichbaum on how to check the significance of the results presented here.
I. Agudo acknowledges financial support from the EU Commission for Science and Research under contract HPRN-CT-2002-00321 (ENIGMA Network). M. Karouzos was supported for this research through a stipend
from the International Max Planck Research School (IMPRS) for Radio
and Infrared Astronomy. We wish to acknowledge the 2cm Survey/MOJAVE team.
This research has made use of data from the University of Michigan Radio Astronomy Observatory which has been supported by the University of Michigan and the National Science Foundation.
This work made use of the VLBA, which is a facility of the National Science Foundation, operated under cooperative agreement by Associated Universities, Inc. We used the European VLBI Network, which is a joint facility of European, Chinese, South African and other radio astronomy institutes funded by their national research councils. This work is also based on observations with the 100m radio telescope of the MPIfR (Max-Planck-Institut f\"ur Radioastronomie) at Effelsberg.
\end{acknowledgements}


\clearpage
\pagebreak
\onecolumn
\tabcolsep0.37mm
\tablecaption{The parameters of the model-fitting.} {\smallskip}
\vspace*{-0.2cm}
\centering
\tablehead{\noalign{\smallskip} \hline \noalign{\smallskip}
$\nu$ & Epoch & S & r & $\theta$ & M.A.& Comp. \\
\multicolumn{1}{c}{\tiny [GHz]} & \multicolumn{1}{c}{} &
\multicolumn{1}{r}{\tiny ~~~~~[mJy]} &
\multicolumn{1}{r}{\tiny [mas]} & \multicolumn{1}{r}{\tiny [deg]} &
\multicolumn{1}{r}{\tiny [mas]} & \multicolumn{1}{c}{}  \\ 
\noalign{\smallskip} \hline \hline \noalign{\smallskip}}
\tabletail{\hline\multicolumn{4}{r}{\small continued in next column}\\}
\tablelasttail{\hline}
\begin{supertabular}{ccrrrrc}
\label{modelfit}
 5.0 & 1992.73 & 600$\pm$30  &     0        &  0               &  0.14$\pm$0.01  &  Core\\
     & 1992.73 & 35$\pm$2    & 0.69$\pm$0.03   & 24.1$\pm$1.2  &  0.01$\pm$0.01   &  a\\
     & 1992.73 & 12$\pm$2    & 1.98$\pm$0.10    & 7.1$\pm$0.4  &  0.01$\pm$0.01   &  c\\
     & 1992.73 & 10$\pm$2    & 4.25$\pm$0.21   & 18.0$\pm$1.0  &  0.53$\pm$0.03  & e\\
     & 1992.73 & 5$\pm$1     & 6.52$\pm$0.33   & 12.0$\pm$0.6   & 0.51$\pm$0.03  & f\\
     \hline
     & 1994.70 & 78$\pm$4    & 0          & 0          & 0.01$\pm$0.01   & Core \\
     & 1994.70 & 231$\pm$12  & 0.32$\pm$0.02   &  1.8$\pm$0.1  & 0.22$\pm$0.01  & a\\
     & 1994.70 & 14$\pm$2    & 1.87$\pm$0.10   & 15.5$\pm$0.1  &  0.77$\pm$0.04  & c\\
     & 1994.70 & 16$\pm$3    & 3.32$\pm$0.20   & 12.9$\pm$0.6   &0.89$\pm$0.04  & d\\
     \hline
     & 1996.63 & 182$\pm$10  &    0       &     0   & 0.03$\pm$0.01 & Core\\
     & 1996.63 & 22$\pm$1    &  1.15$\pm$0.06   &  6.5$\pm$0.3  & 0.01$\pm$0.01     &b\\
     & 1996.63 & 5$\pm$1     & 3.75$\pm$0.20   & 20.2$\pm$1.0   &  0.25$\pm$0.01 &  d\\
     & 1996.63 & 7.0$\pm$0.4   & 5.66$\pm$0.30   & 23.0$\pm$1.2   & 1.78$\pm$0.09  & f\\
     \hline
     & 1999.89 & 552$\pm$30  &    0            &  0        & 0.15$\pm$0.01 &  Core \\
     & 1999.89 & 40$\pm$4    & 1.12$\pm$0.06   &  8.6$\pm$0.4   & 0.01$\pm$0.01      & b\\
     & 1999.89 & 20$\pm$4    &2.30$\pm$0.11    &12.3$\pm$0.6    & 0.62$\pm$0.03 &  c\\
     & 1999.89 & 12$\pm$2    & 3.98$\pm$0.20   & 12.8$\pm$0.6   & 1.70$\pm$0.09 &  d\\
     \hline
     & 2000.82 & 446$\pm$22  &  0              &  0           & 0.07$\pm$0.01  &  Core\\
     & 2000.82 & 48$\pm$2    & 0.51$\pm$0.03   & 26.9$\pm$1.3   &  0.01$\pm$0.01    &  a\\
     & 2000.82 & 25$\pm$2    & 1.36$\pm$0.07   & 13.8$\pm$0.7   & 0.29$\pm$0.01 &  b \\
     & 2000.82 & 12$\pm$2    & 2.23$\pm$0.11   & 11.7$\pm$0.6   & 0.52$\pm$0.03  & c\\
     & 2000.82 & 7.0$\pm$0.4   & 3.25$\pm$0.16   & 11.4$\pm$0.6   & 0.85$\pm$0.04 &  d\\
     & 2000.82 & 4.0$\pm$0.2   & 4.32$\pm$0.22    &11.6$\pm$0.6   & 0.97$\pm$0.05  & e\\
     & 2000.82 & 3$\pm$1     & 6.18$\pm$0.31   & 16.9$\pm$0.8   & 2.17$\pm$0.11  & f\\
     \hline
     \hline
 8.4 & 1994.21 & 261$\pm$14  &    0           &  0          &   0.14$\pm$0.01 &  Core \\
     & 1994.21 & 26$\pm$13   & 0.79$\pm$0.04 &  13.8$\pm$0.7  &   0.45$\pm$0.02 &  a\\
     & 1994.21 & 10$\pm$15   & 1.82$\pm$0.09  &  16.7$\pm$0.8  &   0.75$\pm$0.04 &  b\\
     & 1994.21 & 12$\pm$24   & 3.31$\pm$0.17 &  12.8$\pm$0.6  &   1.15$\pm$0.06 &  d\\
     \hline
     & 1995.65 & 238$\pm$12  & 0              &   0         &   0.07$\pm$0.01 &  Core \\
     & 1995.65 & 47.00$\pm$2.35 & 0.370$\pm$0.02 &   9.9$\pm$0.5  &  0.23$\pm$0.01 &  a\\
     & 1995.65 & 7.0$\pm$0.7   & 1.37$\pm$0.07 &  15.0$\pm$0.8  &   0.49$\pm$0.02 &  b\\
     & 1995.65 & 3.0$\pm$0.6   & 2.63$\pm$0.13 &  19.4$\pm$1.0  &  0.36$\pm$0.02 &  c\\
     & 1995.65 & 4.00$\pm$0.08  & 3.86$\pm$0.19  &  15.5$\pm$0.8  &  1.05$\pm$0.05 &  d\\
     \hline
     & 1997.03 & 191$\pm$10  &   0           &         0    &   0.05$\pm$0.01 &  Core \\
     & 1997.03 & 21.00$\pm$0.11 & 0.52$\pm$0.03  &  22.5$\pm$1.1  &  0.01$\pm$0.01     &  a\\
     \hline
     & 1997.93 & 372$\pm$19  &         0      &        0    &    0.02$\pm$0.01 &  Core\\
     & 1997.93 & 31.0$\pm$1.6  & 0.54$\pm$0.03  &  10.7$\pm$0.5  &   0.36$\pm$0.02 &  a\\
     & 1997.93 & 15.0$\pm$1.5  & 1.60$\pm$0.08   &   8.5$\pm$0.4  &  0.34$\pm$0.02 &  b\\
     & 1997.93 & 7.0$\pm$1.4   & 3.24$\pm$0.16  &  15.8$\pm$0.8  &    0.86$\pm$0.04 &  d\\
     \hline
     & 1999.41 &728.0$\pm$36.4 &      0        &        0     &        0.01$\pm$0.01    &   Core \\
     & 1999.41 &233.0$\pm$11.7 &  0.24$\pm$0.01  &  13.2$\pm$0.7 &   0.12$\pm$0.01 &  a\\
     & 1999.41 & 32$\pm$16   & 0.99$\pm$0.05 &  13.6$\pm$0.7  &   0.30$\pm$0.02 &  b\\
     & 1999.41 & 11.00$\pm$0.22 & 2.05$\pm$0.10 &   9.0$\pm$0.5  &   0.45$\pm$0.02 &  c\\
     & 1999.41 & 7.0$\pm$1.4   &  3.61$\pm$0.18 &  12.6$\pm$0.6 &   0.90$\pm$0.05 &  d\\
     \hline
     & 2002.48 & 663$\pm$33  &     0     &      0          &   0.01$\pm$0.01   &  Core \\
     & 2002.48 & 152$\pm$1   &  0.24$\pm$0.01  &  27.5$\pm$0.2 &   0.15$\pm$0.01 &  a\\
     & 2002.48 & 17.0$\pm$0.5  & 1.15$\pm$0.01  &  18.2$\pm$0.1  &   0.46$\pm$0.02 &   b\\
     & 2002.48 & 7.0$\pm$0.7   & 2.20$\pm$0.01  &  19.7$\pm$0.2  &   0.52$\pm$0.03 &  c\\
     & 2002.48 & 10$\pm$2    &  3.69$\pm$0.18   &  13.6$\pm$0.1 &   1.08$\pm$0.05 &  d\\
     \hline
     & 2003.20 & 1106$\pm$60 &     0       &         0    &    0.12$\pm$0.01 &  Core \\
     & 2003.20 & 31$\pm$2    & 0.58$\pm$0.03  &   32.1$\pm$1.6  &      0.01$\pm$0.01      &  a\\
     & 2003.20 & 23$\pm$2    & 1.55$\pm$0.08  &   19.7$\pm$1.0  &   0.76$\pm$0.04 &  b\\
     & 2003.20 & 4.0$\pm$0.4   & 2.67$\pm$0.13  &   18.2$\pm$0.9  &     0.01$\pm$0.01      &  c\\
     & 2003.20 & 4.0$\pm$0.8   & 3.58$\pm$0.18  &   16.5$\pm$0.8  &    0.70$\pm$0.04 &  d\\
     & 2003.20 & 4.0$\pm$0.5   & 4.52$\pm$0.02  &   22.0$\pm$0.9  &   0.42$\pm$0.02 &  e\\
     \hline
     & 2003.49 & 1518$\pm$80 &     0        &         0     &  0.02$\pm$0.01 &   Core \\
     & 2003.49 & 96$\pm$14   & 0.37$\pm$0.04  &   33.4$\pm$0.3  &     0.01$\pm$0.01       &  a\\
     & 2003.49 & 11$\pm$1    & 1.15$\pm$0.05  &   20.9$\pm$0.4  &   0.01$\pm$0.01 &   b\\
     & 2003.49 & 16$\pm$1    & 2.30$\pm$0.12  &   21.4$\pm$0.4  &   1.20$\pm$0.06 &  c\\ 
     & 2003.49 & 6$\pm$2     & 4.35$\pm$0.06  &   23.2$\pm$1.2  &   2.06$\pm$0.10 &  e\\
     & 2003.49 & 3.0$\pm$0.5   & 6.82$\pm$0.14  &    4.2$\pm$1.3  &   1.25$\pm$0.06 &  x\\
     \hline
     \hline
15.3 & 1994.67 & 321$\pm$20  &     0        &     0        &  0.05$\pm$0.01  & Core \\
     & 1994.67 & 71$\pm$4    & 0.17$\pm$0.01 &  4.6$\pm$0.2    &    0.01$\pm$0.01      & a \\
     & 1994.67 & 20$\pm$1    & 0.80$\pm$0.04  & 34.0$\pm$1.7    &  0.84$\pm$0.04  & b \\
     & 1994.67 & 9.0$\pm$0.9   & 1.57$\pm$0.08  & 11.9$\pm$0.6    &  0.58$\pm$0.03  & c \\
     \hline
     & 1996.53 & 208$\pm$10  &  0         &   0             &  0.05$\pm$0.01  & Core \\
     & 1996.53 & 26.0$\pm$1.3  & 0.23$\pm$0.01  & 17.6$\pm$0.9    &    0.01$\pm$0.01       & a \\   
     & 1996.53 & 19$\pm$1    & 0.67$\pm$0.034 & 6.2$\pm$0.3     & 0.28$\pm$0.01  &  b \\
     & 1996.53 & 3.0$\pm$0.3   & 1.49$\pm$0.08 & 12.6$\pm$0.6    &  0.31$\pm$0.02  & c \\
     & 1996.53 & 1.0$\pm$0.2   & 2.41$\pm$0.12  & 18.6$\pm$0.9    &  0.48$\pm$0.02  & d \\
     & 1996.53 & 2.0$\pm$0.4   & 4.15$\pm$0.21  & 16.3$\pm$0.8    &  0.67$\pm$0.04  & x \\
     \hline
     & 1996.82 & 240$\pm$2   &     0     &     0            & 0.01$\pm$0.01      & Core \\
     & 1996.82 & 13$\pm$1    & 0.38$\pm$0.02  & 17.3$\pm$0.9    &    0.01$\pm$0.01        & a \\
     & 1996.82 & 8.0$\pm$0.8   & 1.48$\pm$0.01  & 11.0$\pm$0.4    &  0.53$\pm$0.03  & c \\
     \hline
     & 1999.55 & 1033$\pm$52 &     0      &    0             &  0.07$\pm$0.01  & Core \\
     & 1999.55 & 174$\pm$9   & 0.18$\pm$0.01 &  27.2$\pm$1.4   &    0.01$\pm$0.01       & a \\
     & 1999.55 & 22$\pm$1    & 0.76$\pm$0.04  &  13.6$\pm$0.7   &  0.09$\pm$0.01  & b \\
     & 1999.55 & 16.0$\pm$1.6  & 1.55$\pm$0.08  &  11.5$\pm$0.6   &  0.36$\pm$0.02  & c \\
     & 1999.55 & 5$\pm$1     & 3.24$\pm$0.16 &  17.8$\pm$0.9   &  0.36$\pm$0.02  & e \\
     \hline
     & 2001.17 & 545$\pm$30  &    0        &    0           &  0.05$\pm$0.01  & Core \\
     & 2001.17 & 72$\pm$22   & 0.42$\pm$0.02 &  21.5$\pm$1.0   &  0.16$\pm$0.01  & a \\
     & 2001.17 & 2.0$\pm$0.5   & 0.89$\pm$0.02 &  16.8$\pm$0.3  &   0.01$\pm$0.01      & b \\
     & 2001.17 & 12$\pm$3    & 1.84$\pm$0.01 &  17.0$\pm$0.1   &  0.61$\pm$0.03  & c \\
     & 2001.17 & 5.0$\pm$1.5   & 3.10$\pm$0.01  &  13.3$\pm$0.7  &  0.69$\pm$0.03  & e \\
     & 2001.17 & 2.0$\pm$0.5   & 5.20$\pm$0.02 &  17.4$\pm$0.9  &   0.01$\pm$0.01     & x \\
     \hline
     & 2003.65 & 2299$\pm$110&    0       &    0            &  0.02$\pm$0.01  & Core \\
     & 2003.65 & 254$\pm$13  & 0.16$\pm$0.01 &  36.0$\pm$1.8   &  0.10$\pm$0.01  & a \\
     & 2003.65 & 15.0$\pm$0.8  & 0.70$\pm$0.04  &  32.7$\pm$1.6   &  0.33$\pm$0.02  & b \\
     & 2003.65 & 10$\pm$2    & 2.20$\pm$0.11  &  18.1$\pm$0.9   &  1.03$\pm$0.05  & d \\
     \hline
     & 2004.44 & 1403$\pm$70 &   0       &    0             &    0.01$\pm$0.01      & Core \\
     & 2004.44 & 419$\pm$20  & 0.17$\pm$0.01 &  29.2$\pm$1.5   &  0.07$\pm$0.01  & a \\
     & 2004.44 & 84.0$\pm$4.2  & 0.56$\pm$0.03 &  31.1$\pm$1.6   &  0.18$\pm$0.01  & b \\
     & 2004.44 & 25$\pm$3    & 1.03$\pm$0.05 &  24.2$\pm$1.2  &  0.25$\pm$0.01  & c \\
     \hline
     & 2004.60 & 1336$\pm$70 &      0       &     0         &   0.01$\pm$0.01           & Core \\
     & 2004.60 & 200$\pm$10  & 0.26$\pm$0.01 &  25.5$\pm$1.3  &    0.01$\pm$0.01      & a \\
     & 2004.60 & 13$\pm$7    & 1.30$\pm$0.07  &  29.9$\pm$1.5  &  0.31$\pm$0.02  & c \\
     \hline
     & 2004.79 & 1841$\pm$90 &    0         &  0           &  0.07$\pm$0.04      & Core \\ 
     & 2004.79 & 15.0$\pm$0.8  & 0.33$\pm$0.02 &  20.8$\pm$1.0  &    0.01$\pm$0.01          & a \\
     & 2004.79 & 100$\pm$5   & 0.69$\pm$0.04 &  21.4$\pm$1.0  &  0.44$\pm$0.02  & b \\
     & 2004.79 & 8.0$\pm$0.8   & 1.46$\pm$0.07 &  17.8$\pm$0.9  &  0.31$\pm$0.02  & c \\
     & 2004.79 & 3.0$\pm$0.5   & 2.02$\pm$0.10  &  21.7$\pm$1.1  &  0.81$\pm$0.04  & d \\
     \hline
     & 2004.97 & 1018$\pm$43 &       0      &     0        &  0.01$\pm$0.02      & Core \\
     & 2004.97 & 288$\pm$40  & 0.16$\pm$0.01  &  24.6$\pm$0.2  &  0.07$\pm$0.01  & a \\
     & 2004.97 & 51.0$\pm$0.5  & 0.80$\pm$0.04  &  18.6$\pm$1.0  &  0.32$\pm$0.02  & b \\
     & 2004.97 & 17$\pm$3    & 1.31$\pm$0.04  &  21.9$\pm$0.3  &  0.09$\pm$0.01  & c \\
     & 2004.97 & 2.0$\pm$0.5   & 1.94$\pm$0.03 &  22.8$\pm$0.4  &  0.01$\pm$0.01  & d \\
     & 2004.97 & 3.0$\pm$0.2   & 3.28$\pm$0.08  &  33.4$\pm$1.7  &  0.01$\pm$0.01  & e \\
     & 2004.97 & 2$\pm$1     & 4.16$\pm$0.02  &  22.8$\pm$0.1  &  0.01$\pm$0.01  & x \\
     \hline
     & 2005.01 & 1406.0$\pm$0.5&      0       &    0         &  0.05$\pm$0.01      & Core \\
     & 2005.01 & 91$\pm$4    & 0.27$\pm$0.04  &  20.4$\pm$0.1  &    0.01$\pm$0.01          & a \\
     & 2005.01 & 16$\pm$8    & 0.43$\pm$0.01  &  18.3$\pm$0.2  &  0.11$\pm$0.01  & b \\
     & 2005.01 & 46.00$\pm$0.05 & 1.05$\pm$0.05  &  23.8$\pm$1.2  &  0.44$\pm$0.02  & c \\
     & 2005.01 & 5.00$\pm$0.03  & 1.99$\pm$0.09  &   8.7$\pm$0.4  &  0.02$\pm$0.01  & d \\
     \hline
     & 2005.09 & 319$\pm$16  &      0       &     0        &    0.01$\pm$0.01           & Core \\
     & 2005.09 & 84.0$\pm$4.2  & 0.24$\pm$0.01 &  14.1$\pm$0.7  &  0.15$\pm$0.01  & a \\
     & 2005.09 & 30.0$\pm$1.5  & 0.86$\pm$0.04 &  25.0$\pm$1.3  &  0.36$\pm$0.02  & b \\
     & 2005.09 & 12.0$\pm$1.2  & 1.46$\pm$0.07 &  19.2$\pm$1.0  &  0.10$\pm$0.01  & c \\
     & 2005.09 & 12$\pm$2    & 2.71$\pm$0.14  &  16.9$\pm$0.8  &  1.65$\pm$0.08  & d \\
     & 2005.09 & 2.0$\pm$0.6   & 5.48$\pm$0.3   &   8.3$\pm$0.4  &  0.01$\pm$0.01  & x \\
     \hline
     & 2005.70 & 868$\pm$50  &        0    &       0      &  0.03$\pm$0.01  & Core \\
     & 2005.70 & 292$\pm$12  & 0.12$\pm$0.01 &   27.5$\pm$1.7 &  0.05$\pm$0.01  & a \\
     & 2005.70 & 98.0$\pm$5.5  & 0.57$\pm$0.03 &   18.9$\pm$0.8 &  0.22$\pm$0.01  & b \\
     & 2005.70 & 9.0$\pm$22.5  & 1.12$\pm$0.06  &  19.2$\pm$0.4  &  0.22$\pm$0.01  & c \\
     & 2005.70 & 5$\pm$2     & 2.00$\pm$0.29  &  19.0$\pm$0.9  &  0.68$\pm$0.03  & d \\
     & 2005.70 & 5.0$\pm$1.5   & 3.43$\pm$0.51  &  16.1$\pm$0.8  &  0.96$\pm$0.05  & e \\
     \hline
     & 2005.72 & 769$\pm$50  &       0      &        0     &    0.01$\pm$0.01          & Core \\
     & 2005.72 & 87.0$\pm$5.5  & 0.57$\pm$0.03 &  20.5$\pm$0.8  &  0.21$\pm$0.01  & b \\
     & 2005.72 & 14.0$\pm$2.5  & 1.00$\pm$0.06  &  19.9$\pm$0.4  &    0.01$\pm$0.01    & c \\
     & 2005.72 & 9$\pm$2     & 2.58$\pm$0.29  &  17.3$\pm$0.9  &  1.44$\pm$0.07 & d \\
     \hline
     & 2006.32 & 498.0$\pm$24.9&   0           &      0       &  0.04$\pm$0.01 & Core \\
     & 2006.32 & 53.0$\pm$2.6  & 0.280$\pm$0.002 &  19.3$\pm$1.0  &  0.10$\pm$0.01  & a \\
     & 2006.32 & 10.0$\pm$0.5  & 0.71$\pm$0.01 &  26.9$\pm$1.3  &  0.01$\pm$0.01  & b \\
     & 2006.32 & 5.0$\pm$0.5   & 1.26$\pm$0.06 &  21.9$\pm$1.1  &  0.12$\pm$0.01  & c \\
     & 2006.32 & 13.0$\pm$1.9  & 1.85$\pm$0.09 &  12.7$\pm$0.6  &  0.61$\pm$0.03  & d \\
     & 2006.32 & 2.0$\pm$0.4   & 3.39$\pm$0.17 &   4.4$\pm$0.2  &    0.01$\pm$0.01    & e \\
     \hline
     \hline
22.2 & 1992.85 &700$\pm$35   &    0         &   0          & 0.13$\pm$0.01   & Core\\
     & 1992.85 & 20$\pm$3    & 1.62$\pm$0.08 &  7.2$\pm$0.4   &  0.01$\pm$0.01   & e\\
     & 1992.85 & 10$\pm$2    & 2.42$\pm$0.12 & 17.9$\pm$0.9   & 0.01$\pm$0.01 & x2\\
     \hline
     & 1993.71 & 248.0$\pm$1.2 &   0          &   0          &  0.01$\pm$0.01            & Core\\
     & 1993.71 & 99$\pm$5    &  0.170$\pm$0.009 & 16.7$\pm$0.8 & 0.10$\pm$0.01  &a\\
     & 1993.71 & 17.0$\pm$0.9  &  0.57$\pm$0.03 & 15.3$\pm$0.8 & 0.13$\pm$0.01 &b\\
     & 1993.71 & 16$\pm$16   &  1.23$\pm$0.06 & 20.3$\pm$1.0 & 0.24$\pm$0.01 &d\\
     \hline
     & 1994.21 & 252.0$\pm$1.3 &   0             &  0           &  0.01$\pm$0.01         & Core\\
     & 1994.21 & 44$\pm$22   & 0.18$\pm$0.01   &22.1$\pm$1.1  &  0.01$\pm$0.01         &a\\
     & 1994.21 & 13.0$\pm$0.7  & 0.80$\pm$0.04   & 18.2$\pm$0.9  & 0.37$\pm$0.02 &c\\
     & 1994.21 &  6.0$\pm$0.9  &  1.80$\pm$0.09  &  15.2$\pm$0.8  &  0.47$\pm$0.02 &e\\
     \hline
     & 1995.15 & 677.0$\pm$33.9&   0           &  0            & 0.05$\pm$0.01 & Core\\
     & 1995.15 & 72$\pm$36   & 0.17$\pm$0.01  & 4.1$\pm$0.2    &0.18$\pm$0.01 &a\\
     & 1995.15 & 4.00$\pm$0.04  & 1.15$\pm$0.06  &13.6$\pm$0.7    &  0.01$\pm$0.01   &c\\
     & 1995.15 & 5$\pm$1     & 2.63$\pm$0.13  &   7.9$\pm$0.4  &  0.01$\pm$0.01  &x3\\
     & 1995.15 & 8$\pm$16    &  3.64$\pm$0.18  & 16.2$\pm$0.8   & 0.14$\pm$0.01 &y\\
     \hline
     & 1995.47 & 173$\pm$9   & 0            &  0            &  0.04$\pm$0.01 & Core\\
     & 1995.47 & 11.0$\pm$0.6  & 0.54$\pm$0.03  & 17.2$\pm$0.9   &  0.01$\pm$0.01 & b\\
     & 1995.47 &7$\pm$14     & 2.12$\pm$0.11  & 15.5$\pm$0.8   & 0.01$\pm$0.01 & x1\\
     \hline
     & 1995.65 &174.0$\pm$8.7  & 0           &  0           &0.06$\pm$0.01 & Core\\
     & 1995.65 & 123.0$\pm$6.2 & 0.14$\pm$0.01  & 10.6$\pm$0.5   & 0.06$\pm$0.01  & a\\
     & 1995.65 & 28$\pm$14   & 0.36$\pm$0.02  & 12.1$\pm$0.6   & 0.23$\pm$0.01 &b\\
     & 1995.65 &4.0$\pm$0.2    & 0.76$\pm$0.04  & 10.2$\pm$0.5   &  0.01$\pm$0.01  &c\\
     & 1995.65 & 5.0$\pm$0.8   & 1.61$\pm$0.08 & 17.3$\pm$0.9    &0.43$\pm$0.02 &e\\
     \hline
     & 1996.34 & 178$\pm$9   &  0          &   0    & 0.01$\pm$0.01 & Core\\
     & 1996.34 & 57.0$\pm$0.3  & 0.18$\pm$0.01  & 10.5$\pm$0.5    &0.01$\pm$0.01 &a\\
     & 1996.34 & 24$\pm$12   & 0.58$\pm$0.03  &  5.1$\pm$0.3   &0.15$\pm$0.01 &b\\
     & 1996.34 & 13.0$\pm$2.6  & 1.99$\pm$0.10 & 23.2$\pm$1.2  & 1.23 $\pm$0.06 &x1\\
     \hline
     & 1996.90 & 218$\pm$11  &  0             &   0            &  0.01$\pm$0.01  & Core\\
     & 1996.90 & 61.0$\pm$3.1  & 0.14$\pm$0.01  & 21.5$\pm$1.1  & 0.13$\pm$0.01  &a\\
     & 1996.90 & 7.00$\pm$0.35  & 0.81$\pm$0.04  &10.8$\pm$0.5   &  0.09$\pm$0.01 & c\\
     & 1996.90 & 6.0$\pm$0.9   & 1.57$\pm$0.08  &10.9$\pm$0.5   & 0.31$\pm$0.01  &e  \\   
     \hline
     & 1997.58 &765.0$\pm$38.3 &   0             &   0           &  0.01$\pm$0.01 & Core\\
     & 1997.58 &164.0$\pm$8.2  & 0.11$\pm$0.01 & 40.7$\pm$2.0    & 0.09$\pm$0.01 & a\\
     & 1997.58 & 22.0$\pm$1.1  & 0.63$\pm$0.03 &  1.6$\pm$0.1   & 0.21$\pm$0.01 & b\\
     & 1997.58 & 4$\pm$3     & 1.60$\pm$0.08   & 11.6$\pm$0.6   &  0.01$\pm$0.01 & e\\
     & 1997.58 & 4.0$\pm$0.8   & 2.21$\pm$0.11 &  2.0$\pm$0.1   &  0.01$\pm$0.01  &x1\\
     \hline
     & 2002.48 & 1343$\pm$12 &  0            &   0                 & 0.01$\pm$0.01  & Core\\
     & 2002.48 & 161$\pm$8   & 0.16$\pm$0.01  & 35.6$\pm$0.2   & 0.09$\pm$0.01  & a\\
     & 2002.48 & 18$\pm$4    & 0.45$\pm$0.03   & 24.2$\pm$1.2   & 0.25$\pm$0.01  & b\\
     & 2002.48 & 9.0$\pm$0.5   & 1.23$\pm$0.04   & 19.0$\pm$0.8   & 0.29$\pm$0.01  & d\\
     & 2002.48 & 3.0$\pm$0.5   & 1.73$\pm$0.03   & 18.5$\pm$0.1   & 0.01$\pm$0.01  &e\\
     & 2002.48 & 3.0$\pm$0.2   &  2.52$\pm$0.03  &  22.2$\pm$4.9  &  0.24$\pm$0.01  & x2\\
     \hline
     & 2003.20 & 1342.0$\pm$6.7&   0          &      0         & 0.02$\pm$0.01 & Core\\
     & 2003.20 & 159$\pm$8   &  0.16$\pm$0.01  & 35.8$\pm$1.8  &  0.09$\pm$0.01  &a\\
     & 2003.20 & 20$\pm$1    &  0.45$\pm$0.02 &  24.3$\pm$1.2  &  0.26$\pm$0.01  &b\\
     & 2003.20 & 9.0$\pm$0.9   &  1.23$\pm$0.06   & 18.3$\pm$0.9  &  0.31$\pm$0.02 & d\\
     & 2003.20 & 3.0$\pm$0.3   & 1.75$\pm$0.09   & 17.9$\pm$0.9  &   0.01$\pm$0.01 & e\\
     & 2003.20 & 4.0$\pm$0.6   &  2.76$\pm$0.14  &   9.7$\pm$0.5  &  0.16$\pm$0.01 & x3\\
     \hline
     & 2003.49 & 2819$\pm$2  &   0             &    0          &   0.01$\pm$0.01          & Core\\
     & 2003.49 & 128$\pm$1   &  0.18$\pm$0.01 &  37.9$\pm$1.9  &  0.04$\pm$0.01 & a\\
     & 2003.49 & 5.0$\pm$0.3   &  0.47$\pm$0.02  &  21.4$\pm$1.0  &   0.01$\pm$0.01 & b\\
     & 2003.49 & 8.0$\pm$0.4   &  0.85$\pm$0.04  & 33.7$\pm$1.7  &  0.04$\pm$0.01 & c\\
     & 2003.49 & 5.0$\pm$0.5   &  1.87$\pm$0.09  &  20.1$\pm$1.0  &  0.25$\pm$0.01 & e\\
     & 2003.49 & 5.0$\pm$0.8   &  2.57$\pm$0.13  &   5.6$\pm$0.3  &  0.22$\pm$0.01 &x2\\
     \hline
     & 2003.88 &2309$\pm$115 &   0           &    0           &  0.04$\pm$0.01 & Core\\
     & 2003.88 &804$\pm$75   &  0.17$\pm$0.01  &30.5$\pm$1.6  & 0.10$\pm$0.01  & a\\
     & 2003.88 & 14.0$\pm$0.5  &  0.88$\pm$0.01  & 31.3$\pm$1.6  &  0.38$\pm$0.02 & c\\
     \hline
     & 2004.60 & 1444$\pm$72&   0         &   0           &  0.02$\pm$0.01 & Core\\
     & 2004.60 & 328$\pm$16  &  0.15$\pm$0.01 &  27.3$\pm$1.4  &  0.05$\pm$0.01 & a\\
     & 2004.60 & 72$\pm$4    &  0.53$\pm$0.03  &  24.6$\pm$1.2   & 0.21$\pm$0.01 & b\\
     & 2004.60 & 30$\pm$2    &  0.98$\pm$0.05  &  28.4$\pm$1.4  &  0.20$\pm$0.01 & c\\
     & 2004.60 & 4.0$\pm$0.4   &  1.86$\pm$0.10  & 27.7$\pm$1.4  &   0.01$\pm$0.01 & e\\
     & 2004.60 & 3.0$\pm$0.6   &  2.86$\pm$0.14  &  21.2$\pm$1.1  &   0.01$\pm$0.01 & x3\\
     \hline
     & 2004.97 & 907$\pm$45  &   0        &    0             & 0.02$\pm$0.01 & Core\\
     & 2004.97 & 475$\pm$24  &  0.10$\pm$0.01  & 26.7$\pm$1.3   & 0.06$\pm$0.01 & a\\
     & 2004.97 & 45$\pm$2    &  0.49$\pm$0.02 &   1.4$\pm$0.1 &  0.22$\pm$0.01 & b\\
     & 2004.97 &29$\pm$3     &  1.08$\pm$0.05 &  25.1$\pm$1.3  &  0.35$\pm$0.02 & c\\
     & 2004.97 & 4.0$\pm$0.4   &  1.90$\pm$0.09  &  21.5$\pm$1.1  &   0.01$\pm$0.01 & e\\
     \hline
     & 2005.09 &376$\pm$20   &   0      &    0              &  0.06$\pm$0.01 & Core\\
     & 2005.09 &41.0$\pm$0.5   &  0.26$\pm$0.01  &  8.5$\pm$0.2  & 0.16$\pm$0.01 & a\\
     & 2005.09 &31$\pm$2     &  0.91$\pm$0.03  & 20.5$\pm$0.2  & 0.64$\pm$0.03 & c\\
     \hline
     \hline
43.2 & 2002.48 &1484.0$\pm$74.2&   0            &    0         &   0.01$\pm$0.01  & Core\\
     & 2002.48 &329.0$\pm$16.5 &  0.040$\pm$0.002  & 51.2$\pm$2.6   &   0.03$\pm$0.01 &  a\\
     & 2002.48 & 130.0$\pm$6.5 &  0.18$\pm$0.01 & 34.7$\pm$1.7    &   0.09$\pm$0.01 &  b\\
     & 2002.48 & 9.0$\pm$4.5   &  0.61$\pm$0.03 &22.1$\pm$1.1     &  0.22$\pm$0.01 &  d\\
     & 2002.48 & 4.0$\pm$0.4   &  1.14$\pm$0.06   & 7.1$\pm$0.4    &  0.10$\pm$0.01 &  e\\
     \hline
     & 2003.20 &1853.0$\pm$92.7&   0           &      0        &    0.01$\pm$0.01        & Core\\
     & 2003.20 & 211.0$\pm$10.6&  0.080$\pm$0.004   &70.2$\pm$3.5   &   0.03$\pm$0.01 & a\\
     & 2003.20 & 127.0$\pm$6.4 &  0.15$\pm$0.01  &38.4$\pm$1.9   &   0.13$\pm$0.01 & b\\
     & 2003.20 & 6.0$\pm$0.3   &  0.54$\pm$0.03   & 27.4$\pm$14   &   0.24$\pm$0.01 &  d\\
     & 2003.20 & 4.0$\pm$0.4   &  1.08$\pm$0.05   & 31.1$\pm$1.6   &   0.01$\pm$0.01 &  e \\
     \hline
     & 2003.49 &3310.0$\pm$16.6&    0           &      0        &   0.01$\pm$0.01 &  Core\\
     & 2003.49 & 294.0$\pm$14.7&  0.080$\pm$0.004   &41.4$\pm$2.1   &   0.01$\pm$0.01 &  a\\
     & 2003.49 & 86.0$\pm$4.3  &  0.18$\pm$0.01    &47.8$\pm$2.4   &   0.10$\pm$0.01 &  b\\
     & 2003.49 & 9.00$\pm$0.45  &  0.33$\pm$0.02   &45.9$\pm$2.3   &    0.01$\pm$0.01 &  c\\
     & 2003.49 & 4.0$\pm$0.2   &  0.64$\pm$0.03  &52.5$\pm$2.6    &    0.01$\pm$0.1 &  d\\
     \hline
     & 2003.88 &2072.0$\pm$10.4&   0           &       0      &   0.02$\pm$0.01 &  Core\\
     & 2003.88 &1636.0$\pm$81.8&  0.040$\pm$0.002  & 12.2$\pm$0.6     &  0.01$\pm$0.01 &  a\\
     & 2003.88 & 570.0$\pm$28.5&  0.13$\pm$0.07  & 29.4$\pm$1.5   &  0.09$\pm$0.01 &  b\\
     & 2003.88 & 57.0$\pm$2.85 &  0.39$\pm$0.02  & 29.9$\pm$1.5   &   0.26$\pm$0.01 &  c\\
     & 2003.88 & 4.0$\pm$0.2   &  0.95$\pm$0.05  &28.2$\pm$1.4   &   0.19$\pm$0.01 &  e\\
     \hline
     & 2004.60 &1871.0$\pm$93.6&   0          &    0          &   0.02$\pm$0.01 &  Core\\
     & 2004.60 & 200$\pm$10  &  0.060$\pm$0.003  &46.8$\pm$2.3   &    0.01$\pm$0.01  &  a\\
     & 2004.60 & 217.0$\pm$10.9&  0.16$\pm$0.01  & 26.0$\pm$1.3   &   0.06$\pm$0.01 &  b\\
     & 2004.60 & 33.0$\pm$1.7  &  0.36$\pm$0.02  &12.3$\pm$6.2    &   0.20$\pm$0.01 &  c\\
     & 2004.60 & 22.0$\pm$1.1  &  0.55$\pm$0.03  & 27.3$\pm$1.4    &  0.09$\pm$0.01 &  d\\
     & 2004.60 & 27.0$\pm$1.4  &  0.83$\pm$0.04  & 29.9$\pm$1.5    &  0.19$\pm$0.01 &  x\\
     & 2004.60 & 4.0$\pm$0.4   &  1.07$\pm$0.05  & 25.5$\pm$1.3    &   0.01$\pm$0.01 &  e\\
     \hline
     & 2004.97 &1382.0$\pm$69.1&   0          &  0              &  0.01$\pm$0.01 &  Core\\
     & 2004.97 & 197.0$\pm$9.9 &  0.11$\pm$0.01  & 23.4$\pm$1.2   &  0.08$\pm$0.01 &  a\\
     & 2004.97 & 12.0$\pm$0.6  &  0.57$\pm$0.03  &26.2$\pm$1.3    &  0.10$\pm$0.01 &  d\\
     & 2004.97 & 18.0$\pm$1.8  &  1.13$\pm$0.06  & 28.0$\pm$1.4    &  0.16$\pm$0.01 &  e\\
     \hline
     & 2005.09 & 325.0$\pm$16.3&   0          &  0           &  0.05$\pm$0.01 &  Core\\
     & 2005.09 & 12.0$\pm$0.6  &  0.33$\pm$0.02   & 29.7$\pm$1.5   &   0.01$\pm$0.01  & c\\
     \hline
\end{supertabular}
\end{document}